\documentstyle[draft]{article}

\input tcilatex

\begin{document}

\author{D.\,D.\,Rabounski and L.\,B.\,Borissova$^*$}

\title{\huge Particles Here and Beyond the Mirror}
\date{}
\maketitle

\thispagestyle{empty}

\begin{center}
\vspace{-6pt}
\parbox{13cm}{\small
\parindent=0.55cm
This study looks at motion of particles using mathematical methods of
chronometric invariants (physical observable values in General Relativity). 
It is shown that aside for
mass-bearing particles and light-like particles ``zero-particles'' can exist
in fully degenerated space-time (zero-space). For a regular observer
zero-particles move instantly, thus transferring long-range action. Further
we show existence of two separate areas in unhomogeneous space-time, where
observable time flows into future and into past, while this duality is not
found in homogeneous space-time. These areas are referred to as our world,
where time flows into future and as the mirror Universe, where time flows in
past. The areas are separated with a space-time membrane, referred to as
zero-space, where observable time stops.\par
This is an electronic version of the book: 
Rabounski D.\,D.\ and Borissova L.\,B. Particles here and beyond the Mirror.
Editorial URSS, Moscow, 2001.\par  \vspace*{-9pt}
\begin{tabbing}
{\footnotesize ISBN 5--8360--0394--7}\`{\footnotesize PACS: 04.20.--q, 04.20.Cv}
\end{tabbing}

}

\vspace*{7pt}
\parbox{11.5cm}{
\centerline{\large\bf Contents}

\vspace{-4pt}
{\small 
\begin{tabbing} 
\hspace{0.5cm}\=\kill 
Foreword\rule[-5pt]{0pt}{0pt}\`{\pageref{fore}} \\ 
\>1\'\,Introduction\`{\pageref{1}} \\ 
\>2\'\,Physical observable values\`{\pageref{2}}\\ 
\>3\'\,Motion of mass-bearing and massless particles\`{\pageref{3}} \\ 
\>4\'\,Degenerated space-time. Zero-particles\`{\pageref{4}} \\ 
\>5\'\,Generalized space-time for particles of three kinds\`{\pageref{5}} \\ 
\>6\'\,Equations of motion \\ 
\>\'\makebox[24pt][r]{6.1}\, General considerations\`{\pageref{6-1}} \\ 
\>\'\makebox[24pt][r]{6.2}\, Generalized space-time\`{\pageref{6-2}} \\ 
\>\'\makebox[24pt][r]{6.3}\, Strictly non-degenerated space-time\`{\pageref{6-3}} \\ 
\>\'\makebox[24pt][r]{6.4}\, Specific case: geodesic equations\`{\pageref{6-4}} \\ 
\>\'\makebox[24pt][r]{6.5}\, Specific case: Newton laws\`{\pageref{6-5}} \\ 
\>\'\makebox[24pt][r]{6.6}\, Resume\`{\pageref{6-6}} \\ 
\>7\'\,Analysis of the equations \\ 
\>\'\makebox[24pt][r]{7.1}\, Space-time and zero-space: limit transitions\`{\pageref{7-1}} \\ 
\>\'\makebox[24pt][r]{7.2}\, Space-time asymmetry and world beyond 
the Mirror\`{\pageref{7-2}} \\ 
\>8\'\,Conditions of direct and reverse flow of time\`{\pageref{8}} \\ 
\>9\'\,Basic introduction into the mirror world\`{\pageref{9}} \\ 
\>10\'\,Motion of particles as a result of space's motion \\ 
\>\'\makebox[24pt][r]{10.1}\, Problem statement\`{\pageref{10-1}} \\ 
\>\'\makebox[24pt][r]{10.2}\, Equations of motion and Killing equations\`{\pageref{10-2}} \\ 
\>\'\makebox[24pt][r]{10.3}\, Calculating mass of the Galaxy\`{\pageref{10-3}} \\ 
\>\'\makebox[24pt][r]{10.4}\, Resume\`{\pageref{10-4}} \\ 
\>11\'\,Who is a super-light observer?\`{ \pageref{11}} \\ 
\>12\'\,World of black holes\`{\pageref{12}} \\ 
\>13\'\,Geometric structure of zero-space\`{\pageref{13}} \\ 
\>14\'\,Conclusions\rule[-5pt]{0pt}{0pt}\`{\pageref{14}} \\ 
Epilogue\`{\pageref{epi}} \\
References\`{\pageref{ref}} \\
Reference expressions\`{\pageref{exp}} \\
\end{tabbing} 
}

}

\parbox{13cm}{\footnotesize
\raisebox{-0.4pt}{\rule{50mm}{0.4pt}}\\ \noindent \rule{3mm}{0pt}
{\footnotesize $^*$Email: rabounski@yahoo.com}

}

\end{center}


\section*{Foreword}

\label{fore}
This book is the generalization of our manuscripts of
1991--1996. These papers were used as the basis for further studies in
1997--99, which were published in our {\it Fields, Vacuum, and the Mirror
Universe}\footnote{%
Borissova~L.\,B.\ and Rabounski~D.\,D. Fields, vacuum, and the mirror
Universe. Editorial URSS, Moscow, 2001 (the~book also is available:
{\it Mathematical Physics Preprint Archive}, mp\_arc:\ 03-206; {\it CERN
Document Server},\linebreak EXT--2003--025).} in March, 2001 as a
contemporary supplement to the famous {\it The Classical Theory of Fields}
by L.\thinspace D.\thinspace Landau and E.\thinspace M.\thinspace Lifshitz.
After this book reached the bookshelves some of our readers began asking for
a detailed account on the previous studies. To meet their request we
prepared our manuscripts of 1991--96 as a book with some recent
amendments. The result is given here for your consideration.
The book is advanced by the transcript of a public lecture that one of the
authors gave in St.\,Petersburg in May, 2001. The lecture contains an
account of the history and main results of our studies in a broad accessible
form.

In conclusion we would like to express hearty thanks to our parents and
whole family for permanent support. Belated thanks go to Dr.~Abram Zelmanov
(1913--1987) and Prof.~Kyril Stanyukovich (1916--1989), our teachers, for
years of friendly discussions. Also we highly appreciate contribution from
our colleague Dr.~Boris Levin, who supported our ideas over the recent
years. We are grateful to Dr.~B.\thinspace A.\thinspace Kotov,
Prof.~V.\thinspace I.\thinspace Sokolov, Prof.~E.\thinspace K.\thinspace
Iordanishvili, and Dr.~I.\thinspace A.\thinspace Kotova, for friendly
discussion and for good questions that improved the manuscript greatly. Many
thanks go to Grigory Semyonov, a friend of ours, for preparing the
manuscript in English. Also we are grateful to our publisher Domingo Mar\'\i
n Ricoy for his interest to our works. Specially we are thankful to
Dr.~Basil~K.~Malyshev who provided his ${\cal B}_{\scriptstyle\cal A}%
{\cal K}_{\scriptstyle\cal O}{\cal M}_{\scriptstyle\cal A}$-{\TeX } system\footnote{%
http://www.tex.ac.uk/tex-archive/systems/win32/bakoma/} for us.

\vspace*{-4pt}
\begin{tabbing}
\`{\it D.\,D.\,Rabounski and L.\,B.\,Borissova}
\end{tabbing}
\medskip

\noindent Dear Ladies and Gentlemen,

\rule{0pt}{13pt} 
In this lecture I would like to present for your kind
attention the history of our research and some of the results we have
obtained over the recent decade. I am going to focus largely on two
subjects. The first is zero-space, zero-particles and possible instant
displacement in space. The second is explanation of anomalous annihilation
of orthopositronium, which was first observed in 1987, by interaction of our
world (world of positive energy of particles) with the mirror Universe
(where particles bear negative energy).

The General Theory of Relativity is built upon view of the world as a
four-dimensional space-time, where any and all objects possess not
three-dimensional volume alone, but their ``longitude'' in time. That is,
any physical body, including ours, is a really existing four-dimensional
instance with the shape of a cylinder elongated in time (cylinder of events
of this body), created by perplexion of other cylinders at the moment of its
``birth'' and split into many other ones at the moment of its ``death''. For
example, for a human the ``temporal length'' is the duration of their life
from conception till death.

The view of the world as a space-time continuum takes its origin from the
historical report by Herman Minkowski, which he gave on 21 September 1908 in
Cologne. There he presented a geometric interpretation of the principle of
invariance of light speed and of Lorentz transformations. Later Albert
Einstein came to the idea that became the cornerstone of his General
Relativity (1908--1916). That was the ``geometric concept of the world'',
according to which geometric structure of space-time defines all properties
of the Universe (both in its observed areas and in areas inaccessible for
observations).

What is space is intuitively clear. But what kind of space, out of those
known in the contemporary mathematics, can comprise three-dimensional space
and time into the single manifold?

Consideration of the problem led Einstein to the fact that the only way to
represent space-time in the way that fits the existing experimental data is
given by {\it four-dimensional pseudo-Riemannian space}. This is a partial
case of the family of Riemannian spaces, i.\thinspace e.\ spaces which
geometry is Riemannian (the square of distance $ds^2$ between infinitely
close points is defined by metric $ds^2{=}g_{\alpha \beta }dx^\alpha
dx^\beta {=}{\it inv}$). In Riemannian space coordinate axes can be of any
kind. Four-dimensional pseudo-Riemannian space is different on the account
of the fact that here there is principal difference between the
three-dimensional space, perceived as space, from the fourth axis --- time.
From mathematical viewpoint this looks as follows: three spatial axes are
real, while the time axis is imaginary (or vice versa), and choice of such
conditions is arbitrary.

A partial case of flat, uniform, and isotropic pseudo-Riemannian space is
referred to as {\it Minkowski space}. This is the space-time of Special
Relativity, an abstracted case where gravitational field, space's rotation
or deformation are absent. In the general case the real pseudo-Riemannian
space is curved, non-uniform and anisotropic. This is General Relativity
space-time where gravitational field, rotation and deformation are present.

Numerous experiments to verify General Relativity showed that
four-dimensional pseudo-\linebreak Riemannian space is our basic space-time, 
within the high precision of up-to-date measurements.

But very soon after A.\,Eddington in 1919 gave the first proof that Sun rays
are curved by its gravitational field, many researchers faced strong
obstacles in fitting together calculations made in the frame of General
Relativity with existing results of observations and experiments.

Here is the problem in a nutshell. All equations in General Relativity are
put down in so-called {\it general covariant form}, which does not depend
upon choice of frame of reference in space-time. The equations, as well as
the variables they contain, are four-dimensional. Which of those
four-dimensional variables are observable in real physical experiments,
i.\,e.\ which components are physical observable values?

Intuitively we may assume that three-dimensional components of
four-dimensional tensor values are observable. But we can not be absolutely
sure that what we observe are three-dimensional components {\it per se}, not
more complicated variables which depend from other factors (e.\thinspace g.\
from properties of physical standards of the space of reference).

Besides, what is the temporal component of a four-dimensional value? In
certain cases this is evident. For instance, three-dimensional components of
four-dimensional potential of electromagnetic field $A^\alpha $ (Greek
indices are four-dimensional, Roman ones are three-dimensional) make up
three-dimensional vector potential $A^i$, while the zero (temporal)
component $A^0{=}\varphi $ is the scalar field potential. But this is not
true for any given vector field.

Further, four-dimensional vector (1st rank tensor) has as few as 4
components (one temporal and three spatial). A 2nd rank tensor, e.\,g.\
rotation or deformation tensor, has 16 components: one temporal, nine
spatial ones and six mixed (time-space) ones. Tensor of higher ranks have
even more components. Are mixed components physical observable values? This
is another question that had no definite answer.

We see that to distinguish physical observable values in General Relativity
is not a trivial problem. Ideally we would like to have a mathematical
technique to calculate physical observable values for tensors of any given
ranks {\it unambiguously}.

Numerous attempts to develop such mathematical technique to define physical
observable values were made in 1930s by outstanding researchers of that
time. The goal was nearly attained by L.\,D.\,Landau and E.\,M.\,Lifshitz in
their {\it The Classical Theory of Fields}, first published in Russian in
1939. Aside for discussing the problem of physical observable values itself,
they introduced interval of physical observable time and three-dimensional
observable interval, which depend upon the properties of the space of
reference and observer's physical standards. But all attempts made in 1930s
were limited to solving certain particular problems. None of them led to
development of a versatile mathematical apparatus.

Such apparatus was developed by A.\thinspace L.\thinspace Zelmanov in
1941--44, who called it {\it theory of chronometric invariants}. Zelmanov
first published it in his dissertation thesis\footnote{%
Zelmanov~A.\thinspace L. On deformation and curvature of accompanying space.
Dissertation thesis, v.\thinspace 1 and v.\thinspace 2. Sternberg Astron.\
Institute, Moscow, 1944 ({\it in Russian}).}.

The essence of Zelmanov's method is simple. As known, components of a tensor
value are defined in so-called {\it ortho-frame}, a system of orthogonal
ideal (straight and uniform) axes, which are tangential to real physical
axes (curved and non-uniform) at the zero point of reference\footnote{%
From mathematical viewpoint in every point of real space-time we choose a
tangential ideal pseudo-Euclidean space. In other words, components of the
tensor are chosen in a tangential pseudo-Euclidean space, but not real
curved and unhomogeneous pseudo-Riemannian space.}. Hence in a real
laboratory we do not measure components of a four-dimensional value, bit
rather its {\it projections} on curved, non-uniform coordinate axes of a
real space of reference.

Real space-time can be imagined as a set of curved and non-uniform spatial
sections, ``stacked'' on the temporal axis (which is non-uniform, too).
Hence a four-dimensional value is projected onto observer's time and their
spatial section. From technical viewpoint projecting onto time and space is
done by means of special tensors ({\it operators of projection}), which were
introduced by Zelmanov and have all properties of operators of projections
in a pseudo-Riemannian space. As a result we obtain physical observable
projections of a four-dimensional value.

Because projection is done in curved and non-uniform structure of real
space, the result depends upon properties of observer's space. Zelmanov
proved that the features of space, which possess properties of physical
observable values (observable projections), are gravitational inertial
force, rotation of space and its deformation. In other words, physical
observable values depend upon gravitational inertial force, rotation and
deformation of observer's space.

Numerous experiments, which have been done since 1950s in various countries,
showed significant impact of space's properties on the measured length and
observed time. The most tremendous out of those experiments were no-landing
flights around the globe in October 1971. During a flight along the Earth's
rotation, observer's space on the plane had more rotation than the space of
the observer who stayed fixed on the ground. During a flight against the
Earth's rotation it was vice versa. Atomic clock on the plane showed
significant variation of observed time depending upon the speed of space's
rotation\footnote{%
Hafele~J.\,C.\ and Keating~R.\,E. Around-the-world atomic clocks: observed
relativistic time gains. {\it Science}, 1972, v.\,177, No.\,4044, 168--170.}.

The only reason to take such an extended introduction was to emphasize that
all our results were obtained using Zelmanov's method of physical observable
values. As you will see later, regular general covariant methods could not
lead to such results.

To summarize, we can say that observable values in real curved, non-uniform
and anisotropic space differ dramatically from so-called ``coordinate
values'', which are defined in an ideal space. This presented two cases,
which could not be studied using general covariant methods: (a)
``split'' of the regular space-time into an area, where time flows from past
into future (our world) and an area, where time flows into the opposite
direction (the mirror Universe); (b) existence of an area of space-time,
where four-dimensional interval, observable three-dimensional interval and
interval of observable time are zeroes (zero-space).

Let us discuss the first case first. According to the theory of chronometric
invariants, interval of physical observable time (projection of
four-dimensional increment $dx^\alpha $ onto time axis) is the value\linebreak
$d\tau {=}\!\left( 1{-}%
\frac{\displaystyle w}{\displaystyle c^2}\right) \!dt{-}\frac{\displaystyle1%
}{\displaystyle c^2}v_idx^i$, i.\thinspace e.\ it depends upon gravitational
potential $w$, upon velocity of space rotation $v_i$ and upon the point of
observation (three-dimensional coordinates $x^i$). We will refer to the
value $\frac{\displaystyle d\tau }{\displaystyle dt}$ as {\it function of
observer's time}, i.\thinspace e.\ it shows where the observer themselves
moves in respect to the time axis $x^0{=}ct$. Hence, at least three
different areas exist, where the observer's function of time bears different
sign. Each of these areas is populated by particles of a specific type:

\begin{enumerate}
\item  \vspace*{-7pt}in the area where $\frac{\displaystyle d\tau }{%
\displaystyle dt}{>}0$ particles move into future in respect to the axis%
\rule{0pt}{9pt} of time coordinate (direct flow of time). This is our world.
It is inhabited by mass-bearing particles (both real sub-light-speed ones
and imaginary tachyons) with positive relativistic masses, as well as
massless (light-like) particles with positive own frequency. In other words,
the area with $\frac{\displaystyle d\tau }{\displaystyle dt}{>}0$ is
inhabited by particles with positive energy $E{=}mc^2{=}\hbar \omega {>}0$;%
\rule{0pt}{9pt}

\item  \vspace*{-7pt}the area where $\frac{\displaystyle d\tau }{%
\displaystyle dt}{<}0$ is inhabited by particles that move into past in
respect to the axis of time (reverse flow of time). This is the mirror
Universe. Particles that inhabit it are the same as those in our world, but
their relativistic masses and frequencies are negative. In other words, the
mirror Universe where $\frac{\displaystyle d\tau }{\displaystyle dt}{<}0$ is
the ``habitat'' of particles with negative energy $E{=}mc^2{=}\hbar \omega {<%
}0$;\rule{0pt}{9pt}

\item  \vspace*{-7pt}the area where $\frac{\displaystyle d\tau }{%
\displaystyle dt}{=}0$ is inhabit by a special kind of particles. Their
relativistic\rule{0pt}{9pt} masses and frequences are zeroes. One may think
of such area as a mere surface that separates our world and the mirror
Universe. But later we will see that a detailed study has proven: the
condition of observable time stop $\frac{\displaystyle d\tau }{\displaystyle %
dt}{=}0$ is true in an entire area of space-time, which is characterized by
its\rule{0pt}{9pt} specific physical properties.
\end{enumerate}

\vspace*{-7pt}Further, chronometrically invariant (observable) equations of
motion for particles with direct flow of time and for those with reverse
flow of time are asymmetric, i.\thinspace e.\ for particles motion either
into past or into future is not the same. This means physical conditions for
motion into past or into future are different. And such asymmetry depends
upon rotation of space or its deformation.

Evidently, if flow of observable time $d\tau $ was not different from flow
of coordinate time $dt$, the very statement of a problem of areas of
space-time with either direct or reverse flow of time would be impossible.

Here we come to a question which I have heard scores of times from my
colleagues. Let's assume four independent coordinate axes --- one temporal
and three spatial. From geometric viewpoint both directions along the time
axis are absolutely equal. But what asymmetry are we speaking about and
isn't it a sort of mistake?

No, it's not a mistake. Of course, if spatial section (three-dimensional
space) is uniform and isotropic, then both directions, into past and into
future, are equal. But as soon as the spatial section becomes rotated or
deformed (like a twisted sheet of paper set on ``the axis of time'' and
spined around it), the space-time becomes anisotropic in respect to past and
future. This anisotropy leads to different physical conditions in motion
into past and into future.

As a result, in real space-time we have two different four-dimensional areas
--- our world with direct flow of time and the mirror Universe with reverse
flow of time. These areas are separated with space-time membrane, on which
time stops from viewpoint of a regular observer $d\tau {=}0$, i.\thinspace
e.\ observer's function of time $\frac{\displaystyle d\tau }{\displaystyle dt%
}{=}0$.

\rule{0pt}{8pt}What sort of membrane is that and isn't it a mere border
surface between our world and the mirror Universe? Our study of the question
using method of physical observable values gave the following results.

So, as well-known two types of trajectories are allowable in four-%
dimensional pseudo-Riemannian space (the basic space-time of General
Relativity):

\begin{enumerate}
\item  \vspace*{-7pt}non-isotropic trajectories, along which
four-dimensional interval is not zero $ds^2{=}c^2d\tau ^2{-}d\sigma ^2{\neq }%
0$, three-dimensional observable interval is not zero $d\sigma {\neq }0$ and
interval of observable time is not zero $d\tau {\neq }0$. These are
trajectories of mass-bearing particles that move at sub-light speed (real
particles) and at super-light speed (tachyons);

\item  \vspace*{-7pt}isotropic trajectories. These are trajectories of
massless light-like particles, which move at the speed of light. Along
isotropic trajectories four-dimensional interval is zero $ds^2{=}0$, but
three-dimensional observable interval and interval of observable time are
not zeroes $d\sigma ^2{=}c^2d\tau ^2{\neq }0$.
\end{enumerate}

\vspace*{-6pt}We have also considered trajectories of third type, along
which space-time interval is zero\linebreak $ds^2{=}c^2d\tau ^2{-}d\sigma ^2{=}0$,
three-dimensional interval and interval of observable time are zeroes too\linebreak
$c^2d\tau ^2{=}d\sigma ^2{=}0$. Mathematically that means {\it full
degeneration} of space-time. What is fully degenerated space-time and does
it contain any particles? From viewpoint of regular general covariant
methods, not related to any frame of reference, here we have absolute zero
and the very statement of the problem is senseless. Therefore we called
fully degenerated space-time {\it zero-space}, and the particles which
inhabit it --- {\it zero-particles}. But the method of physical observable
values, linked to a real frame of reference and its properties, allows an
observer to ``penetrate'' inside zero-space and to see what is going on
there.

As a result, we discovered that zero-space contained the whole world with
its own coordinates, trajectories and particles. But because of the
structure of four-dimensional space-time a regular observer on Earth sees
all zero-space shrunk into a single point where observable time stops. But
that does not mean that the only way to enter zero-space from our world is
through a single special point. To the contrary, entrance is possible at any
point. What is necessary is to create physical condition of degeneration in
the local space of the entering object. The conditions imply a special
combination of gravitational potential $w$, of space rotation velocity $v_i$
and of the penetrating object's own velocity $u^i$, which finally takes the
form $w{+}v_iu^i{=}c^2$. In a partial case, in absence of space's rotation
or if the object rests, the condition of degeneration coincides with the
condition of collapse $w{=}c^2$, i.\thinspace e.\ entering zero-space is
possible also through a black hole.

Because interval of time and spatial interval in zero-space are observed
from our world as zeroes, any displacements of zero-particles are
instantaneous from the viewpoint of a regular observer. We call such way of
interaction {\it long-range action}. Because particles of our world can not
move instantaneously, they can not carry long-range action. But if
interaction is transmitted through zero-space (by means of exchange of
zero-particles), long-range action becomes possible, because the observed
time between emission and reception of signal becomes zero.

Further studies showed that zero-particles also have mass and own frequency,
but to see them the observer must enter zero-space themselves. But how do
zero-particles look like from the viewpoint of our world's observer? Can we
detect them in experiments? We have looked at this problem within de
Broglie's particle-wave concept. As a result we have found that for a
regular observer zero-particles are {\it standing waves}.

From geometric viewpoint trajectories in zero-space are the ultimate case of
full degeneration of light-like (isotropic) trajectories. Hence
zero-particles are the ultimate case of light-like particles, {\it %
light-like standing waves}, or in other words, {\it waves of standing light}
({\it standing-light hologram}). It is possible that ``stop-light
experiments'' done in Harvard by Lene Hau's group and independently by
M.\,Lukin and R.\,Walsworth may be an experimental ``foreword'' to discovery
of zero-particles.

Further study of zero-particles showed that their motion breaks the
relationship between energy and impulse. Geometrically that means that the
square of zero-particle's own vector is not constant in its parallel
transfer along its trajectory in zero-space. As known, the square of vector
of mass-bearing and light-like particles is conserved along their
trajectories. In other words, in ``habitats'' of regular particles (our
world and the mirror Universe) the relationship between energy and impulse
is true, but this is not the case in zero-space.

In up-to-date science the one and only type of particles is known for which
the relationship between energy and impulse is not true. Those are virtual
particles. According to contemporary views based on experimental data,
virtual particles carry interaction between any two observable particles
(either mass-bearing or light-like). This fact allows unambiguous
interpretation of zero-particles and zero-space: (a) zero-particles are
virtual particles that carry interaction between any regular particles, and
(b) zero-space is an area inhabited by virtual particles, and, at the same
time, this is the membrane between our world and the mirror Universe%
\footnote{%
For detailed interpretation of virtual particles as zero-particles see
Chapter~6 in our {\it Fields, Vacuum, and the Mirror Universe}.}.

A natural question arises. If particles of our world interact with each
other by means of exchange of zero-particles in zero-space and particles in
the mirror Universe do the same, why mixed interactions are never observed,
i.\,e.\ why interaction between our world and the mirror Universe is never
observed?

Of course our every-day experience shows no signs of such interaction. But
whether such interaction does not happen at all is a question worth thinking
of.

In relation to this problem our colleague B.\thinspace M.\thinspace Levin
brought our attention to anomalies in annihilation of orthopositronium ($%
\lambda _{{\rm T}}$-anomaly and isotope anomaly) experimentally discovered
in 1987 and still awaiting theoretical explanation. One of these ($\lambda _{%
{\rm T}}$-anomaly) was discovered in Michigan University\footnote{%
Westbrook~C.\,I., Gidley~D.\,W., Conti~R.\,S., and Rich~A. New precision
measurement of the orthopositronium decay rate: a discrepancy with theory. 
{\it Physical Review Letters}, 1987, v.\,58, 1328--1331.} (Ann Arbor, USA)
and caused the measured life span of orthopositronium to be significantly
shorter (by 0.2\%) than the theoretical one as predicted by Quantum
Electrodynamics up to the accuracy of 0.0007\%. Regular annihilation of
orthopositronium produces odd number of $\gamma $-quanta (3,\thinspace
5,\thinspace 7,\dots , largely 3-photon annihilation is observed, because
other modes are very small) and is clearly explained by QED. That means that
in the experiments mentioned in the above a new factor was observed not
explained by QED. The factor appears in $\beta ^{+}$-decay of nuclei, which
was used in these experiments as the source of positrons to produce
positronium in substance. Isotope anomaly was experimentally discovered in
Gatchina (Russia), and showed collective excitation of $^{\text{22}}$Ne
nuclei that lasted through the entire life of orthopositronium\footnote{%
Levin~B.\,M., Kochenda~L.\,M., Markov~A.\,A., and Shantarovich~V.\,P. Time
spectra of annihilation of positrons ($^{\text{22}}$Na) in gaseous neon of
various isotopic compositions. {\it Soviet Journal Nucl.\ Physics}, 1987,
v.\,45(6), 1119--1120.}. In 1995 Levin suggested that $\lambda _{{\rm T}}$%
-anomaly and isotope anomaly were linked to possible 1-photon
orthopositronium annihilation, which caused one $\gamma $-quantum to be
emitted into our world and two $\gamma $-quanta (with the total energy of $%
\sim $3.6$\cdot $10$^{-4}$\thinspace eV) to be emitted into the mirror
Universe thus becoming unavailable for observation\footnote{%
Levin~B.\,M. On the kinematics of one-photon annihilation of
orthopositronium. {\it Physics of Atomic Nuclei}, 1995, v.\,58(2), 332--334.
For this Levin's supposition see also the recent publication: Levin~B.\,M.,
Borissova~L.\,B., and Rabounski~D.\,D. Orthopositronium and space-time
effects. Lomonossov Workshop, Moscow--St.\,Petersburg, 1999 ({\it in Russian}%
).}. Single-photon annihilation of orthopositronium breaks law of
conservation of impulse, but taking into account the emission into the
mirror Universe (3-photon ``mixed'' mode) restores the necessary balance.

Possible interactions with the mirror Universe have been discussed since
long ago\footnote{%
Holdom~B. Two U(1)'s and $\epsilon $ charge shifts. {\it Physics Letters~B},
1986, v.\,166, 196--198. Glashow~S.\,L. Positronium versus the
mirror Universe. {\it Physics Letters~B}, 1986, v.\,167, 35--36.}, but using
methods of quantum theory of fields. No one had approached this problem
using geometric methods of General Relativity before we did so.

First we looked at annihilation {\it per se} as observable interaction
between two particles in our world by means of exchange of zero-particles in
zero-space. As a result we obtained that two types of zero-particles exist,
namely: (a) regular zero-particles, which we called simply {\it virtual
photons}, and (b) zero-particles, i.\thinspace e.\ virtual photons in the
state of collapse ({\it virtual collapsers}). Noteworthy, from the viewpoint
of an observer on the Earth regular virtual photons carry interaction
between particles of our world, while virtual collapsers ``are in charge''
of interactions in the mirror Universe.

Further we looked at the condition of degeneration of space-time for an
electron-positron system, because at the moment of their annihilation
interaction is carried by zero-particles (virtual photons)\footnote{%
For detailed theoretical explanation and the history of orthopositronium
anomalies see Chapter~6 of {\it Fields, Vacuum, and the Mirror Universe}.}.

We found that in annihilation of parapositronium (summary spin is zero)
energy of interaction is transmitted by two regular virtual photons, which
in our world generate two annihilation $\gamma $-quanta. In other words, all
products of annihilation are released into our world.

In annihilation of orthopositronium (spin 1) the process is as follows.
Non-zero spin causes three photons to carry interaction here. Two ``basic''
virtual photons are generated as a result of interaction between own
energies of electron and positron. The third one results from transformation
of spin-energy of orthopositronium, i.\thinspace e.\ of additional energy
carried by orthopositronium thanks to its non-zero inner momentum (spin).
Analysis of the condition of degeneration of space-time shows two possible
channels to convey interaction between electron and positron here:

\begin{enumerate}
\item  \vspace*{-8pt}interaction is carried by three regular virtual
photons. Hence all three $\gamma $-quanta are emitted into our world;

\item  \vspace*{-7pt}two ``basic'' virtual photons collapse to become
virtual collapsers. Hence two $\gamma $-quanta they generate go to the
mirror Universe beyond the reach of an observer. Third zero-particle
produced by spin-energy of orthopositronium is a regular virtual photon and
generates a $\gamma $-quantum that goes to our world. We call this effect
``2+1 split of 3-photon annihilation of orthopositronium''.
\end{enumerate}

\vspace*{-8pt}Experiments show that 99.8\% of atoms of orthopositronium
decay into three $\gamma $-quanta as predicted by standard Quantum
Electrodynamics. Only 0.2\% show either $\lambda _{{\rm T}}$-anomaly (2+1
split of 3-photon annihilation). Hence in annihilation of 99.8\% of atoms
exchange between electron and positron is effected by regular virtual
photons (channel 1). But 0.2\% of atoms decay through channel 2, when
exchange is effected by 1 regular virtual photon and 2 virtual collapsers.
It is these 0.2\% of atoms for which the observed effect of ``anomalous''
1-photon annihilation is possible, when 1 photon is released into our world
and 2 photons are released into the mirror Universe.

To those interested in further information on these and other obtained
results I would like to recommend our books {\it Fields, Vacuum, and the
Mirror Universe} and {\it Particles Here and Beyond the Mirror}. Thank you
very much for your kind attention!

\vspace*{-4pt}
\begin{tabbing}
May 14, 2001\`{\it D.\,D.\,Rabounski}
\end{tabbing}
\newpage


\section{Introduction}

\label{1}\vspace*{6pt}
The main goal of the theory of motion of particles is to define
three-dimensional (spatial) coordinates of a particle at any given moment of
time. To attain the goal one should be aware of three things. First, one
should know in what sort of space-time the events take place. That is, one
should know the geometric structure of space-time, just as one should know
the road conditions to drive on it. Second, one should know the physical
properties of the moving particle. Third, knowledge of equations of motion
of particles of that type is necessary.

The first problem actually leads to choice of a multi-dimensional space from
those known in mathematics, which geometry best fits the geometry of the
observed world. In the early 20th century Albert Einstein proposed
four-dimensional pseudo-Riemannian space with sign-alternating Minkowski
signature (one temporal axis and three spatial axes) as the space-time of
the observed world. Further development of this assumption led to the
General Theory of Relativity, the first geometric theory of space-time and
of motion of particles ever since the dawn of the contemporary science.

Successful experiments to verify General Relativity over the recent 80 years
explicitely say that four-dimensional pseudo-Riemannian space is the basic
space-time of the observed world (as far as the up-to-date measurements'
precision allows to judge). And if inevitable evolution of human
civilization thought and of experimental technology shows that
four-dimensional pseudo-Riemannian space can no-longer explain results of
new experiments, this will mean nothing but a more general space should be
assumed, which will include four-dimensional pseudo-Riemannian space as a
partial case.

This book will focus on motion of particles basing on the geometric concept
of the world's structure: we will assume that the geometry of space-time
defines all properties of the observed world. Therefore, contrary to other
researchers, we are not going to constrain the geometry of space-time by any
limitations and we will solve our problems in the way the geometry of
space-time requires them to be solved.

Hence, any particle in the space-time corresponds to its own ``world'' line,
which sets three-dimensional (spatial) coordinates of the particle at any
given moment of time. Subsequently, our goal to define possible types of
particles evolves to considering all allowable types of trajectories of
motion in four-dimensional space-time.

Generally, referring to equations of motion of free particles in metric
space-time one actually refers to {\it equations of geodesic lines}, which
are four-dimensional equations of trajectories of free particles

\vspace*{-4pt}%
$$
\frac{d^2x^\alpha }{d\rho ^2}+\Gamma _{\mu \nu }^\alpha \frac{dx^\mu }{d\rho 
}\frac{dx^\nu }{d\rho }=0\,,\qquad \alpha ,\mu ,\nu =0,1,2,3,\eqno (1) 
$$

\noindent where $\Gamma _{\mu \nu }^\alpha $ are Christoffel symbols of 2nd
kind and $\rho $ is a parameter of derivation to the geodesic lines. From
geometric viewpoint equations of geodesic lines are equations of {\it %
parallel transfer in Levi-Civita meaning} \cite{bib1} of four-dimensional kinematic
vector $Q^\alpha {=}\frac{\displaystyle dx^\alpha }{\displaystyle d\rho }$

\vspace*{-3pt}%
$$
\frac{DQ^\alpha }{d\rho }=\frac{dQ^\alpha }{d\rho }+\Gamma _{\mu \nu
}^\alpha Q^\mu \frac{dx^\nu }{d\rho }=0\,,\eqno (2) 
$$

\noindent where $D$ is absolute differential. Kinematic vector $Q^\alpha $
is transferred parallel to itself and tangentially to the trajectory of
transfer (a geodesic line). Levi-Civita parallel transfer implies that the
length of the transferred vector is conserved

\vspace*{-5pt}%
$$
Q_\alpha Q^\alpha =g_{\alpha \beta }Q^\alpha Q^\beta =const\,,\eqno (3) 
$$

\vspace*{1pt}\noindent where $g_{\alpha \beta }$ is fundamental metric
tensor.

But the equations of geodesic lines are pure kinematic ones, as they do not
contain physical properties of the moving objects. Therefore to obtain the
full picture of motion of particles we have to build {\it dynamic equations
of motion}, solving which will give us not trajectories of the particles
alone, but their properties (energy, frequency etc.) as well.

To do this we have to define: (1) possible types of trajectories in
four-dimensional space-time (pseudo-Riemannian space); (2) dynamical vector
for each type of trajectory; (3) derivation parameter to each type of
trajectories.

First we consider what types of trajectories are allowable in
four-dimensional pseudo-Riemannian space. Along any geodesic line the
condition $g_{\alpha \beta }Q^\alpha Q^\beta {=}const$ is true. If along
geodesic lines\linebreak $g_{\alpha \beta }Q^\alpha Q^\beta {\neq }0$, such
geodesic lines are referred to as {\it non-isotropic} ones. Along
non-isotropic lines the square of four-dimensional interval is not zero

\vspace*{-4pt}%
$$
ds^2=g_{\alpha \beta }dx^\alpha dx^\beta \neq 0\,.\eqno (4) 
$$

\noindent and the interval $ds$ becomes

\vspace*{-5pt}%
$$
ds=\sqrt{g_{\alpha \beta }dx^\alpha dx^\beta }\text{\qquad if\quad }ds^2>0\,,%
\eqno (5) 
$$

\vspace*{-4pt}%
$$
ds=\sqrt{-g_{\alpha \beta }dx^\alpha dx^\beta }\text{\qquad if\quad }%
ds^2<0\,.\eqno (6) 
$$

If along geodesic lines $g_{\alpha \beta }Q^\alpha Q^\beta {=}0$, such
geodesic lines are referred to as {\it isotropic} ones. Along isotropic
geodesic lines the square of four-dimensional interval is zero

\vspace*{-5pt}%
$$
ds^2=g_{\alpha \beta }dx^\alpha dx^\beta =0\,,\eqno (7) 
$$

\noindent but three-dimensional observable interval and interval of
observable time are not zeroes.

This ends the list of types of trajectories in four-dimensional
pseudo-Riemannian space (the basic space-time of General Relativity).

But other trajectories are theoretically allowable, along which
four-dimensional interval, interval of observable time and three-dimensional
observable interval are zeroes. Such trajectories lay beyond four-%
dimensional pseudo-Riemannian space. These are trajectories in fully
degenerated space-time. We call it degenerated because from the viewpoint of
a regular observer of the Earth all distances and intervals of time in such
space degenerate into zero. Nevertheless, transition into such degenerated
space from the regular space-time is quite possible (provided certain
physical conditions are achieved). And perhaps for the observer, who moves
into such degenerated space-time, the terms ``time'' and ``space'' will not
become void, but will be measured in different units.

Therefore we may consider four-dimensional pseudo-Riemannian space
and degenerated space-time as {\it generalized space-time}, in which both
non-degenerated (isotropic and non-isotropic) and degenerated trajectories
exist.

Hence in generalized four-dimensional space-time, which is the ``extension''
of the basic space-time of General Relativity, three types of trajectories
are allowable:

\begin{enumerate}
\item  \vspace*{-7pt}non-isotropic trajectories (pseudo-Riemannian space).
Along them motion is possible at sub-light and super-light speed;

\item  \vspace*{-7pt}isotropic trajectories (pseudo-Riemannian space).
Motion along them is possible at light speed only;

\item  \vspace*{-7pt}fully degenerated trajectories (zero-trajectories),
which lay in fully degenerated space-time.
\end{enumerate}

\vspace*{-7pt}According to these three types of trajectories three types of
particles can be distinguished, which can exist in four-dimensional
space-time:

\begin{enumerate}
\item  \vspace*{-7pt}mass-bearing particles (rest-mass $m_0{\neq }0$) move
along non-isotropic trajectories ($ds{\neq }0$) at sub-light speed (real
mass-bearing particles) and at super-light speed (imaginary mass-bearing
particles --- tachyons);

\item  \vspace*{-7pt}massless particles (rest-mass $m_0{=}0$) move along
isotropic trajectories ($ds{=}0$) at the speed of light. These are
light-like particles, e.\thinspace g.\ photons;

\item  \vspace*{-7pt}particles of 3rd kind move along trajectories in fully
degenerated space-time.
\end{enumerate}

\vspace*{-7pt}Besides, from pure mathematical viewpoint equations of
geodesic lines contain the same vector $Q^\alpha $ and the same parameter $%
\rho $ irrespective of whether the considered trajectories are isotropic or
non-isotropic. This hints that there must exist equations of motion, which
are the same for mass-bearing and massless particles. We will undertake to
search such generalized equations of motion.

In the next Section we will set forth the basics of mathematical apparatus
of physical observable values (chronometric invariants), which will serve as
our main tool in this book. In Section~3 we will prove existence of
generalized dynamic vector and derivation parameter, which are the same for
mass-bearing and massless particles. Section~4 will focus on the conditions
for degeneration of pseudo-Riemannian space. Section~5 will consider
properties of particles in generalized four-dimensional space-time, which
allows degeneration of metric. In Section~6 chronometrically invariant
dynamic equations of motion valid for all types of particles will be
obtained. In the same Section we will show that Newton's laws of classical
mechanics are partial cases of these equations in pseudo-Riemannian space.
Section~7 will be devoted to two aspects of the obtained equations: (1)
conditions of transformation of generalized space-time into regular
space-time, and (2) asymmetry of motion into future (direct flow of time)
and into past (reverse flow of time). Sections~8 and 9 will focus on the
physical conditions of direct and reverse flow of time. In Section~10 we
will look at motion of particles being carried by motion of space itself.
Sections from 11 to 13 discuss certain specific conditions (super-light
observer, black holes, zero-space). Finally, in Section~14 we will build the
general picture of motion of particles in generalized space-time, basing on
the results obtained in this book.


\section{Physical observable values}

\label{2}
To build a descriptive picture of any physical theory we need to express the
results through real physical values, which can be measured in experiments (%
{\it physical observable values}). In General Relativity this problem is not
a trivial one, because we are looking at objects in four-dimensional
space-time and we have to define, which components of four-dimensional
tensor values are physical observable values.

A mathematical apparatus to calculate physical observable values in
four-dimensional pseudo-Riemannian space was first introduced by
A.\thinspace L.\thinspace Zelmanov and is referred to as {\it theory of
chronometric invariants}. The apparatus was published in 1944 in Zelmanov's
doctorate thesis \cite{bib2} and in his later study of 1956 \cite{bib3}.

Similar results were obtained by C.\thinspace Cattaneo independently from
Zelmanov. Cattaneo published his first study in 1958 \cite{bib4,bib5,bib6,bib7}.

The essence of Zelmanov's apparatus of physical observable values designed
for the four-\linebreak dimensional curved unhomogeneous space-time
(pseudo-Riemannian space), is as follows.

In any point of space-time we can place a local {\it spatial section}
$x^0{=}ct{=}const$ (local space) orthogonal to {\it time line} $x^i{=}const$%
. If the spatial section is everywhere orthogonal to time lines, the space
is referred to as {\it holonomic}. Otherwise the space is referred to as 
{\it non-holonomic}.

Frames of reference of a real observer include coordinate nets spanned over
a real reference body and real clock which can represent a set of real
references to which the observer refer his observations. Therefore physical
observable values should be a result of projection of four-dimensional
values on space and time of the reference body of the observer. Operator of
projection on time is the vector of four-dimensional velocity

\vspace*{-12pt}%
$$
b^\alpha =\frac{dx^\alpha }{ds},\qquad \qquad b_\alpha b^\alpha =+1\,,\eqno %
(8) 
$$

\vspace*{1pt}\noindent of the reference body in respect to the observer.
Operator of projection on spatial section of the observer (his local space)
is defined as four-dimensional symmetric tensor $h_{\alpha \beta }$

\vspace*{-2pt}%
$$
\begin{array}{ll}
h_{\alpha \beta }=-g_{\alpha \beta }+b_\alpha b_\beta \,, & \qquad h^{\alpha
\beta }=-g^{\alpha \beta }+b^\alpha b^\beta , \\ 
h_\alpha ^\beta =-g_\alpha ^\beta +b_\alpha b^\beta ,\rule{0pt}{16pt} & 
\qquad h_i^\alpha h_\alpha ^k=\delta _i^k\,,\quad h_\alpha ^ib^\alpha =0\,. 
\end{array}
\eqno (9) 
$$

If observer rests in respect to his references ({\it accompanying frame of
reference}, $b^i{=}0$) transformation of coordinates only means transition
from one coordinate net to another within the same spatial section.
Therefore physical observable values in accompanying frames of reference
should be invariant in respect to transformation of time, i.\thinspace e.\
should be {\it chronometrically invariant values}. Zelmanov developed the
method to calculate chronometrically invariant projections of any
four-dimensional value and set it forth as a theorem.

{\bf \noindent Zelmanov theorem}\rule{18pt}{0pt}
``We assume that $Q_{00\ldots 0}^{ik\ldots p}$ are
components of four-dimensional tensor $Q_{00\ldots 0}^{\mu \nu
\ldots \rho }$ of $r$-th rank, in which all upper indices are not zero,
while all $m$ lower indices are zeroes. Then tensor values

\vspace*{-12pt}%
$$
T^{ik\ldots p}=\left( g_{00}\right) ^{-\frac m2}Q_{00\ldots 0}^{ik\ldots p}%
\eqno (10) 
$$

\vspace*{2pt}\noindent make up chronometrically invariant three-dimensional
contravariant tensor of ($r{-}m$)-th rank. Hence tensor $T^{ik\ldots p}$ is
a result of $m$-fold projection on time by indices $\alpha
,\beta \ldots \sigma $ and projection on space by $r{-}m$ indices $\mu ,\nu
\ldots \rho $ of the initial tensor $Q_{\alpha \beta \ldots \sigma }^{\mu
\nu \ldots \rho }$''.

According to Zelmanov theorem chronometrically invariant (physical
observable) projections of four-dimensional vector $Q^\alpha $ are values

\vspace*{-5pt}%
$$
b^\alpha Q_\alpha =\frac{Q_0}{\sqrt{g_{00}}},\qquad h_\alpha ^iQ^\alpha =Q^i.%
\eqno (11) 
$$

In accordance to this theorem physical observable projections of symmetric
tensor of the 2nd rank $Q^{\alpha \beta }$ are values

\vspace*{-12pt}%
$$
b^\alpha b^\beta Q_{\alpha \beta }=\frac{Q_{00}}{g_{00}},\ \quad h^{i\alpha
}b^\beta Q_{\alpha \beta }=\frac{Q_0^i}{\sqrt{g_{00}}},\ \quad h_\alpha
^ih_\beta ^kQ^{\alpha \beta }=Q^{ik}.\eqno (12) 
$$

\vspace*{-1pt}Projections of four-dimensional coordinate interval $dx^\alpha 
$ are interval of physical observable time

\vspace*{-2pt}%
$$
d\tau =\sqrt{g_{00}}\,dt+\frac{g_{0i}}{c\sqrt{g_{00}}}dx^i,\eqno (13) 
$$

\vspace*{-2pt}
\noindent and interval of observable coordinates $dx^i$ which are the same
as spatial coordinates. Physical observable velocity of a particle is
three-dimensional chronometrically invariant vector

\vspace*{-6pt}%
$$
{\rm v}^i=\frac{dx^i}{d\tau }.\eqno (14) 
$$

Projecting covariant or conravariant fundamental metric tensor 
on spatial section 
in an accompanying frame of reference

\vspace*{-8pt}%
$$
\begin{array}{l}
h_i^\alpha h_k^\beta g_{\alpha \beta }=g_{ik}-b_ib_k=-h_{ik}\,, \\ 
h_\alpha ^ih_\beta ^kg^{\alpha \beta }=g^{ik}-b^ib^k=g^{ik}=-h^{ik},%
\rule{0pt}{18pt} 
\end{array}
\eqno (15) 
$$

\vspace*{1pt}\noindent we obtain that $h_{ik}{={-}}g_{ik}{+}b_ib_k$ is {\it %
observable metric tensor} using which we can lift and lower indices of
three-dimensional values in accompanying frame of reference. Thus
four-dimensional interval and three-dimensional observable one, represented
through observable values, are

\vspace*{-4pt}%
$$
ds^2=c^2d\tau ^2-d\sigma ^2,\qquad d\sigma ^2=h_{ik}dx^idx^k.\eqno (16) 
$$

The main physical observable properties of the space of reference were deduced
by Zelmanov in his dissertation thesis \cite{bib2} from the property of non-commutativity (non zero 
difference between mixed 2nd derivatives with respect to time and spatial coordinates) 
of his chronometrically invariant operators of derivation
$\frac{\displaystyle^{*}\partial }{\displaystyle\partial t}{=}\frac{%
\displaystyle1}{\displaystyle\sqrt{g_{00}}}\frac{\displaystyle\partial }{%
\displaystyle\partial t}$ and 
$\frac{\displaystyle^{*}\partial }{%
\displaystyle\partial x^i}{=}\frac{\displaystyle\partial }{\displaystyle%
\partial x^i}{-}\frac{\displaystyle g_{0i}}{\displaystyle g_{00}}\frac{%
\displaystyle\partial }{\displaystyle\partial x^0}$

\begin{center}
\vspace*{-23pt}%
$$
\displaystyle\frac{\displaystyle^{*}\partial ^2}{\displaystyle\partial
x^i\partial t}-\displaystyle\frac{\displaystyle^{*}\partial ^2}{\displaystyle%
\partial t\partial x^i}=\displaystyle\frac{\displaystyle1}{\displaystyle c^2}%
F_i\displaystyle\frac{\displaystyle^{*}\partial }{\displaystyle\partial t}%
\,,  \qquad  
\displaystyle\frac{\displaystyle^{*}\partial ^2}{\displaystyle%
\partial x^i\partial x^k}-\displaystyle\frac{^{*}\displaystyle\partial ^2}{%
\displaystyle\partial x^k\partial x^i}=\displaystyle\frac{\displaystyle2}{%
\displaystyle c^2}A_{ik}\displaystyle\frac{\displaystyle^{*}\partial }{%
\displaystyle\partial t}\,.\rule{0pt}{22pt} 
\eqno (17) 
$$
\end{center}

\vspace*{-1pt}
These properties are characterized by three-dimensional
chronometrically invariant tensors: antisymmetric tensor of {\it angular
velocity of rotation of reference's space} $A_{ik}$ and vector of {\it 
gravitational inertial force} $F_i$

\vspace*{-10pt}%
$$
A_{ik}=\frac 12\left( \frac{\partial v_k}{\partial x^i}-\frac{\partial v_i}{%
\partial x^k}\right) +\frac 1{2c^2}\left( F_iv_k-F_kv_i\right) ,\eqno (18) 
$$

\vspace*{-1pt}%
$$
F_i=\frac{c^2}{c^2-w}\left( \frac{\partial w}{\partial x^i}-\frac{\partial
v_i}{\partial t}\right) .\eqno (19) 
$$

Here $w$ and $v_i$ are not chronometrically invariant values that characterize
body of reference and its reference's space. These are gravitational
potential $w{=}c^2(1{-}\sqrt{g_{00}})$  and linear velocity of space's
rotation $v_i{=}{-}c\frac{\displaystyle g_{0i}}{\displaystyle\sqrt{g_{00}}}$.

The necessary and sufficient condition of holonomity of the space should be
equality to zero of the tensor $A_{ik}$. Naturally, if spatial sections
everywhere are orthogonal to time lines in some fixed frame of reference
(holonomic space), then in the frame of reference the values $g_{0i}$ are
zeroes. Because $g_{0i}{=}0$, we have $v_i{=}0$ and $A_{ik}{=}0$ too.
Therefore the tensor $A_{ik}$ also we will refer to as {\it tensor of
non-holonomity of space}.

In quasi-Newtonian approximation, i.\,e.\ in a weak gravitational field at
speeds much lower than the speed of light and in absence of rotation of
space $F_i$ becomes a regular non-relativistic gravitational force $F_i{=}%
\frac{\displaystyle \partial w}{\displaystyle \partial x^i}$\rule[-9pt]{0pt}{21pt}.

In addition, the reference body can be deformed which should be also taken
into account in measurements. This can be done introducing into the
equations a three-dimensional symmetric chronometrically invariant tensor of 
{\it deformation velocities} of the space of reference

\vspace*{-4pt}%
$$
\begin{array}{ll}
D_{ik}=\displaystyle \frac{\displaystyle 1}{\displaystyle 2}\displaystyle 
\frac{\displaystyle ^{*}\partial h_{ik}}{\displaystyle \partial t}\,,\qquad
\quad & D^{ik}=- 
\displaystyle \frac{\displaystyle 1}{\displaystyle 2}\displaystyle \frac{%
\displaystyle ^{*}\partial h^{ik}}{\displaystyle \partial t}\,, \\ D=D_k^k=%
\displaystyle \frac{\displaystyle ^{*}\partial \ln \sqrt{h}}{\displaystyle %
\partial t}\,,\rule{0pt}{24pt} & h=\det \left\| h_{ik}\right\| \,. 
\end{array}
\eqno (20) 
$$

Given these definitions we can generally formulate any geometric object in
Riemannian space with observable parameters of the space of reference. For
instance, having any equations obtained using general covariant methods we
can calculate their chronometrically invariant projections on time and on
spatial section of any particular body of reference and formulate them with
its real physical observable properties. From here we arrive to equations
containing only values measurable in practice.

These are the basics of the mathematical apparatus of physical observable
values --- Zelmanov's chronometric invariants \cite{bib2,bib3,bib8,bib9,bib10}.


\section{Motion of mass-bearing and massless particles}

\label{3}
According to up-to-date physical concepts \cite{bib10} mass-bearing particles are
characterized by four-dimensional vector of impulse $P^\alpha $, while
massless particles are characterized by four-dimensional wave vector $%
K^\alpha $

\vspace*{-12pt}%
$$
P^\alpha =m_0\frac{dx^\alpha }{ds}\,,\qquad K^\alpha =\frac \omega c\frac{%
dx^\alpha }{d\sigma }\,,\eqno (21) 
$$

\noindent where $\omega $ is frequency that characterizes massless particle.
In this case space-time interval $ds$ is taken as the derivation parameter
for mass-bearing particles (non-isotropic trajectories, $ds{\neq }0$). Along
isotropic trajectories $ds{=}0$ (massless particles), but three-dimensional
observable interval $d\sigma {\neq }0$. Therefore $d\sigma $ is taken as the
derivation parameter for massless particles.

The square of impulse vector $P^\alpha$ along trajectories of mass-bearing
particles is not zero

\vspace*{-5pt}%
$$
P_\alpha P^\alpha =g_{\alpha \beta }P^\alpha P^\beta =m_0^2=const\neq 0\,,%
\eqno (22) 
$$

\noindent i.\thinspace e.\ $P^\alpha $ is a non-isotropic vector. The square
of wave vector $K^\alpha $ along trajectories of massless particles is zero,
i.\thinspace e.\ $K^\alpha $ is an isotropic vector

\vspace*{-4pt}%
$$
K_\alpha K^\alpha =g_{\alpha \beta }K^\alpha K^\beta =\frac{\omega ^2}{c^2}%
\frac{g_{\alpha \beta }dx^\alpha dx^\beta }{d\sigma ^2}=\frac{\omega ^2}{c^2}%
\frac{ds^2}{d\sigma ^2}=0\,.\eqno (23) 
$$

Because $ds^2$ in chronometrically invariant form (16) is

\vspace*{-4pt}%
$$
ds^2=c^2d\tau ^2-d\sigma ^2=c^2d\tau ^2\left( 1-\frac{{\rm v}^2}{c^2}\right)
,\qquad \quad \text{{\rm v}}^2=h_{ik}\text{{\rm v}}^i\text{{\rm v}}^k,%
\eqno 
(24) 
$$

\vspace*{-3pt}\noindent we can put $P^\alpha $ and $K^\alpha $ down as

\vspace*{-4pt}%
$$
P^\alpha =m_0\frac{dx^\alpha }{ds}=\frac mc\frac{dx^\alpha }{d\tau }%
\,,\qquad K^\alpha =\frac \omega c\frac{dx^\alpha }{d\sigma }=\frac kc\frac{%
dx^\alpha }{d\tau }\,,\eqno (25) 
$$

\noindent where $k{=}\frac{\displaystyle\omega }{\displaystyle c}$ is the
wave number and $m$ is the relativistic mass. Out of the obtained formulas
we can see that physical observable time $\tau $ can be used as a universal
derivation parameter to both isotropic and non-isotropic trajectories,
i.\thinspace e.\ as the single derivation parameter for mass-bearing and
massless particles.

Calculation of contravariant components of vectors $P^\alpha$ and $K^\alpha$
gives

\vspace*{-4pt}%
$$
P^0=m\frac{dt}{d\tau }\,,\qquad P^i=\frac mc\frac{dx^i}{d\tau }=\frac 1cm%
{\rm v}^i\,,\eqno (26) 
$$

\vspace*{-3pt}%
$$
K^0=k\frac{dt}{d\tau }\,,\qquad K^i=\frac kc\frac{dx^i}{d\tau }=\frac 1ck%
{\rm v}^i\,,\eqno (27) 
$$

\noindent where $m{\rm v}^i$ is three-dimensional vector of
impulse of mass-bearing particle and $k{\rm v}^i$ is three-dimensional wave
vector of massless particle.

The formula for $\frac{\displaystyle dt}{\displaystyle d\tau }$ can be
obtained from the square of vector of four-dimensional velocity of particle $%
U^\alpha $, which for sub-light speed, light speed and super-light speed is,
respectively

\vspace*{-6pt}%
$$
g_{\alpha \beta }U^\alpha U^\beta =+1\,,\qquad U^\alpha =\frac{dx^\alpha }{ds%
}\qquad ds=cd\tau \sqrt{1-\frac{\text{{\rm v}}^2}{c^2}}\,,\eqno (28) 
$$

\vspace*{-6pt}%
$$
g_{\alpha \beta }U^\alpha U^\beta =0\,,\qquad U^\alpha =\frac{dx^\alpha }{%
d\sigma }\qquad ds=0\,,\qquad d\sigma =cd\tau \,,\eqno (29) 
$$

\vspace*{-6pt}%
$$
g_{\alpha \beta }U^\alpha U^\beta =-1\,,\qquad U^\alpha =\frac{dx^\alpha }{%
|ds|}\qquad ds=cd\tau \sqrt{\frac{\text{{\rm v}}^2}{c^2}-1}\,.\eqno (30) 
$$

\vspace*{-2pt}Substituting the definitions for $h_{ik}$, $v_i$ and ${\rm v}%
^i $ into each formula for $g_{\alpha \beta }U^\alpha U^\beta $ we arrive to
three quadratic equations in respect to $\frac{\displaystyle dt}{%
\displaystyle d\tau }$. They are the same for sub-light, light-like and
super-light speeds

\vspace*{-2pt}%
$$
\left( \frac{dt}{d\tau }\right) ^2-\frac{2v_i\text{{\rm v}}^i}{c^2\left( 1-%
\displaystyle\frac{\displaystyle w}{\displaystyle c^2}\right) }\frac{dt}{%
d\tau }+\frac 1{\left( 1-\displaystyle\frac{\displaystyle 
w}{\displaystyle c^2}\right) ^2}\left( \frac 1{c^4}v_iv_k\text{{\rm v}}^i%
\text{{\rm v}}^k-1\right) =0\,.\eqno (31) 
$$

\vspace*{-3pt}This quadratic equation has two solutions

\vspace*{-3pt}%
$$
\left( \frac{dt}{d\tau }\right) _{1,2}\!\!=\frac 1{1-\displaystyle \frac{%
\displaystyle w}{\displaystyle c^2}}\left( \frac 1{c^2}v_i{\rm v}^i\pm
1\right) .\eqno (32) 
$$

\vspace*{-6pt}Function $\frac{\displaystyle dt}{\displaystyle d\tau }$
allows to define what direction in time the particle takes. If $\frac{%
\displaystyle dt}{\displaystyle d\tau }{>}0$ then temporal coordinate $t$
increases, i.\thinspace e.\ the particle moves from past into future (direct
flow of time). If $\frac{\displaystyle dt}{\displaystyle d\tau }{<}0$ then
temporal coordinate decreases, i.\thinspace e.\ the particle moves from
future into past (reverse flow of time).

The value $1{-}\frac{\displaystyle w}{\displaystyle c^2}{=}\sqrt{g_{00}}{>}0$%
, because other cases $\sqrt{g_{00}}{=}0$ and $\sqrt{g_{00}}{<}0$ contradict
the signature conditions $({+}{-}{-}{-})$. Therefore the coordinate time $t$
stops $\frac{\displaystyle dt}{\displaystyle d\tau }{=}0$ provided

\vspace*{-2pt}%
$$
v_i{\rm v}^i=-c^2,\qquad \quad v_i{\rm v}^i=+c^2.\eqno (33) 
$$

The coordinate time $t$ has direct flow $\frac{\displaystyle dt%
}{\displaystyle d\tau }{>}0$ if in the first and in the second solutions,
respectively

\vspace*{-3pt}%
$$
\frac 1{c^2}v_i{\rm v}^i+1>0\,,\qquad \quad \frac 1{c^2}v_i{\rm v}^i-1>0\,.%
\eqno (34) 
$$

\vspace*{-1pt}The coordinate time $t$ has reverse flow $\frac{\displaystyle %
dt}{\displaystyle d\tau }{<}0$ at

\vspace*{-3pt}%
$$
\frac 1{c^2}v_i{\rm v}^i+1<0\,,\qquad \quad \frac 1{c^2}v_i{\rm v}^i-1<0\,.%
\eqno (35) 
$$

For sub-light speed particles $v_i{\rm v}^i{<}c^2$ is always true. Hence
direct flow of time for regularly observed mass-bearing particles takes
place under the first condition from (34) while reverse flow of time takes
place under the second condition from (35).

Noteworthy, we looked at the problem of the direction of coordinate time $t$
assuming that physical observable time is $d\tau {>}0$ always.

Now using formulas (26), (27), and (32) we calculate covariant components $%
P_i$ and $K_i$ as well as projections of four-dimensional vectors $P^\alpha $
and $K^\alpha $ onto time

\vspace*{-5pt}%
$$
P_i=-\frac mc\left( {\rm v}_i\pm v_i\right) ,\qquad K_i=-\frac kc\left( {\rm %
v}_i\pm v_i\right) ,\eqno (36) 
$$

\vspace*{-6pt}%
$$
\frac{P_0}{\sqrt{g_{00}}}=\pm m\,,\qquad \quad \frac{P_0}{\sqrt{g_{00}}}=\pm
k\,,\eqno (37) 
$$

\noindent where values $+m$ and $+k$ take place in observation of particles
that move into future (direct flow of time), while values $-m$ and $-k$ take
place in observation of particles that move into past (reverse flow of time).

Therefore, physical observable values are as follows. For mass-bearing
particles these are relativistic mass $\pm m$ and three-dimensional value $%
\frac{\displaystyle1}{\displaystyle c}m{\rm v}^i$, where $m{\rm v}^i$ is
observable vector of impulse. For massless particles these are wave number $%
\pm k$ and three-dimensional value $\frac{\displaystyle1}{\displaystyle c}k%
{\rm v}^i$, where $k{\rm v}^i$ is observable wave vector.

From the obtained formulas (36) and (37) we can see that observable wave
vector of massless particles $k{\rm v}^i$ is a full analog to observable
vector of impulse of mass-bearing particles $m{\rm v}^i$.

Substituting the obtained values $P^0$, $P^i$, $K^0$, $K^i$, and $g_{ik}$,
expressed through $h_{ik}{=}{-}g_{ik}{+}\frac{\displaystyle1}{\displaystyle %
c^2}v_iv_k$ into the formulas for $P_\alpha P^\alpha $ (22) and $K_\alpha
K^\alpha $ (23) we arrive to the relationships between physical observable
energy and physical observable impulse for mass-bearing particle

\vspace*{-3pt}%
$$
\frac{E^2}{c^2}-m^2{\rm v}_i{\rm v}^i=\frac{E_0^2}{c^2}\,,\eqno (38) 
$$

\vspace*{-3pt}
\noindent and for massless particle

\vspace*{-13pt}%
$$
h_{ik}{\rm v}^i{\rm v}^k=c^2,\eqno (39) 
$$

\vspace*{2pt}
\noindent where $E{=}{\pm }mc^2$ is relativistic energy of the particle, and 
$E_0{=}m_0c^2$ is its rest-energy.

Therefore, by comparing the notations for $P^\alpha $ and $K^\alpha $ we
obtained the universal derivation parameter $\tau $, which is the same for
both mass-bearing and massless particles. But four-dimensional dynamic
vectors $P^\alpha $ and $K^\alpha $ themselves {\it are different}.

Now we are going to calculate the universal dynamic vector, which partial
cases are dynamic vector of mass-bearing particles $P^\alpha $ and dynamic
vector of massless particles $K^\alpha $.

We will tackle the problem assuming that the wave-particle dualism is
peculiar to all particles without any exception. That is, we will consider
motion of massless and mass-bearing particles as propagation of waves in
geometric optics approximation. Four-dimensional wave vector of massless
particles $K^\alpha $ in geometric optics approximation is \cite{bib11}

\vspace*{-3pt}%
$$
K_\alpha =\frac{\partial \psi }{\partial x^\alpha }\,,\eqno (40) 
$$

\noindent where $\psi $ is wave phase (eikonal). In a similar way we put
down the four-dimensional vector of impulse of mass-bearing particles $%
P^\alpha$

\vspace*{-9pt}%
$$
P_\alpha =\frac \hbar c\frac{\partial \psi }{\partial x^\alpha }\,,\eqno %
(41) 
$$

\vspace*{-1pt}\noindent where $\hbar $ is Planck constant, and coefficient $%
\frac{\displaystyle\hbar }{\displaystyle c}$ equalizes the dimensions of
both parts of the equation. From these formulas we arrive to

\vspace*{-4pt}%
$$
\frac{K_0}{\sqrt{g_{00}}}=\frac 1c\frac{^{*}\partial \psi }{\partial t}%
\,,\qquad \frac{P_0}{\sqrt{g_{00}}}=\frac \hbar {c^2}\frac{^{*}\partial \psi 
}{\partial t}\,.\eqno (42) 
$$

Equalizing the values (42) to (37) we obtain

\vspace*{-4pt}%
$$
\pm \,\omega =\frac{^{*}\partial \psi }{\partial t}\,,\qquad \quad \pm
m=\frac \hbar {c^2}\frac{^{*}\partial \psi }{\partial t}\,.\eqno 
(43) 
$$

From here we see that the values $+\omega$ for massless particles and $+m$
for mass-bearing particles take place at wave phase $\psi$ increasing with
time, while $-\omega$ and $-m$ take place at wave phase decreasing with
time. From these expressions we obtain the relationship for particle's
energy, which takes into account its dual (wave-particle) nature

\vspace*{-4pt}%
$$
\pm \,mc^2=\pm \,\hbar \omega =\hbar \frac{^{*}\partial \psi }{\partial t}%
=E\,.\eqno (44) 
$$

Now from formula (41) we obtain the dependence between chronometrically
invariant impulse of particle $p^i$ and its phase $\psi $

\vspace*{-4pt}%
$$
p^i=m{\rm v}^i=-\hbar h^{ik}\frac{^{*}\partial \psi }{\partial x^k}\,,\qquad
p_i=m{\rm v}_i=-\hbar \frac{^{*}\partial \psi }{\partial x^i}\,.\eqno (45) 
$$

Further, the condition $K_\alpha K^\alpha $ can be presented in the form of
\cite{bib11}

\vspace*{-4pt}%
$$
g^{\alpha \beta }\frac{\partial \psi }{\partial x^\alpha }\frac{\partial
\psi }{\partial x^\beta }=0\,,\eqno (46) 
$$

\noindent which is the basic equation of the geometric optics (eikonal
equation). Formulating regular operators of derivation with chronometrically
invariant operators $\frac{\displaystyle^{*}\partial }{\displaystyle\partial
t}$ and $\frac{\displaystyle^{*}\partial }{\displaystyle\partial x^i}$ and
taking into account that

\vspace*{-7pt}%
$$
g^{00}=\frac{\displaystyle1-\frac{\displaystyle1}{\displaystyle 
c^2}v_iv^i}{\displaystyle g_{00}},\quad g^{ik}=-h^{ik},\quad
v^i=h^{ik}v_k=-cg^{0i}\sqrt{g_{00}}\,,\eqno (47) 
$$

\noindent we immediately arrive to eikonal equation for massless particles
in chronometrically invariant form

\vspace*{-4pt}%
$$
\frac 1{c^2}\left( \frac{^{*}\partial \psi }{\partial t}\right) ^2\!\!+h^{ik}%
\frac{^{*}\partial \psi }{\partial x^i}\frac{^{*}\partial \psi }{\partial x^k%
}=0\,.\eqno (48) 
$$

In a similar way we obtain chronometrically invariant eikonal equation for
mass-bearing particles

\vspace*{-4pt}%
$$
\frac 1{c^2}\left( \frac{^{*}\partial \psi }{\partial t}\right) ^2\!\!+h^{ik}%
\frac{^{*}\partial \psi }{\partial x^i}\frac{^{*}\partial \psi }{\partial x^k%
}=\frac{m_0^2c^2}{\hbar ^2}\,,\eqno (49) 
$$

\noindent which at $m_0{=}0$ becomes the same as the former one.

Substituting relativistic mass $m$ (43) into (25) we obtain the dynamic
vector $P^\alpha$ that describes motion of massless and mass-bearing
particles in the geometric optics approximation

\vspace*{-4pt}%
$$
P^\alpha =\frac{\hbar \omega }{c^3}\frac{dx^\alpha }{\partial \tau }%
\,,\qquad P_\alpha P^\alpha =\frac{\hbar ^2\omega ^2}{c^4}\left( 1-\frac{%
{\rm v}^2}{c^2}\right) .\eqno (50) 
$$

The length of the vector is a real value at ${\rm v}{<}c$, is zero at ${\rm v%
}{=}c$ and is an imaginary value at ${\rm v}{>}c$. Therefore the obtained
dynamic vector $P^\alpha $ characterizes motion of particles with any
rest-mass (real, zero or imaginary).

Observable projections of the obtained universal vector $P^\alpha $ are

\vspace*{-4pt}%
$$
\frac{P_0}{\sqrt{g_{00}}}=\pm \frac{\hbar \omega }{c^2}\,,\qquad \quad P^i=%
\frac{\hbar \omega }{c^3}{\rm v}^i,\eqno (51) 
$$

\vspace*{-1pt}
\noindent where the time observable projection has the mass dimension and
the value $p^i{=}cP^i$ has the dimension of impulse.


\section{Degenerated space-time. Zero-particles}

\label{4}
As known, along trajectories of massless particles (isotropic trajectories)
the square of interval of four-dimensional wave is zero

\vspace*{-4pt}%
$$
ds^2=c^2d\tau ^2-d\sigma ^2=0\,,\qquad c^2d\tau ^2=d\sigma ^2=0\,.\eqno (52) 
$$

\vspace*{1pt}
But $ds^2{=}0$ not only at $c^2d\tau ^2{=}d\sigma ^2$, but also when even
stricter condition is true, $c^2d\tau ^2{=}d\sigma ^2{=}0$. The condition $%
d\tau ^2{=}0$ means that physical observable time $\tau $ has the same value
along the entire trajectory. The condition $d\sigma ^2{=}0$ means that all
three-dimensional trajectories have zero lengths. Taking into account the
definitions of $d\tau $ (13), $d\sigma ^2$ (16) and the fact that in an
accompanying system of reference $h_{00}{=}h_{0i}{=}0$, we put down the
conditions $d\tau ^2{=}0$ and $d\sigma ^2{=}0$ as

\vspace*{-2pt}%
$$
cd\tau =\left[ 1-\frac 1{c^2}\left( w+v_iu^i\right) \right] cdt=0\,,\qquad
dt\neq 0\,,\eqno (53) 
$$

\vspace*{-7pt}%
$$
d\sigma ^2=h_{ik}dx^idx^k=0\,,\eqno (54) 
$$

\vspace*{-2pt}\noindent where $u^i{=}\frac{\displaystyle dx^i}{\displaystyle %
dt}$ is three-dimensional coordinate velocity of particle, which is not a
chronometrically invariant (physical observable) value.

As known, the necessary and sufficient condition of full degeneration of
quadratic metric is equality to zero of the determinant of its metric
tensor. For a three-dimensional physical observable metric $d\sigma ^2{=}%
h_{ik}dx^idx^k$ this condition is $h{=}\det ||h_{ik}||{=}0$. But the
determinant of the observable three-dimensional metric tensor $h_{ik}$ has
the form \cite{bib10}

\vspace*{-3pt}%
$$
h=-\frac g{g_{00}}\,,\qquad \quad g=\det ||g_{\alpha \beta }||\,.\eqno (55) 
$$

\vspace*{-1pt}Hence if three-dimensional form $d\sigma ^2$ is degenerated $h{%
=}0$, then four-dimensional form $ds^2$ is also degenerated $g{=}0$. Hence
four-dimensional space-time, where conditions (53) and (54) are true, is 
{\it a~fully degenerated space-time}.

Substituting $h_{ik}{=}{-}g_{ik}{+}\frac{\displaystyle1}{\displaystyle c^2}%
v_iv_k$ into (54) and dividing it by $dt^2$ we obtain {\it physical
conditions of degeneration} (53) and (54) in the final form

\vspace*{-4pt}%
$$
w+v_iu^i=c^2,\qquad g_{ik}u^iu^k=c^2\left( 1-\frac w{c^2}\right) ^2,\eqno %
(56) 
$$

\vspace*{-1pt}\noindent where $v_iu^i$ is the scalar product of space's
rotation velocity $v_i$ and coordinate velocity of particle $u^i$.

If all values $v_i{=}0$ (holonomic space), then $w{=}c^2$ and $\sqrt{g_{00}}{%
=}1{-}\frac{\displaystyle w}{\displaystyle c^2}{{=}0}$. That means that
gravitational potential of the body of reference $w$ is strong enough to
bring the body of reference to gravitational collapse ({\it black hole}).
This case is not discussed here.

In the below we are going to look at degeneration of four-dimensional
space-time, when three-dimensional space is non-holonomic $v_i{\neq}0$,
i.\thinspace e.\ rotates.

Using the definition of interval of observable time $d\tau $ (13), we obtain
the relationship between coordinate velocity of particle $u^i$ and its
observable velocity ${\rm v}^i$

\vspace*{-5pt}%
$$
{\rm v}^i=\frac{u^i}{1-\frac{\displaystyle 1}{\displaystyle 
c^2}\left( w+v_ku^k\right) }\,.\eqno (57) 
$$

\vspace*{-1pt}
Now we can put down $ds^2$ in a form to have the conditions of
degeneration presented explicitly

\vspace*{-4pt}%
$$
ds^2=c^2d\tau ^2\left( 1-\frac{{\rm v}^2}{c^2}\right) =c^2dt^2\left\{ \left[
1-\frac 1{c^2}\left( w+v_ku^k\right) \right] ^2-\frac{u^2}{c^2}\right\} .%
\eqno (58) 
$$

Evidently degenerated space-time can only host the particles
for which physical conditions of degeneration (56) are true.

We will refer to the particles that move in fully degenerated space-time (%
{\it zero-space}) as {\it zero-particles}.


\section{Generalized space-time for particles of three kinds}

\label{5}
Looking at motion of mass-bearing and massless particles we considered a
four-dimensional space-time with strictly non-degenerated metric $g{<}0$.
Now we are going to consider a four-dimensional space-time where
degeneration of metric $g{\leq }0$ is possible.

We already obtained the metric of such generalized space-time in the
previous Section (58). Hence vector of impulse of mass-bearing particle $%
P^\alpha $ in generalized space-time ($g{\leq }0$) takes the form

$$
P^\alpha =m_0\frac{dx^\alpha }{ds}=\frac Mc\frac{dx^\alpha }{dt}\,,\eqno %
(59) 
$$

$$
M=\frac{m_0}{\sqrt{\left[ 1-\frac{\displaystyle 1}{\displaystyle 
c^2}\left( w+v_ku^k\right) \right] ^2-\frac{\displaystyle u^2}{%
\displaystyle 
c^2}}}\,,\eqno (60) 
$$

\noindent where $M$ stands for {\it gravitational rotational mass} of
particle. Gravitational rotational mass $M$ depends not only upon
three-dimensional velocity of particle, but upon gravitational potential $w$
(field of the body of reference) and upon velocity of rotation $v_i$ of the
space itself.

From the obtained formula (59) we see that in four-dimensional space-time
where degeneration of metric is possible ($g{\leq }0$), the generalized
parameter of derivation is coordinate time $t$.

Substituting ${\rm v}^2$ from (57) and $m_0{=}m\sqrt{1{-}{\rm v}^2/c^2}$
into this formula, we arrive to the relationship between relativistic mass
of particle $m$ and its gravitational rotational mass $M$

\vspace*{1pt}%
$$
M=\frac m{1-\frac{\displaystyle 1}{\displaystyle c^2}\left( w+v_iu^i\right)
}\,.\eqno (61) 
$$

\vspace*{1pt}%
From the obtained formula we see that $M$ is a ratio between two values,
each one equal to zero in case of degenerated metric ($g{=}0$), but the
ratio is not zero $M{\neq }0$.

The fact is no surprise. The same is true for relativistic mass $m$ in case
of ${\rm v}^2{=}c^2$, when $m_0{=}0$ and $\sqrt{1{-}{\rm v}^2/c^2}{=}0$, but
their ratio $m{\neq }0$.

Therefore light-like (massless) particles are the ultimate case of
mass-bearing ones at ${\rm v}{\rightarrow }c$. Zero-particles can be
regarded the ultimate case of light-like ones that move in non-holonomic
space at observable velocity ${\rm v}^i$ (57), which depends upon
gravitational potential of the body of reference $w$ and upon the direction
in respect to the velocity of rotation of space.

The temporal component of vector $P^\alpha $ (59) and its physical
observable projection onto time are

\vspace*{1pt}
$$
P^0=M=\displaystyle \frac{\displaystyle m}{\displaystyle 1-\frac{%
\displaystyle 1}{\displaystyle c^2}\left( w+v_iu^i\right) }\,,\eqno (62) 
$$

$$
\displaystyle \frac{\displaystyle P_0}{\displaystyle \sqrt{g_{00}}}=M\left[
1-\displaystyle \frac{\displaystyle 1}{\displaystyle c^2}\left(
w+v_iu^i\right) \right] =m\,,\eqno (63) 
$$

\vspace*{1pt}\noindent while its spatial components are

\vspace*{-7pt}%
$$
P^i=\displaystyle \frac{\displaystyle M}{\displaystyle c}u^i=\displaystyle 
\frac{\displaystyle m}{\displaystyle c}{\rm v}^i,\eqno (64) 
$$

\vspace*{-2pt}%
$$
P_i=-\displaystyle \frac{\displaystyle M}{\displaystyle c}\left[ u_i+v_i-%
\displaystyle \frac{\displaystyle 1}{\displaystyle c^2}v_i\left(
w+v_ku^k\right) \right] .\eqno (65) 
$$

\vspace*{1pt}
Evidently in case of degeneration of space-time, i.\thinspace
e.\ under physical conditions of degeneration (56), these components become

\vspace*{-6pt}%
$$
P^0=M\,,\qquad \quad \frac{P_0}{\sqrt{g_{00}}}=m=0\,,\eqno (66) 
$$

$$
P^i=\frac Mcu^i,\qquad \quad P_i=-\frac Mcu_i\,,\eqno (67) 
$$

\vspace*{1pt}\noindent i.\thinspace e.\ particles that move in degenerated
space-time (zero-particles) bear zero relativistic mass, but their
gravitational rotational mass $M{\neq }0$.

Now we are going to look at motion of mass-bearing particles in generalized
space-time within the dual wave-particle concept. Components of the
universal dynamic vector $P_\alpha {=}\frac{\displaystyle 
\hbar }{\displaystyle 
c}\frac{\displaystyle 
^{*}\partial \psi }{\displaystyle 
\partial x^\alpha }$ (41) here are

\vspace*{2pt}%
$$
\frac{P_0}{\sqrt{g_{00}}}=m=M\left[ 1-\frac 1{c^2}\left( w+v_iu^i\right)
\right] =\frac \hbar {c^2}\frac{^{*}\partial \psi }{\partial t}\,,\eqno (68) 
$$

\vspace*{2pt}%
$$
P_i=\frac \hbar c\left( \frac{^{*}\partial \psi }{\partial x^i}-\frac
1{c^2}v_i\frac{^{*}\partial \psi }{\partial t}\right) ,\eqno (69) 
$$

\vspace{2pt}%
$$
P^i=\frac mc{\rm v}^i=\frac Mcu^i=-\frac \hbar c\hbar ^{ik}\frac{%
^{*}\partial \psi }{\partial x^k}\,,\eqno (70) 
$$

\vspace*{-8pt}%
$$
P^0=M=\frac \hbar {c^2\left( 1-\frac{\displaystyle w}{\displaystyle 
c^2}\right) }\left( \frac{^{*}\partial \psi }{\partial t}-v^i\frac{%
^{*}\partial \psi }{\partial x^i}\right) .\eqno (71) 
$$

Out of these components the following two formulas can be obtained. First
one (72) links gravitational rotational mass $M$ to its corresponding total
energy $E_{total}$. Second one (73) links three-dimensional generalized
impulse $Mu^i$ to change of wave phase $\psi $

$$
Mc^2=\frac 1{1-\frac{\displaystyle 1}{\displaystyle c^2}\left(
w+v_iu^i\right) }\,\hbar \frac{^{*}\partial \psi }{\partial t}=\hbar \Omega
=E_{total}\,,\eqno (72) 
$$

$$
Mu^i=-\,\hbar h^{ik}\frac{^{*}\partial \psi }{\partial x^k}\,,\eqno 
(73) 
$$

\vspace*{1pt}
\noindent where $\Omega $ is gravitational rotational frequency, and $\omega 
$ is regular frequency

\vspace*{-1pt}%
$$
\Omega =\frac \omega {1-\frac{\displaystyle 1}{\displaystyle 
c^2}\left( w+v_iu^i\right) }\,,\qquad \omega =\frac{^{*}\partial \psi }{%
\partial t}\,.\eqno (74) 
$$

The condition $P_\alpha P^\alpha {=}const$ in approximation of geometric
optics (eikonal equation) takes the form (49). For corpuscular form of this
condition in generalized space-time we obtain a chronometrically invariant
formula

\vspace*{-8pt}%
$$
\frac{E^2}{c^2}-M^2u^2=\frac{E_0^2}{c^2}\,,\eqno (75) 
$$

\vspace*{1pt}
\noindent where $M^2u^2$ is the square of generalized three-dimensional
impulse vector, $E{=}mc^2$, and $E_0{=}m_0c^2$. Using this formula we put
down the formula for universal dynamic vector $P^\alpha $, which will
include the degeneration conditions

\vspace*{-7pt}%
$$
P^\alpha =\frac{\hbar \Omega }{c^3}\frac{dx^\alpha }{dt}=\frac{\hbar \frac{%
\displaystyle ^{*}\partial \psi }{\displaystyle 
\partial t}}{c^3\left[ 1-\frac{\displaystyle 1}{\displaystyle c^2}\left(
w+v_iu^i\right) \right] }\frac{dx^\alpha }{\partial t}\,,\eqno (76) 
$$

\vspace*{2pt}
$$
P_\alpha P^\alpha =\frac{\hbar ^2\Omega ^2}{c^4}\left\{ \left[ 1-\frac{%
\displaystyle 1}{\displaystyle c^2}\left( w+v_iu^i\right) \right] ^2-\frac{%
u^2}{c^2}\right\} .\eqno (77) 
$$

For degeneration of space-time we have $m{=}0$, $\omega {=}\frac{%
\displaystyle ^{*}\partial \psi }{\displaystyle \partial t}{=}0$, and
$P_\alpha P^\alpha {=}0$, i.\thinspace e.\ particles\rule[-6pt]{0pt}{8pt}
that move in degenerated space-time from viewpoint of a regular observer
bear zero rest-mass, zero relativistic mass $m$ and zero relativistic
frequency $\omega $, which corresponds to relativistic mass within the
wave-particle concept. Also within this viewpoint the square of
four-dimensional vector (dynamic vector of zero-particles) does conserve $%
P_\alpha P^\alpha {=}0$. Therefore we call such particles {\it zero-particles%
}. For zero-particles gravitational rotational mass $M$ (60), generalized
three-dimensional impulse $Mu^i$ (73) and gravitational rotational frequency 
$\Omega $ (74), which corresponds to mass $M$ within the wave-particle
concept, are {\it not zeroes}.

Zero-space's metric $d\mu ^2$ is not invariant from viewpoint of inner
observer who inhabits zero-space. It can be proven from 2nd condition of
degeneration $d\sigma ^2{=}h_{ik}dx^idx^k$. Substituting here $h_{ik}{=}{-}%
g_{ik}{+}\frac{\displaystyle1}{\displaystyle c^2}v_iv_k$, dividing by $dt^2$%
, and substituting 1st condition of degeneration $w{+}v_iu^i{=}c^2$ we
arrive to inner zero-space's metric

\vspace*{-4pt}%
$$
d\mu ^2=g_{ik}dx^idx^k=\left( 1-\frac w{c^2}\right) ^2c^2dt^2\neq inv\,,%
\eqno (78) 
$$

\vspace*{1pt}
\noindent which is not invariant. Hence, from viewpoint of an observer
within zero-space the square of four-dimensional vector of zero-particles
does not conserve

\vspace*{-2pt}%
$$
U_\alpha U^\alpha =g_{ik}u^iu^k=\left( 1-\frac w{c^2}\right) ^2c^2\neq const.%
\eqno (79) 
$$

\vspace*{-2pt}Equation of eikonal (wave phase) for zero-particles can be
obtained by substituting the conditions $m{=}0$, $\omega {=}\frac{%
\displaystyle ^{*}\partial \psi }{\displaystyle \partial t}{=}0$, and $%
P_\alpha P^\alpha {=}0$ into the eikonal equation (48) or (49). As a result%
\rule{0pt}{9pt} we obtain that the eikonal equation for zero-particles from
viewpoint of a regular observer is

\vspace*{-3pt}%
$$
h^{ik}\frac{^{*}\partial \psi }{\partial x^i}\frac{^{*}\partial \psi }{%
\partial x^k}=0\,,\eqno (80) 
$$

\noindent and is a standing wave equation ({\it information
circle}).

As a result we obtain that in non-holonomic space-time two ultimate
transitions are possible: (a) {\it light barrier}, to overcome which a
particle should exceed the speed of light, and (b) {\it zero-transition} for
which a particle should be in a state of a specific rotation depending upon
particular distribution of matter (conditions of degeneration).


\section{Equations of motion}

\subsection{General considerations}

\label{6-1}
Now we are going to obtain dynamic equations of motion of free particles in
generalized space-time $g{\leq }0$, i.\,e.\ the common equations of motion
for mass-bearing, massless and zero particles.

From geometric viewpoint the equations in question are those of
parallel transfer in the meaning of Levi-Civita of a universal dynamic
vector $P^\alpha $

\vspace*{-4pt}%
$$
DP^\alpha =dP^\alpha +\Gamma _{\mu \nu }^\alpha P^\mu dx^\nu =0\,.\eqno (81) 
$$

\vspace*{2pt}Equations of parallel transfer (81) are in generalized form. To
use them to calculate real physical properties observed in practice, the
equations should contain physical observable values (chronometric
invariants). To bring the equations to the desired form we project the
initial ones onto time and onto space in an accompanying frame of reference

\vspace*{-2pt}%
$$
\begin{array}{l}
b_\alpha DP^\alpha = 
\sqrt{g_{00}}\left( dP^0+\Gamma _{\mu \nu }^0P^\mu dx^\nu \right)
+\displaystyle\frac{\displaystyle g_{0i}}{\displaystyle\sqrt{%
g_{00}}}\left( dP^i+\Gamma _{\mu \nu }^iP^\mu dx^\nu \right) =0\,, \\ 
h_\beta ^iDP^\beta =dP^i+\Gamma _{\mu \nu }^iP^\mu dx^\nu =0\,.%
\rule{0pt}{17pt} 
\end{array}
\eqno (82) 
$$

\vspace*{3pt}But Christoffel symbols of the 2nd kind $\Gamma _{\mu \nu
}^\alpha $ found in these equations are not yet expressed through
chronometrically invariant values. We express 2nd kind Christoffel symbols $%
\Gamma _{\mu \nu }^\alpha $ and 1st kind Christoffel symbols $\Gamma _{\mu
\nu ,\sigma }$, included into them,

\vspace*{-4pt}%
$$
\Gamma _{\mu \nu }^\alpha =g^{\alpha \sigma }\Gamma _{\mu \nu ,\sigma
}\,,\quad \Gamma _{\mu \nu ,\rho }=\frac 12\left( \frac{\partial g_{\mu \rho
}}{\partial x^\nu }+\frac{\partial g_{\nu \rho }}{\partial x^\mu }-\frac{%
\partial g_{\mu \nu }}{\partial x^\rho }\right) \eqno (83) 
$$

\vspace*{-1pt}
\noindent through chronometrically invariant properties of the frame of
reference. Expressing components $g^{\alpha \beta}$ and the first
derivatives from $g_{\alpha \beta}$ through $F_i$, $A_{ik}$, $D_{ik}$, $w$,
and $v_i$ after some algebra we obtain

$$
\Gamma _{00,0}=-\frac 1{c^3}\left( 1-\frac w{c^2}\right) \frac{\partial w}{%
\partial t}\,,\eqno (84) 
$$

\vspace*{-2pt}%
$$
\Gamma _{00,i}=\frac 1{c^2}\left( 1-\frac w{c^2}\right) ^2F_i+\frac 1{c^4}v_i%
\frac{\partial w}{\partial t}\,,\eqno (85) 
$$

\vspace*{-2pt}%
$$
\Gamma _{0i,0}=-\frac 1{c^2}\left( 1-\frac w{c^2}\right) \frac{\partial w}{%
\partial x^i}\,,\eqno (86) 
$$

\vspace*{-2pt}%
$$
\Gamma _{0i,j}=-\frac 1c\left( 1-\frac w{c^2}\right) \left(
D_{ij}+A_{ij}+\frac 1{c^2}F_jv_i\right) +\frac 1{c^3}v_j\frac{\partial w}{%
\partial x^i}\,,\eqno (87) 
$$

$$
\Gamma _{ij,0}=\frac 1c\left( 1{-}\frac w{c^2}\right) \left[ D_{ij}{-}\frac
12\left( \frac{\partial v_j}{\partial x^i}{+}\frac{\partial v_i}{\partial x^j%
}\right) {+}\frac 1{2c^2}\left( F_iv_j{+}F_jv_i\right) \right] ,\eqno (88) 
$$

\vspace*{-7pt}%
$$
\begin{array}{l}
\Gamma _{ij,k}=-\triangle _{ij,k}+ 
\displaystyle\frac{\displaystyle1}{\displaystyle c^2}\left\{
v_iA_{jk}+v_jA_{ik}+\displaystyle\frac{\displaystyle1}{\displaystyle 2}%
v_k\left( \displaystyle\frac{\displaystyle\partial v_j}{\displaystyle%
\partial x^i}+\displaystyle\frac{\displaystyle\partial v_i}{\displaystyle%
\partial x^j}\right) -\right. \\ \qquad \qquad \qquad \qquad \qquad 
\rule{0pt}{17pt}\left. -\displaystyle\frac{\displaystyle1}{\displaystyle2c^2}%
v_k\left( F_iv_j+F_jv_i\right) \right\} +\displaystyle\frac{\displaystyle1}{%
\displaystyle c^4}F_kv_iv_j\,, 
\end{array}
\eqno (89) 
$$

\vspace*{2pt}%
$$
\Gamma _{00}^0=-\frac 1{c^3}\left[ \frac 1{1-\displaystyle \frac{%
\displaystyle w}{\displaystyle c^2}}\frac{\partial w}{\partial t}+\left(
1-\frac w{c^2}\right) v_kF^k\right] ,\eqno (90) 
$$

\vspace*{2pt}%
$$
\Gamma _{00}^k=-\frac 1{c^2}\left( 1-\frac w{c^2}\right) ^2F^k,\eqno 
(91) 
$$

$$
\Gamma _{0i}^0=\frac 1{c^2}\left[ -\frac 1{1-\displaystyle \frac{%
\displaystyle w}{\displaystyle c^2}}\frac{\partial w}{\partial x^i}%
+v_k\left( D_i^k+A_{i\cdot }^{\cdot k}+\frac 1{c^2}v_iF^k\right) \right] ,%
\eqno (92) 
$$

\vspace*{3pt}%
$$
\Gamma _{0i}^k=\frac 1c\left( 1-\frac w{c^2}\right) \left( D_i^k+A_{i\cdot
}^{\cdot k}+\frac 1{c^2}v_iF^k\right) ,\eqno (93) 
$$

\vspace*{3pt}%
$$
\begin{array}{l}
\Gamma _{ij}^0=- 
\displaystyle \frac{\displaystyle 1}{\displaystyle c}\displaystyle \frac{%
\displaystyle 1}{\displaystyle 1-\displaystyle \frac{\displaystyle w}{%
\displaystyle c^2}}\left\{ -D_{ij}+\displaystyle \frac{\displaystyle 1}{%
\displaystyle c^2}v_n  \left[ v_j\left(
D_i^n+A_{i\cdot }^{\cdot n}\right) +v_i\left( D_j^n+A_{j\cdot }^{\cdot
n}\right) + 
\displaystyle \frac{\displaystyle 1}{\displaystyle c^2}v_iv_jF^n\right] + \right. \\ 
\qquad \qquad \qquad \qquad \qquad \qquad \rule{0.2cm}{0pt}
\rule{0pt}{10pt}\left. +\,\displaystyle \frac{\displaystyle 1}{%
\displaystyle 2}\left( \displaystyle \frac{\displaystyle \partial v_i}{%
\displaystyle \partial x^j}+\displaystyle \frac{\displaystyle \partial v_j}{%
\displaystyle \partial x^i}\right) -\displaystyle \frac{\displaystyle 1}{%
\displaystyle 2c^2}\left( F_iv_j+F_jv_i\right) -\triangle _{ij}^nv_n\right\}
, 
\end{array}
\eqno (94) 
$$

\vspace*{3pt}%
$$
\Gamma _{ij}^k=\triangle _{ij}^k-\frac 1{c^2}\left[ v_i\left(
D_j^k+A_{j\cdot }^{\cdot k}\right) +v_j\left( D_i^k+A_{i\cdot }^{\cdot
k}\right) +\frac 1{c^2}v_iv_jF^k\right] .\eqno (95) 
$$

\vspace*{3pt}
Here $\triangle _{jk}^i$ stands for chronometrically invariant Christoffel
symbols, which are defined similarly to $\Gamma _{\mu \nu }^\alpha $. The
only difference is that here instead of $g_{\alpha \beta }$ chronometrically
invariant metric tensor $h_{ik}$ is used, i.\thinspace e.

\vspace*{-12pt}%
$$
\triangle _{jk}^i=h^{im}\triangle _{jk,m}=\frac 12h^{im}\left( \frac{%
^{*}\partial h_{jm}}{\partial x^k}+\frac{^{*}\partial h_{km}}{\partial x^j}-%
\frac{^{*}\partial h_{jk}}{\partial x^m}\right) .\eqno (96) 
$$

Having regular operators of derivation expressed through
chronometrically invariant ones and putting down $dx^0{=}cdt$ through $d\tau 
$ (13), we obtain a chronometrically invariant formula for a regular
differential

\vspace*{-12pt}%
$$
d=\frac \partial {\partial x^\alpha }dx^\alpha =\frac{^{*}\partial }{%
\partial t}d\tau +\frac{^{*}\partial }{\partial x^i}dx^i.\eqno (97) 
$$

\vspace*{3pt}
Now having physical observable components of $P^\alpha $ denoted as

\vspace*{-4pt}%
$$
\frac{P_0}{\sqrt{g_{00}}}=\varphi \,,\qquad \quad P^i=q^i,\eqno 
(98) 
$$

\vspace*{-2pt}\noindent we obtain its other components

\vspace*{-6pt}%
$$
P^0=\frac 1{\sqrt{g_{00}}}\left( \varphi +\frac 1cv_kq^k\right) ,\qquad
P_i=-\frac \varphi cv_i-q_i\,.\eqno (99) 
$$

\vspace*{-1pt}Having the obtained formulas substituted into (82) we arrive to
chronometrically invariant equations of parallel transfer of
vector $P^\alpha $

\vspace*{-3pt}%
$$
\begin{array}{l}
d\varphi + 
\displaystyle \frac{\displaystyle 1}{\displaystyle c}\left( F_iq^id\tau {+}%
D_{ik}q^idx^k\right) =0\,, \\ dq^i+\rule{0pt}{18pt}\left( \displaystyle 
\frac{\displaystyle \varphi }{\displaystyle c}dx^k{+}q^kd\tau \right) \left(
D_k^i{+}A_{k\cdot }^{\cdot i}\right) -\displaystyle \frac{\displaystyle %
\varphi }{\displaystyle c}F^id\tau +\triangle _{mk}^iq^mdx^k=0\,. 
\end{array}
\eqno (100) 
$$

\vspace*{2pt}
From the obtained generalized equations (100) we can make an
easy transition to particular dynamic equations of motion, having $\varphi $
and $q^i$ for different types of particles substituted into the generalized
equations (100) and divided by $dt$.

\subsection{Generalized space-time}

\label{6-2}
The corpuscular and wave forms of the universal dynamic vector $P^\alpha $
for this case were obtained in Section~5.

\subsubsection{Equations of motion of real mass-bearing particles}

From (63) and (64) we obtain for the corpuscular form of the universal
dynamic vector $P^\alpha $ in case of real mass-bearing particles

\vspace*{-8pt}%
$$
\varphi =M\left[ 1-\frac 1{c^2}\left( w+v_ku^k\right) \right] ,\qquad q^i=M%
\frac{u^i}c\,,\eqno (101) 
$$

\vspace*{-4pt}
\noindent where $\frac{\displaystyle u^2}{\displaystyle \left[ \displaystyle %
1{-} \frac{\displaystyle 1}{\displaystyle c^2}\left( w{+}v_ku^k\right)
\right] ^2}{<}c^2$, $d\tau {\neq }0$\thinspace , $dt{\neq }0$.

From\rule{0pt}{8pt} here we immediately arrive to the corpuscular form of
dynamic equations of motion for real mass-bearing particles

\vspace*{-4pt}%
$$
\begin{array}{l}
\displaystyle \frac{\displaystyle d}{\displaystyle dt}\left\{ M\left[ 1{-}%
\displaystyle \frac{\displaystyle 1}{\displaystyle c^2}\left( w{+}%
v_ku^k\right) \right] \right\} {-}\displaystyle \frac{\displaystyle M}{%
\displaystyle c^2}\left[ 1{-}\displaystyle \frac{\displaystyle 1}{%
\displaystyle c^2}\left( w{+}v_ku^k\right) \right] F_iu^i+ 
\displaystyle \frac{\displaystyle M}{\displaystyle c^2}%
D_{ik}u^iu^k=0\,, \\ \displaystyle \frac{\displaystyle d}{\displaystyle dt}%
\left( Mu^i\right) {+}\rule{0pt}{22pt}2M\left[ 1{-}\displaystyle \frac{%
\displaystyle 1}{\displaystyle c^2}\left( w{+}v_ku^k\right) \right] \left(
D_n^i+A_{n\cdot }^{\cdot i}\right) u^n- \\ \qquad \qquad \qquad \rule{2.5cm}{0pt} 
\rule{0pt}{15pt}-M\left[ 1{-}\displaystyle \frac{\displaystyle 1}{%
\displaystyle c^2}\left( w{+}v_ku^k\right) \right] F^i{+}M\triangle
_{nk}^iu^nu^k=0\,, 
\end{array}
\eqno (102) 
$$

\vspace*{-5pt}\noindent where $d{=}\frac{\displaystyle ^{*}\partial }{%
\displaystyle 
\partial t}d\tau {+}\frac{\displaystyle ^{*}\partial }{\displaystyle %
\partial x^i}dx^i$, $\frac{\displaystyle d}{\displaystyle d\tau }{=}\frac{%
\displaystyle 
^{*}\partial }{\displaystyle \partial t}{+}{\rm v}^i\frac{\displaystyle %
^{*}\partial }{\displaystyle \partial x^i}$, and

\vspace*{-1pt}%
$$
\frac d{dt}=\frac{^{*}\partial }{\partial t}\frac{d\tau }{dt}+u^i\frac{%
^{*}\partial }{\partial x^i}=\left[ 1-\frac 1{c^2}\left( w+v_mu^m\right)
\right] \frac{^{*}\partial }{\partial t}+u^i\frac{^{*}\partial }{\partial x^i%
}\,.\eqno (103) 
$$

\vspace*{-1pt}For the wave form of the universal dynamic vector $P^\alpha $
in case of real mass-bearing particles and according to (68) and (70) we
obtain

\vspace*{-6pt}%
$$
\varphi =\frac \hbar {c^2}\frac{^{*}\partial \psi }{\partial t}\,,\qquad
q^i=-\frac \hbar ch^{ik}\frac{^{*}\partial \psi }{\partial x^k}\,,\eqno %
(104) 
$$

\vspace*{-2pt}\noindent where change of wave phase with time $\frac{%
\displaystyle ^{*}\partial \psi }{\displaystyle \partial t}$ is positive for
the particles that move from past into future and is negative for those that
move from future into past. From here we arrive to the wave form of (102),
i.\thinspace e.\ to dynamic equations of propagation of waves, which
correspond to real mass-bearing particles within the wave-particle concept

\vspace*{-13pt}%
$$
\begin{array}{l}
\pm 
\displaystyle \frac{\displaystyle d}{\displaystyle d\tau }\left( 
\displaystyle \frac{\displaystyle ^{*}\partial \psi }{\displaystyle \partial
t}\right) +\left[ 1-\displaystyle \frac{\displaystyle 1}{\displaystyle c^2}%
\left( w+v_pu^p\right) \right] F^i\displaystyle \frac{\displaystyle %
^{*}\partial \psi }{\displaystyle \partial x^i}-D_k^iu^k\displaystyle \frac{%
\displaystyle ^{*}\partial \psi }{\displaystyle \partial x^i}=0\,, \\ 
\rule{0pt}{24pt}\displaystyle \frac{\displaystyle d}{\displaystyle d\tau }%
\left( h^{ik}\displaystyle \frac{\displaystyle ^{*}\partial \psi }{%
\displaystyle \partial x^k}\right) \pm \displaystyle \frac{\displaystyle 1}{%
\displaystyle c^2}\left[ 1-\displaystyle \frac{\displaystyle 1}{%
\displaystyle c^2}\left( w+v_pu^p\right) \right] \displaystyle \frac{%
\displaystyle ^{*}\partial \psi }{\displaystyle \partial t}F^i- \\ \quad \
\ - 
\rule{0pt}{19pt}\left\{ \pm \displaystyle \frac{\displaystyle 1}{%
\displaystyle c^2}\displaystyle \frac{\displaystyle ^{*}\partial \psi }{%
\displaystyle \partial t}u^k-h^{km}\left[ 1-\displaystyle \frac{%
\displaystyle 1}{\displaystyle c^2}\left( w+v_pu^p\right) \right] 
\displaystyle \frac{\displaystyle ^{*}\partial \psi }{\displaystyle \partial
x^m}\right\}
\left( D_k^i+A_{k\cdot }^{\cdot i}\right)
+h^{mn}\triangle _{mk}^iu^k\displaystyle \frac{\displaystyle ^{*}\partial
\psi }{\displaystyle \partial x^n}=0\,. 
\end{array}
\eqno (105) 
$$

\vspace*{-4pt}As you can see, the first term in the temporal component of
the obtained equations (105) and two terms in the spatial components are
positive for particles (waves) that move from past into future. Those
components are negative for motion from future into past.

\subsubsection{Equations of motion of imaginary mass-bearing
particles}

For this case $\varphi $ and $q^i$ in the corpuscular form of $P^\alpha $
will be only different from those presented for real mass-bearing particles
(101) by presence of a multiplier $i{=}\sqrt{-1}$

\vspace*{-6pt}%
$$
\varphi =iM\left[ 1-\frac 1{c^2}\left( w+v_ku^k\right) \right] ,\qquad q^i=iM%
\frac{u^i}c\,,\eqno (106) 
$$

\vspace*{-4pt}\noindent where $\frac{\displaystyle u^2}{\displaystyle \left[ 
\displaystyle 1{-}\frac{\displaystyle 1}{\displaystyle c^2}\left( w{+}%
v_ku^k\right) \right] ^2}{>}c^2$, $d\tau {\neq 0}$\thinspace , $dt{\neq }0$.

Respectively,\rule{0pt}{8pt} the corpuscular form of dynamic equations of
motion for imaginary mass-bearing particles (super-light speed particles ---
tachyons) will differ from similar equations for real (sub-light speed)
particles (102) by presence of coefficient $i$ at the mass term $M$.

The values $\varphi $ and $q^i$ for wave the form of dynamic vector of
imaginary mass-bearing particles are the same as those for real particles
(104). Hence the wave form of dynamic equations of motion of imaginary
particles (within the wave-particle dual concept) will be similar to the
wave form of equations for real particles (105).

\subsubsection{Equations of motion of massless particles}

According to (57) for massless (light-like) particles in generalized
space-time at ${\rm v}^2{=}c^2$

\vspace*{-6pt}%
$$
\frac{u^2}{\left[ 1-\frac{\displaystyle 1}{\displaystyle c^2}\left(
w+v_ku^k\right) \right] ^2}=c^2,\qquad d\tau \neq 0\,,\quad dt\neq 0\,.\eqno %
(107) 
$$

\vspace*{-3pt}Having this formula substituted into $\varphi $ and $q^i$ for
real mass-bearing particles (101) we obtain the same values for the
corpuscular form of the universal dynamic vector $P^\alpha $ of massless
particles

\vspace*{-7pt}%
$$
\varphi =M\frac uc\,,\qquad \quad q^i=M\frac{u^i}c\,.\eqno (108) 
$$

\vspace*{-1pt}Respectively, the corpuscular form of dynamic equations of
motion for massless particles is

\vspace*{-4pt}%
$$
\begin{array}{l}
\displaystyle \frac{\displaystyle d}{\displaystyle dt}\left( Mu\right) +%
\displaystyle \frac{\displaystyle Mu}{\displaystyle c^2}F_iu^i+\displaystyle 
\frac{\displaystyle M}{\displaystyle c}D_{ik}u^iu^k=0\,, \\ \displaystyle 
\frac{\displaystyle d}{\displaystyle dt}\left( Mu^i\right) +\rule{0pt}{19pt}%
2M\displaystyle \frac{\displaystyle u}{\displaystyle c}\left( D_n^i{+}%
A_{n\cdot }^{\cdot i}\right) u^n-M\displaystyle \frac{\displaystyle u}{%
\displaystyle c}F^i+M\triangle _{nk}^iu^nu^k=0\,. 
\end{array}
\eqno (109) 
$$

The values $\varphi $ and $q^i$ for the wave form of massless particles will
be similar to the wave form of $\varphi $ and $q^i$ of mass-bearing
particles (104). Respectively, dynamic equations of propagation of waves,
which correspond to massless particles within de~Broglie's wave-particle
concept, will be similar to formulas (105).

\subsubsection{Equations of motion of zero-particles}

In degenerated space-time, i.\,e.\ under the conditions of degeneration,
components $\varphi $ and $q^i$ in the corpuscular for of generalized
dynamic vector $P^\alpha $ become

\vspace*{-8pt}%
$$
\varphi =0\,,\qquad \quad q^i=M\frac{u^i}c\,,\eqno (110) 
$$

\vspace*{-3pt}\noindent where $w{+}v_ku^k{=}c^2$, $d\tau {=}0$, $dt{\neq }0$%
. From here we can obtain the corpuscular form of dynamic equations of
motion for zero-particles

\vspace*{-5pt}%
$$
\displaystyle \frac{\displaystyle M}{\displaystyle 
c^2}D_{ik}u^iu^k=0\,,\qquad \displaystyle \frac{\displaystyle d}{%
\displaystyle dt}\left( Mu^i\right) +M\triangle _{nk}^iu^nu^k=0\,.\eqno %
(111) 
$$

\vspace*{-2pt}The physical observable components $\varphi $ and $q^i$
for the wave form of generalized dynamic vector $P^\alpha $ in degenerated
space-time are

\vspace*{-11pt}%
$$
\varphi =0\,,\qquad \quad q^i=-\frac \hbar ch^{ik}\frac{^{*}\partial \psi }{%
\partial x^k}\,,\eqno (112) 
$$

\vspace*{-1pt}\noindent from which we arrive to the wave form of dynamic
equations of motion of zero-particles

\vspace*{-3pt}%
$$
D_k^mu^k\frac{^{*}\partial \psi }{\partial x^m}=0\,,\qquad \frac d{dt}\left(
h^{ik}\frac{^{*}\partial \psi }{\partial x^k}\right) +h^{mn}\triangle
_{mk}^iu^k\frac{^{*}\partial \psi }{\partial x^n}=0\,,\eqno (113) 
$$

\vspace*{-2pt}\noindent i.\thinspace e.\ dynamic equations of propagation of
waves that correspond to zero-particles within de~Broglie's wave-particle
concept.

\subsection{Strictly non-degenerated space-time}

\label{6-3}
For this case the corpuscular and the wave forms of the universal dynamic
vector $P^\alpha $ were obtained earlier in Section~3.

\subsubsection{Equations of motion of real mass-bearing particles}

According to (37) and (26) for the corpuscular form of the universal dynamic
vector $P^\alpha $ for real mass-bearing particles we have

\vspace*{-9pt}%
$$
\varphi =\pm m\,,\qquad \quad q^i=\frac 1cm{\rm v}^i,\eqno (114) 
$$

\vspace*{-2pt}
\noindent where ${\rm v}^2{<}c^2$, $d\tau {\neq }0$, $dt{\neq }0$. Using
these values we obtain dynamic equations of motion of particles with
positive relativistic mass $m{>}0$ (which move from past into future)

\vspace*{-4pt}%
$$
\begin{array}{l}
\displaystyle \frac{\displaystyle dm}{\displaystyle d\tau }-\displaystyle 
\frac{\displaystyle m}{\displaystyle c^2}F_i\text{{\rm v}}^i+\displaystyle 
\frac{\displaystyle m}{\displaystyle c^2}D_{ik}\text{{\rm v}}^i\text{{\rm v}}%
^k=0\,, \\ \displaystyle \frac{\displaystyle d\left( m\text{{\rm v}}%
^i\right) }{\displaystyle d\tau }+\rule{0pt}{20pt}2m\left( D_k^i+A_{k\cdot
}^{\cdot i}\right) \text{{\rm v}}^k-mF^i+m\triangle _{nk}^i\text{{\rm v}}^n%
\text{{\rm v}}^k=0\,, 
\end{array}
\eqno (115) 
$$

\noindent as well as for particles with negative mass $m{<}0$
(which move into past)

\vspace*{-4pt}%
$$
\begin{array}{l}
- 
\displaystyle \frac{\displaystyle dm}{\displaystyle d\tau }-\displaystyle 
\frac{\displaystyle m}{\displaystyle c^2}F_i\text{{\rm v}}^i+\displaystyle 
\frac{\displaystyle m}{\displaystyle c^2}D_{ik}\text{{\rm v}}^i\text{{\rm v}}%
^k=0\,, \\ \displaystyle \frac{\displaystyle d\left( m\text{{\rm v}}%
^i\right) }{\displaystyle d\tau }+mF^i+m\triangle _{nk}^i\text{{\rm v}}^n%
\text{{\rm v}}^k=0\,.\rule{0pt}{20pt} 
\end{array}
\eqno (116) 
$$

For the wave form of the universal dynamic vector $P^\alpha $ from (42) and
(45) we obtain formulas that are similar to those obtained earlier for $%
\varphi $ and $q^i$ in generalized space-time (104)

\vspace*{-4pt}%
$$
\varphi =\frac \hbar {c^2}\frac{^{*}\partial \psi }{\partial t}\,,\qquad
\quad q^i=-\frac \hbar ch^{ik}\frac{^{*}\partial \psi }{\partial x^k}\,,%
\eqno (117) 
$$

\vspace*{-5pt}\noindent 
where\rule[-6pt]{0pt}{0pt} $\frac{\displaystyle ^{*}\partial \psi }{%
\displaystyle \partial t}$ i.\thinspace e.\ change of phase with time, is
positive for motion of particles from\rule{0pt}{8pt} past into future and is
negative for motion of particles into past. Taking into account that the
chronometrically invariant equations of parallel transfer of $P^\alpha $
(100) in strictly non-degenerated space-time are divided by interval of
observable time $d\tau $, we obtain the wave form of dynamic equations of
motion of mass-bearing real particles

\vspace*{-10pt}%
$$
\begin{array}{l}
\pm 
\displaystyle \frac{\displaystyle d}{\displaystyle d\tau }\left( 
\displaystyle \frac{\displaystyle ^{*}\partial \psi }{\displaystyle \partial
t}\right) +F^i\displaystyle \frac{\displaystyle ^{*}\partial \psi }{%
\displaystyle \partial x^i}-D_k^i\text{{\rm v}}^k\displaystyle \frac{%
\displaystyle ^{*}\partial \psi }{\displaystyle \partial x^i}=0\,, \\ 
\displaystyle \frac{\displaystyle d}{\displaystyle d\tau }\rule{0pt}{22pt}%
\left( h^{ik}\displaystyle \frac{\displaystyle ^{*}\partial \psi }{%
\displaystyle \partial x^k}\right) -\left( D_k^i+A_{k\cdot }^{\cdot
i}\right) \left( \pm \displaystyle \frac{\displaystyle 1}{\displaystyle c^2}%
\displaystyle \frac{\displaystyle ^{*}\partial \psi }{\displaystyle \partial
t}\text{{\rm v}}^k-h^{km}\displaystyle \frac{\displaystyle ^{*}\partial \psi 
}{\displaystyle \partial x^m}\right) \pm \\ \qquad \qquad \qquad \qquad
\quad \quad \rule{0pt}{19pt}\pm \displaystyle \frac{\displaystyle 1}{%
\displaystyle c^2}\displaystyle \frac{\displaystyle ^{*}\partial \psi }{%
\displaystyle \partial t}F^i+h^{mn}\triangle _{mk}^i\text{{\rm v}}^k%
\displaystyle \frac{\displaystyle ^{*}\partial \psi }{\displaystyle \partial
x^n}=0\,. 
\end{array}
\eqno (118) 
$$

The first term of the temporal projection and the first two
terms of the spatial projections are positive for motion of particles from
past into future. These terms are negative for motion from future into past.

\subsubsection{Equations of motion of imaginary mass-bearing
particles}

In this case the corpuscular forms of $\varphi $ and $q^i$ will be different
from those of $\varphi $ and $q^i$ for real mass-bearing particles (114) by
presence of $i{=}\sqrt{-1}$

\vspace*{-5pt}%
$$
\varphi =\pm im\,,\qquad \quad q^i=i\frac 1cm{\rm v}^i,\eqno (119) 
$$

\vspace*{-1pt}\noindent where ${\rm v}^2{>}c^2$, $d\tau {\neq }0$, $dt{\neq }%
0$. Respectively, the corpuscular form of dynamic equations of motion of
imaginary (super-light speed) particles will be different from those we have
obtained for real (sub-light speed) particles by presence of coefficient $i$
and mass term $m$.

The wave forms of $\varphi $ and $q^i$ for imaginary mass-bearing particles
will be similar to the wave forms of $\varphi $ and $q^i$ for real
mass-bearing particles (117). Respectively, dynamic equations of propagation
of waves, which correspond to imaginary mass-bearing particles, will be
similar to dynamic equations of propagation of waves, which correspond to
real mass-bearing particles (118).

We see that from the viewpoint of the wave concept there is no difference at
what speed a mass-bearing particle travels (or the wave propagates) ---
below the speed of light or above that. To the contrary, from the viewpoint
of the corpuscular concept there is difference, because the corpuscular
equations of motion of super-light speed (imaginary) particles differ from
those of sub-light speed ones by presence of coefficient $i$ at the mass
term $m$.

\subsubsection{Equations of motion of massless particles}

In this case the corpuscular forms of $\varphi $ and $q^i$ becomes

\vspace*{-5pt}%
$$
\varphi =\pm \frac \omega c=\pm k\,,\qquad \quad q^i=\frac 1ck{\rm v}%
^i=\frac 1ckc^i,\eqno (120) 
$$

\vspace*{-1pt}\noindent where ${\rm v}^2{=}c^2$, $d\tau {\neq }0$, $dt{\neq }%
0$ and chronometrically invariant (physical observable) vector of
three-\linebreak dimensional velocity of particle ${\rm v}^i$ equals to
chronometrically invariant (observable) three-dimensional vector of light
velocity $c^i$

\vspace*{-15pt}%
$$
{\rm v}^i=\frac{dx^i}{d\tau }=c^i.\eqno (121) 
$$

Respectively, the corpuscular dynamic equations of motion of massless
particles are: for massless particles that bear positive relativistic
frequency $\omega {>}0$ and travel from past into future

\vspace*{-4pt}%
$$
\begin{array}{l}
\displaystyle \frac{\displaystyle d\omega }{\displaystyle d\tau }-%
\displaystyle \frac{\displaystyle \omega }{\displaystyle c^2}F_ic^i+%
\displaystyle \frac{\displaystyle \omega }{\displaystyle c^2}%
D_{ik}c^ic^k=0\,, \\ \displaystyle \frac{\displaystyle d\left( \omega
c^i\right) }{\displaystyle d\tau }+\rule{0pt}{20pt}2\omega \left(
D_k^i+A_{k\cdot }^{\cdot i}\right) c^k-\omega F^i+\omega \triangle
_{nk}^ic^nc^k=0\,.\rule{0pt}{21pt} 
\end{array}
\eqno (122) 
$$

\vspace*{1pt}
For massless particles that bear $\omega {<}0 $ and travel from future into past

\vspace*{-4pt}%
$$
\begin{array}{l}
- 
\displaystyle \frac{\displaystyle d\omega }{\displaystyle d\tau }-%
\displaystyle \frac{\displaystyle \omega }{\displaystyle c^2}F_ic^i+%
\displaystyle \frac{\displaystyle \omega }{\displaystyle c^2}%
D_{ik}c^ic^k=0\,, \\ \displaystyle \frac{\displaystyle d\left( \omega
c^i\right) }{\displaystyle d\tau }+\omega F^i+\omega \triangle
_{nk}^ic^nc^k=0\,.\rule{0pt}{21pt} 
\end{array}
\eqno (123) 
$$

The wave forms of $\varphi $ and $q^i$ for massless particles
are similar to those of $\varphi $ and $q^i$ for mass-bearing particles
(117). Respectively, dynamic equations of propagation of waves, which
correspond to massless (light-like) particles within the wave-particle dual
concept, will be similar too. The only difference will be in vector of
observable velocity of light $c^i$, which will replace vector of observable
velocity of particle ${\rm v}^i$.

\subsection{Specific case: geodesic equations}

\label{6-4}
What are geodesic equations? As we mentioned in Section~1, those are
kinematic equations of motion of particles along the shortest (geodesic)
trajectories. From geometric viewpoint geodesic equations are those of
Levi-Civita parallel transfer

\vspace*{-3pt}%
$$
\frac{DQ^\alpha }{d\rho }=\frac{dQ^\alpha }{d\rho }+\Gamma _{\mu \nu
}^\alpha Q^\mu \frac{dx^\nu }{d\rho }=\frac{d^2x^\alpha }{d\rho ^2}+\Gamma
_{\mu \nu }^\alpha \frac{dx^\mu }{d\rho }\frac{dx^\nu }{d\rho }=0\eqno (124) 
$$

\vspace*{-3pt}\noindent of four-dimensional kinematic vector of particle $%
Q^\alpha {=}\frac{\displaystyle dx^\alpha }{\displaystyle d\rho }$, which is
tangential to the trajectory at its every point. Respectively, non-isotropic
geodesic equations (trajectories of mass-bearing free particles) and
isotropic geodesic equations (massless free particles) are

\vspace*{-3pt}%
$$
\frac{DQ^\alpha }{ds}=\frac{d^2x^\alpha }{ds^2}+\Gamma _{\mu \nu }^\alpha 
\frac{dx^\mu }{ds}\frac{dx^\nu }{ds}=0\,,\quad \qquad Q^\alpha =\frac{%
dx^\alpha }{ds}\,,\eqno (125) 
$$

\vspace*{-10pt}%
$$
\frac{DQ^\alpha }{d\sigma }=\frac{d^2x^\alpha }{d\sigma ^2}+\Gamma _{\mu \nu
}^\alpha \frac{dx^\mu }{d\sigma }\frac{dx^\nu }{d\sigma }=0\,,\quad \qquad
Q^\alpha =\frac{dx^\alpha }{d\sigma }\,.\eqno (126) 
$$

\vspace*{2pt}But any kinematic vector, similar to dynamic vector $P^\alpha $
of mass-bearing particles and to wave vector $K^\alpha $ of massless
particles, is a partial case of an arbitrary vector $Q^\alpha $, for which
we have obtained universal equations of motion. Hence having values $\varphi 
$ and $q^i$ for kinematic vector of mass-bearing particles substituted into
the universal equations of motion (100), we will immediately arrive to
non-isotropic geodesic equations in chronometrically invariant form.
Similarly, having substituted $\varphi $ and $q^i$ for kinematic vector of
massless particles we will arrive to chronometrically invariant isotropic
geodesic equations. This is what we are going to do now.

For kinematic vector of mass-bearing particles we have

\vspace*{-3pt}%
$$
\begin{array}{l}
\varphi = 
\displaystyle \frac{\displaystyle Q_0}{\displaystyle \sqrt{g_{00}}}=%
\displaystyle \frac{\displaystyle g_{0\alpha }Q^\alpha }{\displaystyle \sqrt{%
g_{00}}}=\pm \displaystyle \frac{\displaystyle 1}{\displaystyle \sqrt{1-{\rm %
v}^2/c^2}}\,, \\ q^i=\rule{0pt}{21pt}Q^i=\displaystyle \frac{\displaystyle %
dx^i}{\displaystyle ds}=\displaystyle \frac{\displaystyle 1}{\displaystyle 
\sqrt{1-{\rm v}^2/c^2}}\frac{\displaystyle dx^i}{\displaystyle cd\tau }=%
\displaystyle \frac{\displaystyle 1}{\displaystyle c\sqrt{1-{\rm v}^2/c^2}}%
{\rm v}^i. 
\end{array}
\eqno (127) 
$$

For massless particles, taking into account that 
for isotropic trajectories $d\sigma {=}cd\tau $ we have

\vspace*{-6pt}%
$$
\begin{array}{l}
\varphi = 
\sqrt{g_{00}}\displaystyle \frac{\displaystyle dx^0}{\displaystyle d\sigma }+%
\displaystyle \frac{\displaystyle 1}{\displaystyle c\sqrt{g_{00}}}%
g_{0i}c^i=\pm 1\,, \\ 
q^i=\displaystyle \frac{\displaystyle dx^i}{%
\displaystyle d\sigma }\rule{0pt}{21pt}=\displaystyle \frac{\displaystyle %
dx^i}{\displaystyle cd\tau }=\displaystyle \frac{\displaystyle 1}{%
\displaystyle c}\,c^i. 
\end{array}
\eqno (128) 
$$

Having these values substituted into generalized equations of
motion (100), we obtain {\it chronometrically invariant geodesic equations
for mass-bearing particles} (non-isotropic geodesic equations)

\vspace*{-14pt}%
$$
\begin{array}{l}
\pm 
\displaystyle \frac{\displaystyle d}{\displaystyle d\tau }\left( \frac 1{%
\sqrt{1-{\rm v}^2/c^2}}\right) -\displaystyle \frac{\displaystyle F_i{\rm v}%
^i}{\displaystyle c^2\sqrt{1-{\rm v}^2/c^2}}+\displaystyle \frac{%
\displaystyle D_{ik}{\rm v}^i{\rm v}^k}{\displaystyle c^2\sqrt{1-{\rm v}%
^2/c^2}}=0\,, \\ \displaystyle \frac{\displaystyle d}{\displaystyle d\tau }%
\rule{0pt}{23pt}\left( \displaystyle \frac{\displaystyle {\rm v}^i}{%
\displaystyle \sqrt{1-{\rm v}^2/c^2}}\right) \mp \displaystyle \frac{%
\displaystyle F^i}{\displaystyle \sqrt{1-{\rm v}^2/c^2}}+\displaystyle \frac{%
\displaystyle \triangle _{nk}^i{\rm v}^n{\rm v}^k}{\displaystyle \sqrt{1-%
{\rm v}^2/c^2}}\, 
\rule{0pt}{18pt}+\displaystyle \frac{\displaystyle \left( 1\pm 1\right) }{%
\displaystyle \sqrt{1-{\rm v}^2/c^2}}\left( D_k^i+A_{k\cdot }^{\cdot
i}\right) {\rm v}^k=0\,, 
\end{array}
\eqno (129) 
$$

\vspace*{-3pt}\noindent as well as equations for {\it massless particles}
(isotropic geodesic equations)

\vspace*{-1pt}%
$$
\begin{array}{l}
D_{ik}c^ic^k-F_ic^i=0\,, \\ 
\displaystyle \frac{\displaystyle dc^i}{\displaystyle d\tau }\rule{0pt}{18pt}%
\mp F^i+\triangle _{nk}^ic^nc^k+\left( 1\pm 1\right) \left( D_k^i+A_{k\cdot
}^{\cdot i}\right) c^k=0\,. 
\end{array}
\eqno (130) 
$$

\vspace*{1pt}
The upper sign in the alternating terms in these equations stands for motion
of particles 
into past, while the lower sign stands for
motion into past. As seen, we again have the asymmetry of motion in time.
The same asymmetry was observed in dynamic equations of motion. We see that
this asymmetry does not depend upon physical properties of particles
themselves, but rather upon properties of the observer's space (body) of
reference, i.\thinspace e.\ upon $F^i$, $A_{ik}$, and $D_{ik}$. In
absence of gravitational inertial forces, rotation or deformation of the
space of reference, the asymmetry is absent too.

\subsection{Specific case: Newton laws}

\label{6-5}
In this Section we are going to prove that chronometrically invariant
dynamic equations of motion of mass-bearing particles are four-dimensional
generalization of Newton's 1st and 2nd laws in space-time, where
gravitational inertial force $F^i$, rotation $A_{ik}$, or deformation $%
D_{ik} $ are present.

At low speed $m{=}m_0$ and dynamic equations of motion become

\vspace*{-5pt}%
$$
\frac{DP^\alpha }{ds}=m_0\frac{d^2x^\alpha }{ds^2}+m_0\Gamma _{\mu \nu
}^\alpha \frac{dx^\mu }{ds}\frac{dx^\nu }{ds}=0\,,\qquad P^\alpha =\frac{%
dx^\alpha }{ds}\,,\eqno (131) 
$$

\noindent having the equations divided by $m_0$, dynamic
equations turn immediately into kinematic ones, i.\thinspace e.\ regular
non-isotropic geodesic equations.

These are dynamic equations of motion of so-called ``free particles'',
i.\,e.\ particles that fall freely under action of gravitational field.

Motion of particles under action of an additional force $R^{\alpha }$ not of
gravitational nature, is not geodesic and their equations of motion become

\vspace*{-5pt}%
$$
m_0\frac{d^2x^\alpha }{ds^2}+m_0\Gamma _{\mu \nu }^\alpha \frac{dx^\mu }{ds}%
\frac{dx^\nu }{ds}=R^\alpha .\eqno (132) 
$$

But these are dynamic equations of motion of particles in {\it four-%
dimensional space-time}, while Newton laws were set forth for 
{\it three-dimensional space}. In particular, the derivation parameter in
these equations is space-time interval, not applicable to three-dimensional
space.

Let us now look at chronometrically invariant dynamic equations of motion of
mass-bearing particles. At low speed these are

\vspace*{-6pt}%
$$
\frac{m_0}{c^2}\left( D_{ik}{\rm v}^i{\rm v}^k-F_i{\rm v}^i\right) =0\,,%
\eqno (133) 
$$

\vspace*{-7pt}%
$$
m_0\frac{d^2x^i}{d\tau ^2}-m_0F^i+m_0\triangle _{nk}^i{\rm v}^n{\rm v}%
^k+2m_0\left( D_k^i+A_{k\cdot }^{\cdot i}\right) {\rm v}^k=0\,,\eqno (134) 
$$

\noindent where spatial observable projections (134) are
dynamic equations of motion in three-dimensional space.

In space-time with three-dimensional Euclidean metric all values
$h_i^k{=}\delta _i^k$ and tensor of space deformation velocities $D_{ik}{=}%
\frac{\displaystyle1}{\displaystyle2}\frac{\displaystyle^{*}\partial h_{ik}}{%
\displaystyle\partial t}{=}0$. In this case $\triangle _{kn}^i{=}0$ and
hence the term $m_0\triangle _{nk}^i{\rm v}^n{\rm v}^k{=}0$. If also $F^i{=}%
0 $ and $A_{ik}{=}0$, spatial observable projections of equations of motion
become

\vspace*{-4pt}%
$$
m_0\frac{d^2x^i}{d\tau ^2}=0\,,\eqno (135) 
$$

\vspace*{-2pt}
\noindent or in another form

\vspace*{-13pt}%
$$
{\rm v}^i=\frac{dx^i}{d\tau }=const.\eqno (136) 
$$

\vspace*{1pt}
Hence four-dimensional generalization of {\it Newton's 1st law} for
mass-bearing particles can be set forth as follows:

\vspace*{6pt} \parshape=3 0.55cm 14.4cm 0.55cm 14.4cm 0.55cm 14.4cm
\noindent {\small {\bf If a particle is free from action of
gravitational inertial forces (or such forces are balanced) and at the same
time rotation or deformation of space is absent, such particle will
experience straight and even motion.}}

\vspace*{6pt} \noindent Such condition, as seen from formulas for
Christoffel symbols (90--95), is only possible when all $\Gamma _{\mu \nu
}^\alpha {=}0$, because any of Christoffel symbols are functions of at least
one of the values $F^i$, $A_{ik}$, or $D_{ik}$.

Now let us assume that $F^i{\neq }0$, but $A_{ik}{=}0$ and $D_{ik}{=}0$. In
this case three-dimensional dynamic equations of motion of particles become

\vspace*{-7pt}%
$$
\frac{d^2x^i}{d\tau ^2}=F^i.\eqno (137) 
$$

But gravitational potential and force $F^i$ (and values $%
A_{ik} $, $D_{ik}$) by definition describe the body of reference itself,
which is the source of gravitational field. Hence $F^i$ describes
gravitational field itself, i.\thinspace e.\ is a force that acts on a
unit-mass particle. Gravitational inertial force that acts on a particle
with mass $m_0$ is

\vspace*{-13pt}%
$$
\Phi ^i=m_0F^i,\eqno (138) 
$$

\noindent and three-dimensional dynamic equations of motion become

\vspace*{-5pt}%
$$
m_0\frac{d^2x^i}{d\tau ^2}=\Phi ^i.\eqno (139) 
$$

\vspace*{1pt}
Respectively, four-dimensional generalization of {\it Newton's 2nd law} for
mass-bearing particles can be set forth as follows:

\vspace*{6pt} \parshape=3 0.55cm 14.4cm 0.55cm 14.4cm 0.55cm 14.4cm
\noindent {\small {\bf Acceleration that a particle gains from
gravitational field is proportional to gravitational inertial force that
acts on the particle and is reciprocal to its mass in absence of deformation
or rotation of space.}}

\newpage
\noindent Having any particular value of gravitational
inertial force $\Phi ^i$ substituted into three-dimensional equations of
motion (134)

\vspace*{-8pt}%
$$
m_0\frac{d^2x^i}{d\tau ^2}+m_0\triangle _{nk}^i{\rm v}^n{\rm v}^k+2m_0\left(
D_k^i+A_{k\cdot }^{\cdot i}\right) {\rm v}^k=\Phi ^i,\eqno (140) 
$$

\noindent we can solve them to obtain trajectory of motion of particle in
three-dimensional space, i.\,e.\ dependance of its three-dimensional
coordinates from time. But presence of gravitational inertial force is not
mandatory to make motion curved and uneven. As seen from the equations this
happens if at least one of the values $F^i$, $A_{ik}$, or $D_{ik}$ is not
zero. Hence theoretically a particle can be in state of uneven and curved
motion in absence of gravitational inertial forces, but in presence of
rotation or deformation of space or both.

If particle moves under joint action of gravitational inertial force $\Phi
^i $ and another non-gravitational force $R^i$, its three-dimensional
dynamic equations of motion become

\vspace*{-6pt}%
$$
m_0\frac{d^2x^i}{d\tau ^2}+m_0\triangle _{nk}^i{\rm v}^n{\rm v}^k+2m_0\left(
D_k^i+A_{k\cdot }^{\cdot i}\right) {\rm v}^k=\Phi ^i+R^i.\eqno (141) 
$$

\vspace*{-1pt}
In a flat three-dimensional space, as we have mentioned in the above, $%
\triangle _{kn}^i{=}0$ and the second term in these equations will be zero.

If in a {\it flat} three-dimensional space rotation or deformation are
absent, dynamic equations of motion of particles become very simple

\vspace*{-6pt}%
$$
m_0\frac{d^2x^i}{d\tau ^2}=\Phi ^i,\qquad \qquad m_0\frac{d^2x^i}{d\tau ^2}%
=\Phi ^i+R^i,\eqno (142) 
$$

\noindent to describe motion under action of gravitational inertial force $%
\Phi ^i$ (the first equations) as well as motion under joint action of $\Phi
^i$ and another non-gravitational force $R^i$ which deviates particles from
geodesic line (the second equations).

We obtained, that motion under action of gravitational inertial forces is
possible in either curved or flat space. Why?

As known, curvature of space-time is characterized by Riemann-Christoffel
tensor of curvature $R_{\alpha \beta \gamma \delta }$, which depends upon
the second derivatives of $g_{\alpha \beta }$ and the first derivatives of $%
g_{\alpha \beta }$. The condition $R_{\alpha \beta \gamma \delta }{\neq }0$
is the necessary and sufficient condition of curved space. To have non-zero
curvature of space-time it is necessary and sufficient that the second
derivatives of metric $g_{\alpha \beta }$ are non-zeroes.

But we also know, that the first derivatives of space-time metric
$g_{\alpha \beta }$ in a flat space-time may be not equal to zero.

Our chronometrically invariant equations of motion contain values $\triangle
_{kn}^i$, $F^i$, $A_{ik}$, and $D_{ik}$, which depend upon the first
derivatives of metric tensor only. Therefore at $R_{\alpha \beta \gamma
\delta }{=}0$ (flat space-time) gravitational inertial force $F^i$, rotation 
$A_{ik}$, and deformation of space $D_{ik}$, which are functions of the
first derivatives of metric $g_{\alpha \beta }$ may be not equal to zero.

\subsection{Resume}

\label{6-6}
So, did we get?

First, using method of chronometric invariants we have obtained, that in
General Relativity space-time particles may move not only into future (in
respect to us), but into past as well. Besides, physical observable values
for mass-bearing particles are relativistic mass $\pm m$ and
three-dimensional value $\frac{\displaystyle1}{\displaystyle c}m{\rm v}^i$,
where plus stands for motion of particle into future, while minus stands for
motion into past. For massless particles physical observable values are wave
number $\pm k{=}{\pm }\frac{\displaystyle\omega }{\displaystyle c}$ and
three-dimensional value $\frac{\displaystyle1}{\displaystyle c}kc^i$.
Besides, motions of particles into future and into past are asymmetric in
respect to each other. This asymmetry depends upon properties of space only
(i.\thinspace e.\ gravitational inertial force $F^i$, rotation $A_{ik}$, and
deformation $D_{ik}$).

Further, looking at motion of particles as propagation of waves
(within de~Broglie's wave-particle dual concept), we observe no asymmetry.
From observer's viewpoint propagation of waves is the same in both
directions in time, while movement of particle ``balls'' is not.

Second, we have shown that chronometrically invariant dynamic
equations of motion are generalization of Newton's laws in four-%
dimensional space-time, where gravitational inertial force $F^i$, rotation $%
A_{ik}$, or deformation $D_{ik}$ are present. Motion of particles is
straight and even only if $F^i{=}0$, $A_{ik}{=}0$, and $D_{ik}{=}0$. If any
of the above values is not zero, motion of a particle is no longer straight
and even. Besides, curved and uneven motion may take place in a flat space
too, of course provided at least one of the values $F^i$, $A_{ik}$, or $%
D_{ik}$ is not zero.

All the above results have been obtained exclusively thanks to Zelmanov's
method of chronometric invariants (the mathematical apparatus of physical
observable values). Regular general covariant methods would be of no use
here.

As a result we can see that not all physical effects in General Relativity
are yet known in contemporary science. Further development of experimental
physics and astronomy will discover new phenomena, related, in particular,
to acceleration, rotation and deformation of space of reference.


\section{Analysis of the equations}

\subsection{Space-time and zero-space: limit transitions}

\label{7-1}
As we can see, at $w{=}{-}v_iu^i$ in our formulas values in generalized
space-time ($g{\leq }0$) are replaced by those in non-degenerated space-time
($g{<}0$)

\vspace*{-2pt}%
$$
d\tau =\left[ 1-\frac 1{c^2}\left( w+v_iu^i\right) \right] dt=dt\,,\eqno %
(143) 
$$

\vspace*{-2pt}%
$$
u^i=\frac{dx^i}{dt}=\frac{dx^i}{d\tau }=\text{{\rm v}}^i,\eqno (144) 
$$

\vspace*{-2pt}%
$$
M=\frac m{1-\frac{\displaystyle1}{\displaystyle c^2}\left( w+v_iu^i\right)
}=m\,,\eqno (145) 
$$

\vspace*{-2pt}%
$$
P^0=M=m\,,\qquad P^i=\frac 1cMu^i=\frac 1cm\text{{\rm v}}^i,\eqno 
(146) 
$$

\noindent and in this transition coordinate time $t$ coincides to observable
time $\tau $.

Of course this transformation also occurs under a narrower condition $w{=}{-}%
v_iu^i{=}0$, when $w{\rightarrow }0$ (weak gravitational field) and $v_i{=}0$
(no rotation of space) at the same time. But in the observed part of the
Universe there is hardly an area where rotation and gravitational filed are
absolutely absent. Therefore we see that transition to regular (strictly
non-degenerated) space-time always happens at

\vspace*{-4pt}%
$$
w=-v_iu^i=-v_i\text{{\rm v}}^i.\eqno (147) 
$$

\vspace*{1pt}
Corpuscular equations of motion of mass-bearing and massless particles in
generalized space-time transform into the equations in regular (strictly
non-degenerated) space-time only in case of motion from past into future
(direct flow of time). That is, only for real mass-bearing particles with $m{%
>}0$, imaginary mass-bearing particles with $im{>}0$ and for massless
particles with $\omega {>}0$.

Equations of motion in wave form in generalized space-time transform into
the equations in regular space-time in the same way for particles with $m{>}%
0 $, $im{>}0$ and $\omega {>}0$ (direct flow of time), and for particles
with $m{<}0$, $im{<}0$ and $\omega {<}0$ (reverse flow of time).

Later we are going to find out {\it why} does that happen.

In regular space-time ($g{<}0$) we obtained $P^0$ (26), which after
substitution of $\frac{\displaystyle dt}{\displaystyle d\tau }$ (32) and the
transition conditions $w{=}{-}v_iu^i{=}{-}v_i{\rm v}^i$ takes
sign-alternating relativistic mass, i.\,e.\ values\rule{0pt}{7pt} $+m$ and $%
-m$

\vspace*{-3pt}%
$$
P^0=m\frac{dt}{d\tau }=\frac m{1-\frac{\displaystyle 
w}{\displaystyle c^2}}\left( \frac 1{c^2}v_i\text{{\rm v}}^i\pm 1\right)
=\pm m\,.\eqno 
(148) 
$$

\vspace*{-2pt}In generalized space-time $g{\leq }0$ we obtained $P^0{=}M$,
but through another method (62) without using $\frac{\displaystyle dt}{%
\displaystyle d\tau }$, due to which the formula accepts alternating sign.

But in such case the component $P^0{=}{\pm }m$ in regular space-time (148),
taking two values, can not be a partial case of a single value $P^0{=}M$ in
generalized space-time.

To understand the reason we turn from sign-alternating $P^0{=}{\pm }m$ in
regular space-time to $P^0{=}M$ in generalized space-time. This can be
easily done by substituting the already known relationship between physical
observable velocity ${\rm v}^i$ and its coordinate velocity $u^i$ (57), into
sign-alternating formula $P^0{=}{\pm }m$ (148).

As a result we obtain expanded relationship for the component $P^0$ in
generalized space-time

\vspace*{-5pt}%
$$
P^0=\frac m{1-\frac{\displaystyle w}{\displaystyle c^2}}\left[ \frac 1{c^2}%
\frac{v_iu^i}{1-\frac{\displaystyle1}{\displaystyle c^2}\left(
w+v_iu^i\right) }\pm 1\right] ,\eqno (149) 
$$

\vspace*{-1pt}\noindent which evidently accepts alternating sign. For
particles that move in generalized space-time from past into future $P^0$
becomes

\vspace*{-6pt}%
$$
P^0=\frac m{1-\frac{\displaystyle1}{\displaystyle c^2}\left( w+v_iu^i\right)
}=+M\,,\eqno (150) 
$$

\vspace*{-2pt}\noindent which is similar to (62). For particles that travel
from future into past, $P^0$ becomes

\vspace*{-5pt}%
$$
P^0=\frac{m\left[ \frac{\displaystyle1}{\displaystyle c^2}\left(
2v_iu^i+w\right) -1\right] }{\left( 1-\frac{\displaystyle w}{\displaystyle %
c^2}\right) \left[ 1-\frac{\displaystyle1}{\displaystyle c^2}\left(
w+v_iu^i\right) \right] }=-M\,.\eqno (151) 
$$

\vspace*{-1pt}
The obtained $P^0$ values are final generalized equations, because in
transition to regular space-time the first value $P^0{=}{+}M$ (150)
unambiguously transforms into $P^0{=}{+}m$, while the second value $P^0{=}{-}%
M$ (151) transforms into $P^0{=}{-}m$.

Noteworthy, our remarks in respect to sign-alternating value $P^0$ do not
affect correctness of the obtained equations of motion, because those
include gravitational rotational mass in general notation $M$ without any
respect to its particular composition. The only difference is that
sequential substitution of the two values of $M$ into equations of motion in
generalized space-time will produce independent equations: for particles
that move from past into future and for those that move from future into
past.

Let us now return to physical condition $w{=}{-}v_iu^i$ (147), which marks
the transition from dynamic equations of motion in generalized space-time to
those in regular space-time. We have also seen that under this condition $%
d\tau {=}dt$ (143). But we know that in regular space-time the equality $%
d\tau {=}dt$ is not imperative. To the contrary, in the observed Universe
interval of physical observable time $d\tau $ is almost always a bit
different from interval of coordinate time $dt$.

Therefore transition from generalized space-time to regular space-time
occurs under physical conditions $w{=}{-}v_iu^i$, which are merely a partial
case of physical conditions in regular space-time. But from here we see that
strictly non-degenerated space-time ($g{<}0$) is {\it not} an area of
generalized space-time $g{\leq }0$ we have looked at.

Does that contain a contradiction between equations of motion in regular
space-time and in generalized space-time?

No it doesn't. All laws applicable to regular space-time ($g{<}0$) are as
well true in non-degenerated area ($g{<}0$) of generalized space-time $g{%
\leq }0$. Those two non-degenerated areas {\it are not} the same. That is,
degenerated space-time added to regular space-time produces two absolutely
separate manifolds. Generalized space-time is a different manifold
absolutely independent from either strictly non-degenerated space-time or
degenerated one. And there is no surprise in that transition from one to
another occurs under very limited partial conditions.

The only question is what configuration of those manifolds exists in the
observable Universe. Two options are possible here:

\begin{enumerate}
\item  \vspace*{-7pt}non-degenerated space-time ($g{<}0$) and degenerated
space-time ($g{=}0$) exist as two separate manifold (regular
space-time of General Relativity with a small ``add-on'' of zero-space);

\item  \vspace*{-7pt}non-degenerated space-time and degenerated space-time
exist
as two internal areas of the same manifold --- generalized space-%
time ($g{\leq }0$).
\end{enumerate}

\vspace*{-7pt}In either case transition from non-degenerated space-time into
degenerated one occurs under physical condition of degeneration (56). Future
experiments will show which one of these two options exists in reality.

\subsection{Space-time asymmetry and world beyond the Mirror}

\label{7-2}
Let us now compare corpuscular equations of motion for particles with $m{>}0$
(115) and $\omega {>}0$ (122) with those for particles with $m{<}0$ (116)
and $\omega {<}0$ (123).

Even the first look at the equations shows that corpuscular equations of
motion for particles with positive relativistic mass or frequency (which
travel from past into future) are different from those for particles with
negative mass or frequency (which travel into past). The same asymmetry
exists for the wave form of such equations of motion. Why?

Asymmetry of equations of motion for particles that move into future or into
past says that in four-dimensional uneven space-time there exists a
fundamental asymmetry of directions from past into future and from future
into past.

To understand the reasons for such fundamental asymmetry let us consider an
example.

We assume that in four-dimensional space-time there exists a {\it Mirror}
which coincides with the spatial section and hence separates past from
future. We also assume that the Mirror reflects particles and waves that
move either from past into future or from future into past. Then particles
that travel from past into future ($m{>}0$, $im{>}0$, and $\omega {>}0$)
always hit the Mirror and bounce back in time, i.\thinspace e.\ into past.
Consequently their properties reverse ($m{<}0$, $im{<}0$, and $\omega {<}0$%
). And {\it vice versa} particles and waves that travel from past future
into past ($m{<}0$, $im{<}0$, and $\omega {<}0$) hitting the Mirror change
the sign of their properties ($m{>}0$, $im{>}0$, and $\omega {>}0$) to
bounce back into future.

Now everything becomes easy to understand. Let us look at the wave form of
dynamic equations of motion (118). After reflection from the Mirror the
value $\frac{\displaystyle^{*}\partial \psi }{\displaystyle\partial t}$%
\rule[-6pt]{0pt}{0pt}
changes its sign. Hence the equations of propagation of wave into future
(``plus'' in the equations) become those of propagation of the same wave
into past (``minus'' in the equations). And {\it vice versa} the equations
of propagation into past (``minus'') after reflection become the equations
of propagation into future (``plus'').

Noteworthy, equations of propagation of waves from past into future and from
future into past transform into each other {\it in full}, i.\,e.\ no terms
are contracted and no terms are added. Hence the wave form of matter {\it %
fully reflects} from the Mirror.

But this is not the case for corpuscular equations. After reflection from
the Mirror the values $\varphi {=}{\pm }m$ for mass-bearing particles and $%
\varphi {=}{\pm }k{=}{\pm }\frac{\displaystyle\omega }{\displaystyle c}$ for
massless particles change their signs. But here equations of motion from
past into future transform into those from future into past {\it not in full}%
.

In spatial projections of equations of motion from past into future there is
an additional term present. The term is not found in spatial projections of
equations of motion from past into future. For mass-bearing and massless
particles the term is, respectively

\vspace*{-2pt}%
$$
2m\left( D_k^i+A_{k\cdot }^{\cdot i}\right) {\rm v}^k,\qquad \quad 2k\left(
D_k^i+A_{k\cdot }^{\cdot i}\right) c^k.\eqno (152) 
$$

\vspace*{2pt}Hence we see that particle that moves from future into past
hits the Mirror and bounces back to acquire an additional term in dynamic
equations of motion. And {\it vice versa}, particle that travels from past
into future bounces from the Mirror to loose a term in its dynamic
equations. Therefore the Mirror itself affects trajectories of particles!

As a result, particles with negative masses or frequencies {\it are not}
simple mirror reflections of particles with positive masses or frequencies.
Either in case of motion of particles or in case of propagation of waves we
do not deal with simple reflection or bouncing from the Mirror, but with 
{\it penetration} through the Mirror into the {\it mirror world}.

In this mirror world all particles bear negative masses or frequencies and
move (from viewpoint of our world's observer) from future into past. The
wave form of matter from our world has no effect on events in the mirror
world, while the wave form of matter in the mirror world has no effect on
events in our world. To the contrary, particles in our world may affect
events in the mirror world and particles in the mirror world may have effect
on events in our world.

Full isolation of our world from the mirror world, i.\,e.\ absence of mutual
influence between particles of both worlds takes place under an evident
condition

\vspace*{-1pt}%
$$
D_k^i{\rm v}^k=-A_{k\cdot }^{\cdot i}{\rm v}^k,\eqno (153) 
$$

\vspace*{2pt}\noindent when the auxiliary term (152), which causes asymmetry
in corpuscular equations of motion, is zero. This happens when $D_k^i{=}0$
and $A_{k\cdot }^{\cdot i}{=}0$, i.\thinspace e.\ with full absence of
deformation or rotation of the space of the body of reference.

Noteworthy, if particles with positive masses (or frequencies) co-existed in
our world with those with negative masses (or frequencies), they would
interfere to destroy each other inevitably and no particles would be left in
our world. But we observe nothing of the kind.

Therefore in the second part of analysis of the obtained equations of motion
we can make the following conclusions.

\begin{enumerate}
\item  \vspace*{-7pt}Fundamental asymmetry of directions from past into
future and from future into past is due to existence of a certain mirror
medium (the Mirror) in space-time, which fills spatial sections. All
particles or waves that travel either from past or from future reflect from
it. This mirror medium is degenerated space-time (zero-space);

\item  \vspace*{-7pt}Space-time falls apart into our world and the mirror
world. In our world $m{>}0$, $im{>}0$, $\omega {>}0$, and $\frac{%
\displaystyle^{*}\partial \psi }{\displaystyle\partial t}{>}0$,%
\rule{0pt}{14pt} and all
particles travel from past into future. In the mirror world\rule{0pt}{9pt} $m%
{<}0$, $im{<}0$, $\omega {<}0$, and $\frac{\displaystyle^{*}\partial \psi }{%
\displaystyle\partial t}{<}0$ and particles move from future into past;

\item  \vspace*{-7pt}Particles with $m{<}0$, $im{<}0$, $\omega {<}0$, and
waves with $\frac{\displaystyle^{*}\partial \psi }{\displaystyle\partial t}{<%
}0$, which travel from future into past are particles of our world%
\rule{0pt}{9pt} that penetrated into the mirror world through the Mirror;

\item  \vspace*{-7pt}We can not observe particles with negative masses or
frequencies not waves with negative phases because they exist in the mirror
world, i.\thinspace e.\ beyond the Mirror. Particles or waves that we can
observe on the exit from the Mirror (or when bouncing the Mirror, as it
seems to us) have positive properties as they have come from the mirror
world into our world and travel from past into future.
\end{enumerate}


\section{Conditions of direct and reverse flow of time}

\label{8}
In this Section we are going to look at physical conditions under which: (a)
time has direct flow, i.\thinspace e.\ from past into future, (b) time has
reverse flow, i.\thinspace e.\ from future into past, and (c) time stops.

In contemporary physics time is defined as the fourth coordinate $x^0{=}ct$
of four-dimensional space-time, where $c$ is speed of light and $t$ is
coordinate time. The structure of the formula itself says that $t$ changes
evenly with the speed of light and does not depend upon physical conditions
of observation. Hence coordinate time is also referred to as {\it ideal time}%
. Aside for ideal time there is observer's {\it real time} $\tau $, which
depends upon conditions of observation. Theory of chronometric invariants
defines interval of physical observable time as projection of increment of
four-dimensional coordinates $dx^\alpha $ on time

\vspace*{-4pt}%
$$
d\tau =\frac 1cb_\alpha dx^\alpha .\eqno (154) 
$$

In the frame of reference of sub-light speed (substantional) observer, which
accompanies their body of reference, interval of observable time according
to (13) is

\vspace*{-4pt}%
$$
d\tau =\left( 1-\frac w{c^2}\right) dt-\frac 1{c^2}v_idx^i=dt-\frac
1{c^2}wdt-\frac 1{c^2}v_idx^i.\eqno (155) 
$$

From here we see that $d\tau $ consists of three parts: (a)
interval of coordinate time $dt$, (b) interval of ``gravitational'' time $%
dt_g{=}\frac{\displaystyle1}{\displaystyle c^2}wdt$, and (3) interval of
``rotational'' time $dt_{rot}{=}\frac{\displaystyle1}{\displaystyle c^2}%
v_idx^i$. The stronger is gravitational field of the body of reference and
the faster rotates the space of reference, the slower flows observer's time.
Theoretically strong enough gravitational field and fast enough rotation of
space may stop observer's physical time.

We define the {\it mirror world} as space-time where time flows backward in
respect to that in reference space-time. Direction of coordinate
time $t$, which describes displacement along temporal coordinate axis $x^0{=}%
ct$, is defined by the sign of derivative $\frac{\displaystyle dt}{%
\displaystyle d\tau }$. Respectively, direction of observable time $\tau $
is defined by the sign of derivative $\frac{\displaystyle d\tau }{%
\displaystyle dt}$.

The formula for $\frac{\displaystyle dt}{\displaystyle d\tau }$ was obtained
in Section~3 from the condition of conservation of four-dimensional velocity
of particle along its four-dimensional trajectory as (28--30). But it can be
also obtained in another way by presenting the square of space-time interval 
$ds^2{=}c^2d\tau ^2{-}d\sigma ^2$ as

\vspace*{-4pt}%
$$
ds^2=\left( 1-\frac w{c^2}\right) ^2c^2dt^2-2\left( 1-\frac w{c^2}\right)
v_idx^idt+g_{ik}dx^idx^k.\eqno (156) 
$$

From here we see that the square of elementary distance
between two infinitely close points in space-time is the sum of the square
of three-dimensional distance $g_{ik}dx^idx^k$ and of two terms, which
depend upon physical properties of space.

The value $\left( 1{-}\frac{\displaystyle w}{\displaystyle c^2}\right)
^{\!2}\!\!c^2dt^2$ is a term in $ds^2$ caused by presence of the fourth
dimension (time) and presence of space of reference's own gravitational
field $w$. In absence of gravitational field temporal coordinate $x^0{=}ct$
changes evenly with speed of light. But if $w{\neq }0$, coordinate $x^0$
changes ``slower'' by value $\frac{\displaystyle w}{\displaystyle c^2}$. The
stronger is gravitational potential $w$, the slower flows coordinate time.
At $w{=}c^2$ coordinate time stops at all. As well-known, such condition is
implemented in gravitational collapse (black~hole).

The value $\left( 1{-}\frac{\displaystyle w}{\displaystyle c^2}\right)
\!v_idx^idt$ is a term in $ds^2$ that is due to joint action of
gravitational and space rotation. This term is not zero only when $w{\neq }%
c^2$ (no gravitational collapse) and $v_i{\neq }0$ (three-dimensional space
rotates, i.\thinspace e.\ is non-holonomic).

\vspace*{-3pt}Having both parts of (156) divided by $ds^2{=}c^2d\tau
^2\!\left( 1{-}\frac{\displaystyle{\rm v}^2}{\displaystyle c^2}\right) $ we
obtain quadratic equation in respect to $\frac{\displaystyle dt}{%
\displaystyle d\tau }$ (31), which has two solutions similar to (32). From
this formula we see that coordinate time increases $\frac{\displaystyle dt}{%
\displaystyle d\tau }{>}0$, stops $\frac{\displaystyle dt}{\displaystyle %
d\tau }{=}0$ and decreases $\frac{\displaystyle dt}{\displaystyle d\tau }{<}%
0 $ under following conditions

\vspace*{-1pt}%
$$
\frac{dt}{d\tau }>0\qquad \text{if}\quad v_i{\rm v}^i>\pm c^2,\eqno (157) 
$$

\vspace*{-5pt}%
$$
\frac{dt}{d\tau }=0\qquad \text{if}\quad v_i{\rm v}^i=\pm c^2,\eqno (158) 
$$

\vspace*{-5pt}%
$$
\frac{dt}{d\tau }<0\qquad \text{if}\quad v_i{\rm v}^i<\pm c^2.\eqno (159) 
$$

\vspace*{3pt}
As known, regular real particles' velocities are lower than the speed of
light. Therefore the condition of coordinate time stop $v_i{\rm v}^i{=}{\pm }%
c^2$ (158) can not be true in the world of substance, but is not impossible
for other states of matter (for light-like matter, for instance).

Coordinate time increases $\frac{\displaystyle dt}{\displaystyle d\tau }{>}0$
(157) at $v_i{\rm v}^i{>}{\pm }c^2$. In a regular laboratory rotational
velocities are also below\rule{0pt}{9pt} the speed of light. Hence in
regular laboratory conditions $v_i{\rm v}^i{>}{-}c^2$ (the angle $\alpha $
between the rotation velocity and the observable velocity is within the
limits ${-}\frac \pi 2{<}\alpha {<}\frac \pi 2$). In this case flow of
coordinate time is direct, i.\thinspace e.\ from past into future.

Coordinate time decreases $\frac{\displaystyle dt}{\displaystyle d\tau }{<}0$
at $v_i{\rm v}^i{<}{\pm }c^2$.

Until\rule{0pt}{9pt} now we have looked at flow of coordinate time $t$ only.
Now we are going to analyze possible directions of physical observable time $%
\tau $, which depends upon the sign of derivative $\frac{\displaystyle d\tau 
}{\displaystyle dt}$. We obtain this value by dividing the formula for $%
d\tau $ (155) by $dt$

\vspace*{-3pt}%
$$
\frac{d\tau }{dt}=1-\frac 1{c^2}\left( w+v_iu^i\right) .\eqno (160) 
$$

\vspace*{2pt}
By definition, regular observer's clock always counts positive intervals of
time irrespective of in what direction clock's hands rotate. Therefore in a
regular laboratory on Earth physical observable time may increase or stop,
but it never decreases. Nevertheless decrease of observable time $\frac{%
\displaystyle d\tau }{\displaystyle dt}{<}0$ is possible in certain
circumstances.

From (160) we see that observable time increases $\frac{\displaystyle d\tau 
}{\displaystyle dt}{>}0$, stops $\frac{\displaystyle d\tau }{\displaystyle dt%
}{=}0$ or decreases $\frac{\displaystyle d\tau }{\displaystyle dt}{<}0$
under the following conditions

\vspace*{-9pt}%
$$
\frac{d\tau }{dt}>0\qquad \text{if}\quad w+v_iu^i<c^2,\eqno (161) 
$$

\vspace*{-12pt}%
$$
\frac{d\tau }{dt}=0\qquad \text{if}\quad w+v_iu^i=c^2,\eqno (162) 
$$

\vspace*{-7pt}%
$$
\frac{d\tau }{dt}<0\qquad \text{if}\quad w+v_iu^i>c^2.\eqno (163) 
$$

\vspace*{-1pt}
Evidently, the condition of observable time stop $w{+}v_iu^i{=}c^2$ is also
the condition of degeneration of space-time (56). In a partial case, when
rotation of space is absent, observable time stops in collapse~$w{=}c^2$.

Generally zero-space is described by whole spectrum of physical conditions
represented as\linebreak $w{+}v_iu^i{=}c^2$. Black holes ($w{=}c^2$) is only a
partial case of such conditions in absence of space's rotation $v_i{=}0$. In
other words, the {\it mirror membrane} between the world with direct flow of
time and the world with reverse flow of time (the mirror world) are not
black holes in zero-space alone, but the zero-space in general.

So what is flow of coordinate time $t$ and what is flow of physical
observable time $\tau $?

In the function of coordinate time $\frac{\displaystyle dt}{\displaystyle %
d\tau }$ we assume that observer's measured time $\tau $ is the standard, in
respect\rule{0pt}{9pt} to which time coordinate $t$ is defined. Here we are
linked to the observer themselves and from their point of view we define
where do they travel (into past, into future or rests). That is, the
function of coordinate time $\frac{\displaystyle dt}{\displaystyle d\tau }$
defines {\it observable motion}\rule{0pt}{9pt} of the observer along time
axis $x^0{=}ct$ from their own viewpoint.

In the function of observable time $\frac{\displaystyle d\tau }{%
\displaystyle dt}$ change of observer's temporal coordinate $t$ is the
standard. That is,\rule{0pt}{9pt} observer's measured time $\tau $ is
defined in respect to motion of the whole observer's spatial section along
the axis of time, which occurs evenly at the speed of light. Therefore
function of observable time $\frac{\displaystyle d\tau }{\displaystyle dt}$
gives a view of the observer from aside, showing their {\it true motion} in
respect\rule{0pt}{9pt} to time axis.

In other words, the function of coordinate time $\frac{\displaystyle dt}{%
\displaystyle d\tau }$ shows the membrane between the world with direct flow
of time and\rule{0pt}{9pt} that with the reverse flow of time from
``inside'', from viewpoint of the observer who travels into future or into
past. The function of observable time $\frac{\displaystyle d\tau }{%
\displaystyle dt}$ gives a look at the membrane from ``outside'', from
viewpoint of space-time\rule{0pt}{9pt} itself. This means that it is the
function of observable time $\frac{\displaystyle d\tau }{\displaystyle dt}$
that shows the {\it true structure} of space-time membrane between the worlds%
\rule{0pt}{9pt} with direct and reverse flow of time.


\section{Basic introduction into the mirror world}

\label{9}
To obtain a more detailed view of space-time membranes we are going to use 
{\it local geodesic frame of reference}. Fundamental metric tensor within
infinitesimal vicinities of any point of such frame is

\vspace*{-6pt}%
$$
\tilde g_{\mu \nu }=g_{\mu \nu }+\frac 12\left( \frac{\partial ^2\tilde
g_{\mu \nu }}{\partial \tilde x^\rho \partial \tilde x^\sigma }\right)
\left( \tilde x^\rho -x^\rho \right) \left( \tilde x^\sigma -x^\sigma
\right) +\ldots \ ,\eqno (164) 
$$

\vspace*{-1pt}\noindent i.\thinspace e.\ values of its components in the
vicinities of a point is different from the those at this point itself are
only different by figures of 2nd order of smallness and less which can be
neglected. Therefore at any point of local geodesic frame of reference
fundamental metric tensor (up within the 2nd order of smallness figures) is
a constant, while the first derivatives of the metric, i.\thinspace e.\
Christoffel symbols, are zeroes \cite{bib10}.

Evidently within infinitesimal vicinities of any point of Riemannian space a
local geodesic frame of reference can be set. Subsequently at any point of
local geodesic frame of reference a tangential flat space can be set so that
local geodesic frame of reference of the Riemannian space is global geodesic
one for that flat space. Because in a flat space metric tensor is constant,
in the vicinities of a point of Riemannian space the values $\tilde g_{\mu
\nu }$ converge to values of that tensor $g_{\mu \nu }$ in tangential flat
space. That means that in a tangential flat space we can build a system of
basic vectors $\vec e_{(\alpha )}$ tangential to curved coordinate lines of
the Riemannian space. Because coordinate lines of Riemannian space are
generally curved and in non-holonomic space are not even orthogonal to each
other, lengths of basic vectors are sometimes substantially different from
unit length.

Let $d\vec r$ be a four-dimensional vector of infinitesimal displacement
$d\vec r{=}(dx^0,dx^1,dx^2,dx^3)$. Then $d\vec r{=}\vec e_{(\alpha
)}dx^\alpha $, where the components are

\vspace*{-3pt}%
$$
\begin{array}{ll}
\vec e_{(0)}=(e_{(0)}^0,0,0,0), & \qquad \vec e_{(1)}=(0,e_{(1)}^1,0,0), \\ 
\vec e_{(2)}=(0,0,e_{(2)}^2,0),\rule{0pt}{16pt} & \qquad \vec
e_{(3)}=(0,0,0,e_{(3)}^3). 
\end{array}
\eqno (165) 
$$

Scalar product of vector $d\vec r$ with itself gives $d\vec rd\vec r{=}ds^2$%
, i.\thinspace e.\ the square of four-dimensional interval. On the other
hand $ds^2{=}g_{\alpha \beta }dx^\alpha dx^\beta $. Hence

\vspace*{-4pt}%
$$
g_{\alpha \beta }=\vec e_{(\alpha )}\vec e_{(\beta )}=e_{(\alpha )}e_{(\beta
)}\cos (x^\alpha ;x^\beta ),\eqno (166) 
$$

\vspace*{1pt}
\noindent which facilitates better understanding of geometric structure of
different areas within Riemannian space and even beyond. According to the
formula

\vspace*{-7pt}%
$$
g_{00}=e_{(0)}^2\,,\eqno (167) 
$$

\noindent  and on the other hand $\sqrt{g_{00}}{=}1{-}\frac{\displaystyle w}{%
\displaystyle c^2}$. Hence length of temporal basic vector $\vec e_{(0)}$
tangential to coordinate line of time $x^0{=}ct$ is

\vspace*{-12pt}%
$$
e_{(0)}=\sqrt{g_{00}}=1-\frac w{c^2}\eqno (168) 
$$

\vspace*{-2pt}\noindent and is the lesser than one the greater is the
gravitational potential $w$. In case of collapse ($w{=}c^2$) the
length of temporal basic vector $\vec e_{(0)}$ becomes zero.

According to (166) the value $g_{0i}$ is

\vspace*{-6pt}%
$$
g_{0i}=e_{(0)}e_{(i)}\cos (x^0;x^i),\eqno (169) 
$$

\noindent on the other hand $g_{0i}{=}{-}\frac{\displaystyle1}{\displaystyle %
c}v_i\left( 1{-}\frac{\displaystyle w}{\displaystyle c^2}\right) {=}{-}\frac{%
\displaystyle1}{\displaystyle c}v_ie_{(0)}$. Hence

\vspace*{-2pt}%
$$
v_i=-ce_{(i)}\cos (x^0;x^i).\eqno (170) 
$$

\vspace*{1pt}Then according to general formula (166)

\vspace*{-4pt}%
$$
g_{ik}=e_{(i)}e_{(k)}\cos (x^i;x^k),\eqno (171) 
$$

\vspace*{-1pt}\noindent we obtain that observable metric tensor $h_{ik}{=}{-}%
g_{ik}{+}\frac{\displaystyle 1}{\displaystyle c^2}v_iv_k$ takes the form

\vspace*{-3pt}%
$$
h_{ik}=e_{(i)}e_{(k)}\left[ \cos (x^0;x^i)\cos (x^0;x^k)-\cos
(x^i;x^k)\right] .\eqno (172) 
$$

From (170) we see that geometrically $v_i$ is a projection (scalar product)
of spatial basic vector $\vec e_{(i)}$ onto temporal basic vector $\vec
e_{(0)}$, multiplied by speed of light. If spatial sections are everywhere
orthogonal to lines of time (holonomic space), then $\cos (x^0;x^i){=}0$ and 
$v_i{=}0$. In non-holonomic space spatial sections are not orthogonal to
lines of time and $\cos (x^0;x^i){\neq }0$. Generally $|\!\cos (x^0;x^i)|{%
\leq }1$ hence velocity of space rotation $v_i$ (170) can not exceed speed
of light.

If $\cos (x^0;x^i){=}{\pm }1$, then velocity of space rotation is

\vspace*{-5pt}%
$$
v_i=\mp ce_{(i)}\,,\eqno (173) 
$$

\vspace*{1pt}
\noindent and temporal basic vector $\vec e_{(0)}$ coincides with spatial
basic vectors $\vec e_{(i)}$ (time ``falls'' into space). At $\cos (x^0;x^i){%
=}{+}1$ temporal basic vector is co-directed with the spatial ones $\vec
e_{(0)}{\uparrow }{\uparrow }\vec e_{(i)}$. In case $\cos (x^0;x^i){=}{-}1$
temporal and spatial basic vectors are oppositely directed $\vec e_{(0)}{%
\uparrow }{\downarrow }\vec e_{(i)}$.

Let us have a closer look at the condition $\cos (x^0;x^i){=}{\pm }1$. If
any spatial basic vector is co-directed (or oppositely directed) to the
temporal basic vector the space is degenerated. Maximum degeneration occurs
when all three vectors $\vec e_{(i)}$ coincide with each other and with the
temporal basic vector $\vec e_{(0)}$.

The condition of stop of coordinate time $v_i{\rm v}^i{=}{\pm }c^2$
presented through the basic vectors is

\vspace*{-5pt}%
$$
e_{(i)}v^i\cos (x^0;x^i)=\mp c\eqno (174) 
$$

\vspace*{1pt}
\noindent and becomes true when $e_{(i)}{=}1$, ${\rm v}{=}c$ and $\cos
(x^0;x^i){=}{\pm }1$. In this case if the velocity of rotation reaches the
speed of light the angle between the time line and the spatial lines becomes
either zero or $\pi $ depending upon the rotation direction.

Let us illustrate this with a few examples.

\medskip
\noindent {\bf 1. Space does not rotate, i.\thinspace e.\ is holonomic}

\smallskip
\noindent In this case $v_i{=}0$ and spatial sections are everywhere
orthogonal to lines of time and the angle between them is $\alpha {=}\frac
\pi 2$. Hence in absence of space rotation temporal basic vector $\vec
e_{(0)}$ is orthogonal to all spatial basic vectors $\vec e_{(i)}$. That
means that all clocks can be synchronized and will display the same time
(synchronization of clocks at different points in space does not depend upon
ways of synchronization). Velocity of space rotation $v_i{=}{-}ce_{(i)}\cos
\alpha {=}0$. At $v_i{=}0$

\vspace*{-1pt}%
$$
d\tau =\left( 1-\frac w{c^2}\right) cdt\,,\qquad \qquad h_{ik}=-g_{ik}\,,%
\eqno (175) 
$$

\vspace*{-1pt}\noindent and metric of space-time $ds^2{=}c^2d\tau ^2{-}%
d\sigma ^2$ becomes

\vspace*{-4pt}%
$$
ds^2=\left( 1-\frac w{c^2}\right) ^2c^2dt^2+g_{ik}dx^idx^k,\eqno (176) 
$$

\vspace*{-1pt}\noindent i.\thinspace e.\ pace of observable time depends
only upon gravitational potential~$w$. Two options are possible here. (a)
Gravitational inertial force $F_i{=}0$ and space's rotation $v_i{=}0$. Then
according to definitions of $F_i$ and $v_i$ (see Section\thinspace 2) we
have $\sqrt{g_{00}}{=}1{-}\frac{\displaystyle w}{\displaystyle c^2}{=}1$ and 
$g_{0i}{=}{-}\frac{\displaystyle 1}{\displaystyle c}\sqrt{g_{00}}\,v_i{=}0$.
Equality to zero of gravitational potential $w$ means, in particular, that
it does not depend upon three-dimensional coordinates. In this case motion
of observer across space where no rotation is present leaves pace of
different clocks the same (global synchronization is preserved with time).
(b) If $F_i{\neq }0$ and $v_i{=}0$ then in the formula for $F_i$ (19) the
derivative $\frac{\displaystyle \partial w}{\displaystyle \partial x^i}{\neq 
}0$. That means that gravitational potential depends upon three-dimensional
coordinates, i.\thinspace e.\ pace of time is different at different points
of space. Hence at $F_i{\neq }0$ synchronization of clocks at different
point of space where no rotation is present does not preserve with time.

In a space where no rotation is present collapsed matter may exist (black
holes, $w{=}c^2$) only if $F_i{\neq }0$. If $F_i{=}0$ then according to
definition of $F_i$ (19) in space where no rotation is present $w{=}0$ and
gravitational collapse is not possible.

\medskip
\noindent {\bf 2. Space rotates at sub-light velocity}

\smallskip
\noindent Consequently in this case spatial sections are not orthogonal to
lines of time $v_i{=}{-}ce_{(i)}\cos \alpha {\neq }0$. Because ${-}1{\leq }%
\cos \alpha {\leq }{+}1$, then ${-}c{\leq }v_i{\leq }{+}c$. Hence $v_i{>}0$
at $\cos \alpha {>}0$ and $v_i{<}0$ at $\cos \alpha {<}0$.

\medskip
\noindent {\bf 3. Space rotates at light velocity (first case)}

\smallskip
\noindent The lesser is $\alpha $ the greater is $v_i$. In the ultimate case
when $\alpha {=}0$ velocity of space rotation $v_i{=}{-}c$. Consequently
spatial basic vectors $\vec e_{(i)}$ coincide with temporal basic vector $%
\vec e_{(0)}$ (space coincides with time).

\medskip
\noindent {\bf 4. Space rotates at light velocity (second case)}

\smallskip
\noindent If $\alpha {=}\pi $, then $v_i{=}{+}c$ and temporal basic vector $%
\vec e_{(0)}$ also coincides with spatial basic vectors $\vec e_{(i)}$ but
is oppositely directed. The case may be interpreted as space that coincides
with ``anti-time'' which flows from future into past.


\section{Motion of particles as a result of space's motion}

\subsection{Problem statement}

\label{10-1}
Having substituted gravitational potential $w$ and velocity of space
rotation $v_i$ into definition of interval of observable time $d\tau $ (13),
we obtain the expression (13) as

\vspace*{-5pt}%
$$
\left( 1+\frac 1{c^2}v_i{\rm v}^i\right) d\tau =\left( 1-\frac w{c^2}\right)
dt\,.\eqno (177) 
$$

\vspace*{-1pt}From here we see that significant difference between $d\tau $
and $dt$ may result from either strong gravitational field or velocities
comparable to speed of light. Hence in everyday life the difference between
observable time $\tau $ and coordinate time $t$ is not great.

Physical observable time coincides with the time coordinate $dt{=}d\tau $
only under the condition

\vspace*{-5pt}%
$$
w=-v_i{\rm v}^i.\eqno (178) 
$$

\vspace*{1pt}Actually such condition implies that attraction of particle by
the central body is fully compensated by rotation of space and rotation of
the particle itself. That is (178) is mathematical interpretation of the 
{\it weightlessness condition}. Having gravitational potential substituted
according to Newton' formula we obtain

\vspace*{-14pt}%
$$
\frac{GM}r=v_i{\rm v}^i.\eqno (179) 
$$

\vspace*{-1pt}
If orbital velocity of particle equals to velocity of rotation of the
central body in this orbit the condition of weightlessness for the particle
becomes

\vspace*{-5pt}%
$$
\frac{GM}r=v^2,\eqno (180) 
$$

\vspace*{-1pt}
\noindent i.\thinspace e.\ the more distant is the orbit from the central
body, the lesser is the velocity of a satellite in this~orbit.

Is this statement confirmed by experimental data? The Table in the below
gives orbital velocities of the Moon and the planets s measured in
astronomical observations as well as those calculated from the condition of
weightlessness.

\vspace*{-3pt}
\begin{center}
{\small 
\begin{tabular}{l|r@{.}l|r@{.}l} \hline
  \multicolumn{1}{l|}{\raisebox{-7pt}{\bf Planet}} &
  \multicolumn{4}{|c}{\raisebox{-2pt}{\bf Orbital velocity, km/s}} \\ \cline{2-5}
  \multicolumn{1}{l|}{} & \multicolumn{2}{|l|}{\footnotesize\bf Measured} &
  \multicolumn{2}{|c}{\footnotesize\bf Calculated\rule{0pt}{9pt}} \\ \hline  
  Mercury\rule{0pt}{10pt} &  ~\quad 47 &  9  & ~\quad 47 &  9   \\
  Venus   &  35 &  0  &  35 &  0   \\
  Earth   &  29 &  8  &  29 &  8   \\
  Mars    &  24 &  1  &  24 &  1   \\
  Jupiter &  13 &  1  &  13 &  1   \\
  Saturn  &   9 &  6  &   9 &  6   \\
  Uran    &   6 &  8  &   6 &  8   \\
  Neptune &   5 &  4  &   5 &  4   \\
  Pluto   &   4 &  7  &   4 &  7   \\
  Moon    &   1 &  0  &   1 &  0   \\ \hline
\end{tabular}}
\end{center}

\vspace{-2pt} \noindent From the Table we see that our condition of
weightlessness is true for any satellite that orbits a central body.
Noteworthy, the condition is true when orbital velocity of a planet equals
to (or is close to) the velocity of rotation of the space of the central
body in this orbit (180). That means that rotating space of the central body 
{\it carries} any and all bodies around making them rotate.

If the space of the central body rotated like a solid body i.\thinspace e.\
without any deformation its angular velocity would be constant $\omega {=}%
const$ while orbital velocities ${\rm v}{=}\omega r$ of the carried
satellites would grow along with radiuses of their orbits. But as we have
just seen from the example of the Solar system planets, rotation velocity
declines along with distance from the Sun. That means that in reality space
of the central body (space of reference) does not rotate like a solid body,
but rather like a viscous and deformable medium, where layers distant from
the center do not rotate as far as those closer to the center. As a result
space of the central body is {\it twisted} and the profile of orbital
velocities simply repeats the structure of twisted space.

Hence we see that orbital motion of particles in gravitational field results
from {\it rotation of the space of the attracting body itself}.

What are possible consequences for mathematical theory of motion of
particles of the conclusions we have just made? We are going to see that in
the below.

Let us assume a metric space. Evidently, motion of the space allows to match
any of its points to vector of motion of such point $Q^\alpha $. It is also
evident that all points of the space will experience the same motion as the
space itself. Hence  $Q^\alpha $ can be regarded the vector of
motion of the space itself (in a given point). As a result we obtain a
vector field that describes motion of the whole space.

Of course if length of vector $Q^\alpha $ is constant the space will move so
that its metric will stay the same too. Hence if in such space vector of
motion $Q^\alpha $ is set in a given point then metric of the space can be
found processing from motion of the point (along with motion of the space)

A way to solve the problem was paved in the late 19th century by
S.\thinspace Lie \cite{bib12}. He obtained equations of derivative from space metric 
$g_{\alpha \beta }$ to trajectory of motion of vector $Q^\alpha $, which
contained components of $Q^\alpha $ as fixed coefficients. The number of the
equations as qual to the number of components of the metric. Hence having
vector $Q^\alpha $ fixed, i.\thinspace e.\ having motion of the space set,
we can solve the equations to find all components of metric $g_{\alpha \beta
}$ proceeding from the components $Q^\alpha $.

Later Van~Danzig suggested to call such derivative of metric {\it Lie
derivative}.

Now we are going to look at a partial case of motion of space which leaves
its metric constant. The case was studied by W.\thinspace Killing \cite{bib13}.
Evidently such motion will make Lie derivative equal to zero ({\it Killing
equations}). Hence if motion of space leaves its metric the same and we know
vector $Q^\alpha $ for any of its points (i.\thinspace e.\ motion of the
space at this point is set), motion of the point can be used to obtain
metric of the space from Killing equations.

On the other hand motion of particles is described by dynamic equations of
motion. To the contrary, these equations leave metric of space fixed and the
problem here is to find dynamic vector of motion of particle $Q^\alpha $.
Fixed metric in dynamic equations of motion makes Christoffel symbols, which
are functions of metric components $g_{\alpha \beta }$, appear in the
equations as fixed coefficients. Hence as soon as particular metric of space
is set we can use dynamic equations of motion to obtain vector $Q^\alpha $
for the particle in such space.

Therefore we arrive to the following. Because $g_{\alpha \beta }$ is a
symmetric tensor ($g_{\alpha \beta }{=}g_{\beta \alpha }$) only 10 its
components out of 16 have different values. In Killing equations (10
equations) vector of motion of a point in space is fixed, while components
of the metric are unknown (10 unknowns). Dynamic equations of motion of free
particles (4 equations), to the contrary, leave metric fixed, but components
of vector of motion of particles (4 components) are unknown. Then as soon as
we look at free motion of particle as motion of the point in space carried
by motion of the space itself, we can make a system of 10 Killing equations
(equations of motion of space) and of 4 dynamic equations of motion of
particle. The system of 14 equations will have 14 unknowns, 10 out of which
are unknown components of metric and 4 are unknown components of dynamic
vector of particle. Hence having this system solved we obtain motion of
particle in space and metric of the space at the same time.

In particular, solving the system we can find motions of particles which
result from motion of space itself. For such type of motion knowledge of
motion of a certain particle can produce metric of space.

For instance, having Killing equations and dynamic equations of motion
solved for a satellite (or a planet) we can use its motion to find metric of
the space of the central body.

In the next Section we are going to obtain Killing equations in
chronometrically invariant form.

\subsection{Equations of motion and Killing equations}

\label{10-2}
Let us assume a space (not necessarily a metric one) that moves. Evidently
vector of motion of any point of the space $Q^\alpha $ is vector of motion
of the space itself at this point. Motion of space is described by {\it Lie
derivative}

\vspace*{-10pt}%
$$
\stackunder{L}{\delta }g_{\alpha \beta }=Q^\alpha \frac{\partial g_{\alpha
\beta }}{\partial x^\sigma }+g_{\alpha \sigma }\frac{\partial Q^\sigma }{%
\partial x^\beta }+g_{\beta \sigma }\frac{\partial Q^\sigma }{\partial
x^\alpha }\,,\eqno (181) 
$$

\noindent which is derivative of the metric of the space to
direction of motion of vector $Q^\alpha $ (direction of motion of the space
itself).

We are now looking to the picture as follows. We assume a point in space. If
space moves the point will be subjected to action of carrying vector $%
Q^\alpha $ which is vector of motion of the space itself. For the point
itself the space will rest and only ``the wind'' produced by motion of space
as vector $Q^\alpha $ will disclose motion of the whole space.

Generally Lie derivative is not zero, that is motion of space alters its
metric. But in Riemannian space metric is fixed by definition and length of
vector being transferred parallel to itself is constant. That means that
parallel transfer of vector across ``non-smooth'' structure of Riemannian
space will alter the vector along with configuration of the space. As a
result Lie derivative of metric in Riemannian space will be zero

\vspace*{-10pt}%
$$
\stackunder{L}{\delta }g_{\alpha \beta }=0\,.\eqno (182) 
$$

Lie equations in Riemannian space were first studied by Killing and are
referred to as Killing equations. Later A.\thinspace Z.\thinspace Petrov
showed \cite{bib14} that Killing equations for any point are the necessary and
sufficient condition for the motion of the point to be motion of Riemannian
space itself. In other words if a point is carried by motion of Riemannian
space and moves along, Killing equations must be true for that point.

Evidently to obtain components of metric out of Killing equations we need to
set a particular vector of motion of a point $Q^\alpha $. Then we will have
10 Killing equations vs.\ 10 unknown metric components and will able to
solve the system.

Generally there might be different kinds of motion of Riemannian space. We
will set vector of motion $Q^\alpha $ as to fit the needs of our problem.

There exists {\it free} ({\it geodesic}) {\it motion} in which a point moves
along geodesic (the shortest) trajectory. We assume that any point of
Riemannian space carried by motion of the space itself moves along geodesic
trajectory. Hence motion of the entire Riemannian space will be geodesic as
well. Then we can match motion of a point carried by motion of space to
motion of a free particle.

We will call a motion {\it geodesic motion of space} if free motion of
particles results from their carrying by moving space.

Let us look at a system of dynamic equations of motion of free particles and
Killing equations

\vspace*{-3pt}%
$$
\left. 
\begin{array}{l}
\displaystyle \frac{\displaystyle DQ^\alpha }{\displaystyle d\rho }=0 \\ 
\stackunder{L}{\delta }g_{\alpha \beta }=0 
\end{array}
\right\} \,,\eqno (183) 
$$

\vspace*{-1pt}\noindent where $Q^\alpha $ stands for dynamic vector of
motion of particle, $\rho $ stands for derivation parameter to trajectory of
motion while Lie derivative can be expressed through {\it Lie differential}
as

\vspace*{-5pt}%
$$
\stackunder{L}{\delta }g_{\alpha \beta }=\frac{\stackunder{L}{D}g_{\alpha
\beta }}{d\rho }\,.\eqno (184) 
$$

\vspace*{-1pt}Actually this system of equations means that motion of free
particle is geodesic one (equations of motion of particles) and at the same
time results from carrying of particle by geodesic motion of the space
(Killing equations). The system solves as a set of components of the dynamic
vector $Q^\alpha $ and components of metric $g_{\alpha \beta }$ for which
geodesic motion of particles results from geodesic motion of the space
itself.

To solve the problem correctly we need to present Killing equations in
chronometrically invariant form thus presenting them through physical
properties of space (standards). It is especially interesting to know which
of physical standards result from motion of space itself.

According to theory of chronometric invariants (12) physical observable
values are projection of Killing equation on time (1 component),
mixed projection (3 components) and spatial projection (6
components)

\vspace*{-11pt}%
$$
\frac{\stackunder{L}{\delta }g_{00}}{g_{00}}=0\,,\eqno (185) 
$$

\vspace*{-3pt}%
$$
\frac{\stackunder{L}{\delta }g_0^i}{\sqrt{g_{00}}}=\frac{g^{i\alpha }%
\stackunder{L}{\delta }g_{0\alpha }}{\sqrt{g_{00}}}=0\,,\eqno (186) 
$$

$$
\stackunder{L}{\delta }g^{ik}=g^{i\alpha }g^{k\beta }\stackunder{L}{\delta }%
g_{\alpha \beta }=0\,.\eqno (187) 
$$

Here we are looking at motion of space and particles from the
viewpoint of a regular sub-light-speed observer.

Having presented derivatives of metric in Lie derivative through
chronometric invariant operators and having substituted short
notation of observable components of dynamic vector of particle $Q^\alpha $
as $\varphi {=}\frac{\displaystyle Q_0}{\displaystyle\sqrt{g_{00}}}$ and $q^i%
{=}Q^i$, we arrive to {\it chronometrically invariant Killing equations}

$$
\frac{^{*}\partial \varphi }{\partial t}-\frac 1cF_iq^i=0\,,\eqno (188) 
$$

\vspace*{-10pt}%
$$
\frac 1c\frac{^{*}\partial q^i}{\partial t}-h^{im}\frac{^{*}\partial \varphi 
}{\partial x^m}-\frac \varphi {c^2}F^i+\frac 2cA_{k\cdot }^{\cdot i}q^k=0\,,%
\eqno (189) 
$$

\vspace*{-3pt}%
$$
\frac{2\varphi }cD^{ik}+h^{im}h^{kn}q^l\frac{^{*}\partial h_{mn}}{\partial
x^l}+h^{im}\frac{^{*}\partial q^k}{\partial x^m}+h^{km}\frac{^{*}\partial q^i%
}{\partial x^m}=0\,.\eqno (190) 
$$

\vspace*{1pt}
If vector $Q^\alpha $ at the same time complies
chronometrically invariant Killing equations and chronometrically invariant
dynamic equations of motion of particle, then such particles moves being
carried by geodesic motion of space.

Joint solution of the equations in general form is problematic and so we
will limit ourselves to a single partial case, which is still of great
importance. Let dynamic vector of motion of space $Q^\alpha $ be dynamic
vector of motion of mass-bearing particles

\vspace*{-4pt}%
$$
Q^\alpha =m_0\frac{dx^\alpha }{ds}=\frac mc\frac{dx^\alpha }{d\tau }\,,\eqno %
(191) 
$$

\noindent and let observer accompany the particle (${\rm v}^i{=}0$). Then

\vspace*{-3pt}%
$$
\varphi =m_0=const,\qquad q^i=\frac mc{\rm v}^i,\eqno (192) 
$$

\vspace*{-2pt}
\noindent and Killing equations (188--190) are simplified to

\vspace*{-6pt}%
$$
F^i=0\,,\qquad \qquad D^{ik}=0\,.\eqno (193) 
$$

\vspace*{-1pt}
According to (20) $D^{ik}{=}0$ means stationary state of observable metric $%
h^{ik}{=}const$. The condition $F^i{=}0$ means that for the following
equalities to become only true by transformation of time coordinate

\vspace*{-5pt}%
$$
g_{00}=1\,,\qquad \qquad \frac{\partial g_{0i}}{\partial t}=0\,.\eqno (194) 
$$

\vspace*{-2pt}Besides, $F^i$ and $A_{ik}$ are related through Zelmanov
identity \cite{bib10}

\vspace*{-3pt}%
$$
\frac 12\left( \frac{^{*}\partial F_k}{\partial x^i}-\frac{^{*}\partial F_i}{%
\partial x^k}\right) +\frac{^{*}\partial A_{ik}}{\partial t}=0\,,\eqno (195) 
$$

\vspace*{-3pt}\noindent from which we see that $F^i{=}0$ means

\vspace*{-12pt}%
$$
\frac{^{*}\partial A_{ik}}{\partial t}=0\,,\eqno (196) 
$$

\vspace*{-2pt}\noindent i.\thinspace e.\ rotation of space of reference is 
{\it stationary}.

Further, as seen from Killing equations (193) tensor of deformation
velocities of space of reference is zero, hence {\it stationary rotation}
does not alter structure of space. Equality to zero of gravitational
inertial force in Killing equations means that from the viewpoint of carried
particle (${\rm v}^i{=}0$) it weighs nothing and is not attracted to
anything (weightlessness state). This does not contradict with the
weightlessness condition $w{=}{-}v_i{\rm v}^i$ obtained earlier because from
the viewpoint of carried particle gravitational potential of the body of
reference $w{=}0$ and $F^i{=}0$ as well.

Hence if $Q^\alpha $ is a vector of motion of mass-bearing particles, then 
{\it geodesic motion of space} along this vector is {\it stationary rotation}%
.

As we see geodesic motion of mass-bearing particles is stationary rotation.
And such stationary rotation results from carrying by the space of the
gravitating body. But we know that the basic type of motion in the Universe
is orbiting. Hence the basic motion in the Universe is a geodesic motion
which results from carrying of objects by stationary (geodesic) rotation of
spaces of their central bodies.

\subsection{Calculating mass of the Galaxy}

\label{10-3}
We will proceed from the fact that orbital motion of the Sun around the
center of the Galaxy complies with the condition of weightlessness. If space
of our Galaxy {\it fully} carries the Sun then the orbital velocity of the
Sun ${\rm v}^i$ will be equal to velocity of rotation of the Galactic space $%
v_i$ in the same orbit. Hence the condition of weightlessness for the Sun
will be quite simple (180). After substituting distance from the center of
our Galaxy to the Sun $r{=}3{\cdot }10^{22}$\thinspace cm and orbital
velocity of the Sun 250\thinspace km/s we obtain the mass of our Galaxy

\newpage
\vspace*{-14pt}%
$$
M=\frac{rv_i{\rm v}^i}G=\frac{rv^2}G=\text{2.8}\cdot \text{10}^{44}\text{%
\thinspace g}.\eqno (197) 
$$

\vspace*{-1pt}According to astronomical data the mass of our Galaxy is $3{%
\cdot }10^{44}$\thinspace g (within the sphere of 15,000 parsec). The true
mass of our Galaxy is a bit over the value. A small difference from the
figure 2.8${\cdot }10^{44}$\thinspace g we have obtained can be explained by
the fact that the Sun is {\it not fully} carried by rotation of the Galactic
space. Therefore it seems reasonable to introduce so-called {\it coefficient
of carrying} 

\vspace*{-3pt}%
$$
k=\frac{{\rm v}}v\,,\eqno (198) 
$$

\vspace*{-5pt}\noindent and the condition of weightlessness becomes

\vspace*{-8pt}%
$$
\frac{GM}r=\frac{{\rm v}^2}k\,.\eqno (199) 
$$

\vspace*{-2pt}Having substituted the observed mass of the Galaxy $3{\cdot }%
10^{44}$\thinspace g into here, we obtain the coefficient of carrying of the
Sun by rotation of the Galactic space (because we have substituted the least
estimate $3{\cdot }10^{44}$\thinspace g the true value of the coefficient
should be even lower)

\vspace*{-6pt}%
$$
k=\frac{r{\rm v}^2}{GM}=\text{0.94}\,.\eqno (200) 
$$

\subsection{Resume}

\label{10-4}
So what is the space if it bears gravitational potential $w$ can be deformed
and in rotation behaves like viscous media? Noteworthy, if we place a
particle into the space, the moving space will carry it just like current in
the ocean carries a tiny boat and a giant iceberg.

The answer is as follows: according to the results we have obtained in the
above, the space of reference of a body and its gravitational field are the
same thing. And physically speaking, points of the space of reference are
particles of gravitational field of the body of reference.

If the space of reference does not rotate, a satellite will fall on the body
of reference under action of gravitational force. But in presence of
rotation the satellite will be under action of carrying force. The force
will act like wind of oceanic stream to push the satellite forward, not
allowing it to fall down and making it orbit the central body along with the
rotating space (of course additional velocity given to the satellite will
make it move faster than the rotating space).


\section{Who is a super-light observer?}

\label{11}
We can outline a few types of frames of reference which may
exist in General Relativity space-time. Particles (including the observer
themselves), which travel at sub-light speed (``inside'' the light cone),
bear real relativistic mass. In other words, the particles, the body of
reference and the observer are in the state of matter commonly referred to
as ``substance''. Therefore any observer whose frame of reference is
described by such monad will be referred to as {\it sub-light speed }({\it %
substantional}){\it \ observer}.

Particles and the observer that travel at the speed of light (i.\thinspace
e.\ over the surface of light hypercone) bear $m_0{=}0$ but their
relativistic mass (mass of motion) $m{\neq }0$. They are in light-like state
of matter. Hence we will call an observer whose frame of reference is
characterized by such monad a {\it light-like observer.}

Accordingly, we will call particles and the observer that travel at
super-light speed {\it super-light} particles and observer. They are in the
state of matter for which $m_0{\neq }0$ but their relativistic mass is
imaginary.

It is intuitively clear who a sub-light speed observer is, the term requires
no further explanations. Same more or less applies to light-like observer.
From point of view of light-like observer the world around looks like
colorful system of light waves. But who is a super-light observer? To
understand this let us give an example.

Imagine a new supersonic jet plane to be commissioned into operation. All
members of the commission are inborn blind. And so is the pilot. Thus we may
assume that all information about the surrounding world the pilot and the
members of the commission gain from sound, that is from transversal waves in
air. It is sound waves that build a picture that those people will perceive
as their ``real world''.

Now the plane took off and began to accelerate. As long as its speed is less
than the speed of sound, the blind members of the commission will match its
``heard'' position in the sky to the one we can see. But once the sound
barrier is overcome, everything changes. Blind members of the commission
will still perceive the speed of the plane equal to the speed of sound
regardless to its real speed. For the speed of propagation of sound waves in
the air will be the {\it maximum speed of propagation of information} while
the real supersonic jet plane will be beyond their ``real world'' in the
world of ``imaginary objects'' and all its properties will be imaginary too.
The blind pilot will hear nothing as well. Not a single sound will reach him
from the past reality and only local sounds from the cockpit (which also
travels at the supersonic speed) will break the silence. Once the speed of
sound is overcome, the blind pilot leaves the subsonic world for a new
supersonic one. From his new viewpoint (supersonic frame of reference) the
old subsonic fixed world tat contains the airport and the members of the
commission will simply disappear to become an area of ``imaginary values''.

What is light? Transversal waves that run across a certain medium at a
constant speed. We perceive the world around through sight, receiving light
waves from other objects. It is waves of light that build our picture of the
``true real world''.

Now imagine a spaceship that accelerates faster and faster to eventually
overcome the light barrier at still growing speed. From pure mathematical
viewpoint this is quite possible in the space-time of General Relativity.
For us the speed of the spaceship will be still equal to the speed of light
whatever is its real speed. For us the speed of light will be the maximum
speed of propagation of information and the real spaceship for us will stay
in another ``unreal'' world of super-light speeds where all properties are
imaginary. The same is true for the spaceship's pilot. From his viewpoint
having the light barrier overcome brings him into a new super-light world
that becomes his ``true reality''. And the old world of sub-light speeds is
gone to the area of ``imaginary reality''.


\section{World of black holes}

\label{12}
We will call {\it black hole} (gravitational collapser) an area of
space-time where the condition $g_{00}{=}0$ is true \cite{bib10,bib11}. Because
according to the theory of chronometric invariants $\sqrt{g_{00}}\,{=}1{-}%
\frac{\displaystyle w}{\displaystyle c^2}$, the condition of collapse $g_{00}%
{=}0$ also implies $w{=}c^2$. We will look at black holes from outside, from
viewpoint of a regular observer who stays above the surface of collapser.

We put down the formula for four-dimensional interval so that it contains an
explicit ratio of $w$ and $c^2$

\vspace*{-15pt}%
$$
ds^2=\left( 1-\frac w{c^2}\right) ^2c^2dt^2-2\left( 1-\frac w{c^2}\right)
v_idx^idt+g_{ik}dx^idx^k.\eqno (201) 
$$

\vspace*{-3pt}Having substituted $w{=}c^2$ into here we obtain the metric on
the surface of black hole

\vspace*{-5pt}%
$$
ds^2=g_{ik}dx^idx^k.\eqno (202) 
$$

From here we see that the collapse in four-dimensional
space-time can be correctly defined only if three-dimensional space does not
rotate i.\thinspace e.\ is holonomic.

As a matter of fact the denominator of velocity of space rotation

\vspace*{-5pt}%
$$
v_i=-c\frac{g_{0i}}{\sqrt{g_{00}}}=-c\frac{g_{0i}}{1-\displaystyle \frac{%
\displaystyle w}{\displaystyle c^2}}\eqno (203) 
$$

\vspace*{-5pt}\noindent turns zero in case of collapse ($w{=}c^2$) and the
rotation velocity becomes infinite. To avoid this we assume $g_{0i}{=}0$.
Then metric (201) becomes

\vspace*{-5pt}%
$$
ds^2=\left( 1-\frac w{c^2}\right) ^2c^2dt^2+g_{ik}dx^idx^k,\eqno (204) 
$$

\noindent and the problem of peculiar state of space will be automatically
removed. Proceeding from this the metric on the surface of black hole (202)
is

\vspace*{-2pt}%
$$
ds^2=-d\sigma ^2=-h_{ik}dx^idx^k,\qquad \quad h_{ik}=-g_{ik}\,.\eqno (205) 
$$

From here we see that four-dimensional potential on the surface of
gravitational collapser is space-like --- elementary distance between two
point on the surface of black hole is imaginary

\vspace*{-5pt}%
$$
ds=id\sigma =i\sqrt{h_{ik}dx^idx^k}\,.\eqno (206) 
$$

\vspace*{1pt}
If $ds{=}0$ three-dimensional observable distance $d\sigma $ between two
points on the surface of collapser also becomes zero. Because in absence of
rotation of space the interval of observable space equals\linebreak 
$d\tau {=\!}\sqrt{g_{00}}%
\,dt{=}\!\left( 1{-}\frac{\displaystyle w}{\displaystyle c^2}\right) \!dt$,
on the surface of collapser $d\tau {=}0$ (observable time stops).

Now we are going\rule{0pt}{8pt} to look at collapse in different areas of 
four-dimensional space-time.

\medskip
\noindent {\bf 1. Collapse in sub-light area}

\smallskip
\noindent
Within this area $ds^2{>}0$. This is the habitat of regular real particles
that travel at sub-light speeds. Hence collapser in this area consists of
collapsed substance ({\it substantional black hole}). On the surface of such
collapser metric is space-like: here $ds^2{<}0$ and particles bear imaginary
relativistic masses. Of course metric on the surface of real black hole is
not degenerated.

\medskip
\noindent {\bf 2. Collapse in light-like area}

\smallskip
\noindent
Within this area $ds^2{=}0$. This is isotropic space of light-like
(massless) particles. Collapser in this area is made of collapsed light-like
matter ({\it light-like black hole}). Metric (205) on its surface is $%
d\sigma ^2{=}{-}g_{ik}dx^idx^k{=}0$. This can be true provided that: (a)
surface of light-like collapser shrinks to a point (all $dx^i{=}0$), or (b)
three-dimensional spatial metric is degenerated $g_{(3D)}{=}\det ||g_{ik}||{=%
}0$ (because four-dimensional metric is degenerated too light-like collapser
is zero-space in this case).

\medskip
\noindent {\bf 3. Collapse in degenerated space-time (zero-space)}

\smallskip
\noindent
Degenerated matter of zero-space can collapse too. We will call such
collapsers {\it degenerated black holes}. As a matter of fact, from the
condition of degeneration of space-time

\vspace*{-5pt}%
$$
w+v_iu^i=c^2,\qquad \quad g_{ik}dx^idx^k=\left( 1-\frac w{c^2}\right)
^2c^2dt^2,\eqno (207) 
$$

\vspace*{-1pt}\noindent we see that in case of collapse ($w{=}c^2$)

\vspace*{-5pt}%
$$
v_iu^i=0\,,\qquad \quad g_{ik}dx^idx^k=0\,.\eqno (208) 
$$

\vspace*{1pt}
Hence collapse in zero-space also occurs in absence of rotation ($v_i{=}0$).
And because conditions (208) are true at the same time the surface of
degenerated black hole is shrunk into a point.


\section{Geometric structure of zero-space}

\label{13}
Regular real observer perceives zero-space as an area defined
by the observableconditions $d\tau {=}0$ and $d\sigma ^2{=}h_{ik}dx^idx^k{=}0$.

Physical sense of the first condition $d\tau {=}0$ is that real observer
perceives any two events in zero-space as simultaneous, whatever distant
from them they are. Such way of instantaneous spread of information is
referred to as {\it long-range action}.

The second condition $d\sigma ^2{=}0$ means absence of observable distance
between the event and the observer. Such ``superposition'' of observer and
observed object is only possible if we assume that our regular four-%
dimensional pseudo-Riemannian space is ``stuffed'' with
zero-space.

Let us now turn to mathematical interpretation of degeneration conditions.

The value $cd\tau $ is a projection of four-dimensional interval $dx^\alpha $
onto time $cd\tau {=}b_\alpha dx^\alpha $. Monad vector of the observer $%
b^\alpha $ by definition is not zero and $dx^\alpha $ are not zeroes too.
Then $d\tau {=}0$ is true at $d\sigma ^2{=}0$ only if four-dimensional
metric $d\sigma ^2{=}c^2d\tau ^2{-}d\sigma ^2{=}g_{\alpha \beta }dx^\alpha
dx^\beta $ is degenerated, i.\thinspace e.\ determinant of fundamental
metric tensor is zero

\vspace*{-9pt}%
$$
g=\det \left\| g_{\alpha \beta }\right\| =0\,.\eqno (209) 
$$

\vspace*{2pt}Similarly the condition $d\sigma ^2{=}h_{ik}dx^idx^k{=}0$ means
that in that area three-dimensional observable metric is degenerated too

\vspace*{-9pt}%
$$
h=\det \left\| h_{ik}\right\| =0\,.\eqno (210) 
$$

Having the conditions of degeneration of space-time $w{+}%
v_iu^i{=}c^2$\ and $g_{ik}dx^idx^k{=}\!\left( 1{-}\frac{\displaystyle w}{%
\displaystyle c^2}\right) ^2\!c^2dt^2$ substituted into $ds^2$ we obtain
formula for the metric of zero-space\rule{0pt}{8pt}

\vspace*{-4pt}%
$$
ds^2=\left( 1-\frac w{c^2}\right) ^2c^2dt^2-g_{ik}dx^idx^k=0\,.\eqno (211) 
$$

Hence inside zero-space three-dimensional space is holonomic while rotation
is present in the temporal component of its metric

\vspace*{-9pt}%
$$
\left( 1-\frac w{c^2}\right) ^2c^2dt^2=\left( \frac{v_iu^i}{c^2}\right)
^2c^2dt^2.\eqno (212) 
$$

\vspace*{-3pt}If $w{=}c^2$ (gravitational collapse) metric of zero-space
(211) becomes

\vspace*{-5pt}%
$$
ds^2=-g_{ik}dx^idx^k=0\,,\eqno (213) 
$$

\vspace*{-1pt}
\noindent i.\thinspace e.\ becomes purely spatial. And
three-dimensional space becomes degenerated too

\vspace*{-5pt}%
$$
g_{(3D)}=\det \left\| g_{ik}\right\| =0\,.\eqno (214) 
$$

\vspace*{1pt}
Here the condition $g_{(3D)}{=}0$ is obtained from the fact
that quadratic form $g_{ik}dx^idx^k$ is sign-definite and can only become
zero provided the determinant of metric tensor $g_{ik}$ equals to zero.

Because in zero-space $w{+}v_iu^i{=}c^2$ in case of gravitational collapse
the condition $v_iu^i{=}0$ also becomes true.

The value $v_iu^i{=}vu\cos (v_i;u^i)$, which is scalar product of velocity
of space rotation and coordinate velocity of particle will be referred to as 
{\it spirality} of zero-particle.

If $v_iu^i{>}0$\rule{0pt}{10pt} then the angle between $v_i$ and $u^i$
is within the range
of $\frac{3\pi }2{<}\alpha {<}\frac \pi 2$. Because the second condition of
degeneration of space-time $g_{ik}u^iu^k{=}\!\left( 1{-}\frac{\displaystyle w%
}{\displaystyle c^2}\right) ^2$ implies that $u{=}c\left( 1{-}\frac{%
\displaystyle w}{\displaystyle c^2}\right) $ then potential $w{<}c^2$
(regular gravitational field).

If $v_iu^i{<}0$\rule{0pt}{10pt} then $\alpha $ is within 
$\frac \pi 2{<}\alpha {<}\frac{3\pi 
}2$ and $w{>}c^2$ (superstrong gravitational field).

The condition $v_iu^i{=}0$\rule{0pt}{10pt} is only true when 
$\alpha {=}\frac \pi 2;\frac{%
3\pi }2$ or if $w{=}c^2$ (gravitational collapse).

Hence\rule{0pt}{10pt} spirality of zero-particle is zero if either velocity 
of particle is orthogonal to velocity of space rotation or gravitational 
collapse occurs (because in this case module of coordinate velocity of 
particle equals zero $u{=}0$).

Because $w{=}c^2(1{-}e_{(0)})$ and $v_i{=}{-}ce_{(i)}\cos (x^0;x^i)$ then
condition of space degeneration $w{+}v_iu^i{=}c^2$ becomes

\vspace*{-14pt}%
$$
ce_{(0)}=-e_{(i)}u^i\cos (x^0;x^i).\eqno (215) 
$$

Dimension of space is defined by the number of linearly
independent basic vectors. In our formula (215), which is basic notation of
the condition $w{+}v_iu^i{=}c^2$, temporal basic vector is linearly
dependent from all spatial basic vectors. This means degeneration of
space-time. Hence formula (215) can be regarded the geometric condition of
degeneration.

In case of gravitational collapse ($w{=}c^2$) the length of temporal basic
vector $e_{(0)}{=}1{-}\frac{\displaystyle w}{\displaystyle c^2}$ becomes
zero. In absence of gravitational field ($w{=}0$, or in weak gravitational
field $w{\rightarrow }0$) the value $e_{(0)}$ takes its maximum length equal
to one. In intermediate cases $e_{(0)}$ becomes shorter as gravitational
field becomes stronger.

As known in any point of four-dimensional space there exists so-called {\it %
isotropic cone} --- a hypersurface which metric is

\vspace*{-13pt}%
$$
g_{\alpha \beta }dx^\alpha dx^\beta =0\,.\eqno (216) 
$$

Geometrically this is an area of space-time that hosts light-like particles.
Because the square of interval in this area is zero, all directions inside
are equal, i.\thinspace e.\ are isotropic. Hence the area that hosts
light-like particles is commonly referred to as {\it isotropic cone} or the 
{\it light cone}.

Because in zero-space metric is also equal to zero (211) an isotropic cone
can be set in any of its points. But such will be described by a somewhat
different equation

\vspace*{-4pt}%
$$
\left( 1-\frac w{c^2}\right) ^2c^2dt^2-g_{ik}dx^idx^k=0\,.\eqno (217) 
$$

The difference between this isotropic cone and the light cone is that it
satisfies the condition

\vspace*{-6pt}%
$$
1-\frac w{c^2}=\frac{v_iu^i}{c^2}\,,\eqno (218) 
$$

\vspace*{-1pt}\noindent which is only typical for degenerated space-time
(zero-space). Hence we will call it {\it degenerated isotropic cone}. And
because the specific term (218) is the direct function of space rotation,
degenerated isotropic cone is a cone of rotation.

Under gravitational collapse ($w{=}c^2$) the first term in (217) becomes
zero (the point of stop of coordinate time), while the remaining second term 
$g_{ik}dx^idx^k{=}0$ describes three-dimensional degenerated hypersurface.
But if $w{=}0$ then $v_iu^i{=}0$ and the equation of degenerated isotropic
cone (217) becomes

\vspace*{-4pt}%
$$
c^2dt^2-g_{ik}dx^idx^k=0\,,\eqno (219) 
$$

\vspace*{-4pt}
\noindent i.\,e.\ coordinate time flows evenly.

The greater is gravitational potential $w$ the ``heavier'' is degenerated
cone and the closer it is to the spatial section. In the ultimate case when $%
w{=}c^2$ degenerated cone under action of its own ``super-weight'' becomes
flattened over three-dimensional space (collapses). The ``lightest'' cone $w{%
=}0$ is the most distant one from spatial section.

Hence black hole in zero-space is similar to zero-space observed by regular
real observer. In other words zero-space for us is degenerated state of
regular space-time, while for zero-observer black hole is degenerated state
of zero-space. That means that isotropic light cone contains degenerated
isotropic cone of zero-space, which in its turn contains collapsed
degenerated isotropic cone of the space inside black holes. This is
illustration of fractal structure of the world presented here as a system of
isotropic cones found inside each other.


\section{Conclusions}

\label{14}
Now we can build the general picture of motion of particles in
four-dimensional space-time.
Mass-bearing particles with positive relativistic masses $m{>}0$ inhabit our
world with direct flow of time (``inner'' part of the light cone) and move
from past into future in respect to a regular observer. Particles with
negative relativistic masses $m{<}0$ inhabit the mirror world and move from
future into past in respect to us. In the mirror world time has reverse flow
in respect to that in our world.

Inside the ``walls'' of the cone the condition $c^2d\tau ^2{=}d\sigma ^2$ is
true, i.\thinspace e.\ the temporal and the spatial projections of
four-dimensional coordinates are equal, while the space-time interval is
degenerated $ds^2{=}c^2d\tau ^2{-}d\sigma ^2{=}0$. This is the habitat of
light-like (massless) particles. Light-like particles of our world with
positive own frequencies $\omega {>}0$ move from pat into future. In the
mirror world light-like particles with negative own frequencies $\omega {<}0$
move from future into past in respect to us.

Inside the membrane, which separates our world from the mirror world, a more
strict condition is true $c^2d\tau ^2{=}d\sigma ^2{=}0$, i.\thinspace e.\
observable time is degenerated $d\tau {=}0$ and observable three-dimensional
metric is degenerated too $d\sigma ^2{=}h_{ik}dx^idx^k{=}0$. This is fully
degenerated space-time inhabited by zero-particles. Regular relativistic
mass of zero-particles is zero, but their gravitational rotational mass $M$
is not zero (60). Besides, within the wave-particle dual concept equation of
wave phase (eikonal) for zero particles is a standing wave equation (79). In
other words from viewpoint of regular observer zero-particles are light-like
standing waves (waves of ``stopped'' light).

As long as gravitational potential $w$ grows we ``descend'' into the funnel
of zero-space deeper and deeper until $w$ finally becomes equal to $c^2$ and
we found ourselves in a black hole (gravitational collapser). The surface of
a black hole in zero-space shrinks to a point $g_{ik}dx^idx^k{=}0$.

It is the matter of degenerated black holes (which contrary to regular black
holes are collapsed matter of zero-space) through which interaction between
our-world particles and mirror-world particles is possible.

There is another interesting fact. Zero-space can only exist in presence of
rotation under condition $w{+}v_iu^i{=}c^2$. In absence of rotation
zero-space always collapses, i.\thinspace e.\ black hole expands to occupy
the whole zero-space. If both gravitational field and rotation are absent
penetration inside the membrane becomes impossible and any connection
between our world and the mirror world is lost.

The picture we have built so far is a static one. To ``vitalize'' it we need
to look at the nature of the mechanism that ``moves the world''.

Motion of real particles is determined by their four-dimensional vector of
impulse $P^\alpha {=}m_0\frac{\displaystyle dx^\alpha }{\displaystyle ds}$,
which length is

\vspace*{-12pt}%
$$
\sqrt{P_\alpha P^\alpha }=\pm m_0\,.\eqno (220) 
$$

This says that real particle with positive rest-mass $m_0{>}0$, which exists
in regular world with direct flow of time, always has a dual (twin) particle
with rest-mass $m_0{<}0$, which exists in the mirror world. We can suppose
that particle with opposite mass ``charges'' attract each other in the same
manner as particles with opposite electric charges.

The condition (220) also means that vector $P^\alpha $ is dual by its
physical nature and falls apart into two parts, one of which stays in
regular world, while another stays in the mirror world. Opposite mass
``charges'' evolve and move ``towards'' each other, penetrate zero-space
producing standing waves there, and then get into a black hole (each one
from its own side). Once they meet in a zero-point ($t_0{=}0$) they
annihilate. Annihilation energy throws them away from the black hole first
into zero-space and then into regular world and the mirror world. Then the
whole process repeats again and again. Exit of particle into our world is 
{\it materialization}, its penetration into zero-space is {\it %
dematerialization}. Observed life span of particles is actually the period
of time between materialization and dematerialization.

The same is true for particles with zero relativistic masses (light-like
particles). Duality of their dynamic vector $K^\alpha {=}\frac{\displaystyle%
\omega }{\displaystyle c}\frac{\displaystyle dx^\alpha }{\displaystyle %
d\sigma }$ makes its projection onto time sign-alternating

\vspace*{-5pt}%
$$
\frac{K_0}{\sqrt{g_{00}}}=\pm \frac \omega c\,.\eqno (221) 
$$

\vspace*{-1pt}In general, such pure geometric viewpoint allows to see that
collapsed particles of black holes, regular particles of zero-space, as well
as particles of light-like and substantional worlds are a sequence of states
of the same space-time structure.\smallskip


\section*{Epilogue}

\label{epi}
In {\it Rainbow Far Away}, written by Arcady and Boris
Strugatsky over 30 years ago a character recalls that\dots

``Being a schoolboy he was surprised by the problem: move things across vast
spaces in no time. The goal was set to contradict any existing views of
absolute space, space-time, kappa-space\dots\ At that time they called it
`punch of Riemannian fold'. Later it would be dubbed `hyper-infiltration',
`sigma-infiltration', or `zero-contraction'. At length it was named
zero-transportation or `zero-T' for short. This produced `zero-T-equipment',
`zero-T-problems', `zero-T-tester', `zero-T-physicist'.

--- What do you do?

--- I'm a zero-physicist.

A look full of surprise and admiration.

--- Excuse me, could you explain what zero-physics is? I don't understand a
bit of it.

--- Well\dots\ Neither I do''.

This passage might be a good afterword to our study. In the early 1960s
words like ``zero-space'' or ``zero-transportation'' sounded science-fiction
or at least something to be brought to life generations from now.

But science is progressing faster then we think it does. The results
obtained in this book suggest that the variety of existing particles and
types of their interaction is not limited to those known to contemporary
physics. We should expect that further advancement of experimental technique
and technology will discover zero-particles, which inhabit degenerated
space-time (zero-space) and can be observed as waves of ``stopped'' light.
From viewpoint of a regular observer zero-particles move instantly, thus
they can realize zero-transportation.

We think it's a mistake to believe that most Laws of Nature have been
already discovered by contemporary science. More likely we are just at the
very beginning of a long road to the Unknown~World.


{\small\label{ref}


\section*{Reference expressions}

\label{exp}
\noindent Components of the operator of projection on time (monad
$b^\alpha$) in an accompanying frame of reference ($b^i{=}0$)

\vspace*{-4pt}%
$$
\begin{array}{l}
b^0= 
\displaystyle \frac{\displaystyle 1}{\displaystyle \sqrt{g_{00}}}\,,\quad
b_0=g_{0\alpha }b^\alpha =\sqrt{g_{00}}\,,\quad b_i=g_{i\alpha }b^\alpha =%
\displaystyle \frac{\displaystyle g_{i0}}{\displaystyle \sqrt{g_{00}}}=-%
\displaystyle \frac{\displaystyle 1}{\displaystyle c}v_i\,, \\ \sqrt{g_{00}}%
=1-\displaystyle \frac{\displaystyle w}{\displaystyle c^2}\,,\quad \quad
v_i=-c\displaystyle \frac{\displaystyle g_{0i}}{\displaystyle \sqrt{g_{00}}}%
\,.\rule{0pt}{16pt} 
\end{array}
$$

\vspace*{-1pt}\noindent Components of observable metric tensor $h_{\alpha \beta }$
(the operator of projection on spatial section) in an accompanying frame of reference

\vspace*{-10pt}%
$$
\begin{array}{l}
h_{00}=h_{0i}=0\,,\qquad \quad h_{ik}=-g_{ik}+ 
\displaystyle \frac{\displaystyle g_{0i}g_{0k}}{\displaystyle g_{00}}%
=-g_{ik}+\displaystyle \frac{\displaystyle 1}{\displaystyle c^2}v_iv_k\,, \\ 
\rule{0pt}{24pt}h^{00}=-g^{00}+\displaystyle \frac{\displaystyle 1}{%
\displaystyle g_{00}}=-\displaystyle \frac{\displaystyle 1-\displaystyle 
\frac{\displaystyle 1}{\displaystyle c^2}v_iv^i}{\displaystyle 1-%
\displaystyle \frac{\displaystyle w}{\displaystyle c^2}}+\displaystyle \frac{%
\displaystyle 1}{\left( \displaystyle 1-\displaystyle \frac{\displaystyle w}{%
\displaystyle c^2}\right) ^2}\,, \\ h^{0i}=-g^{0i}= 
\displaystyle \frac{\displaystyle 1}{\displaystyle c\sqrt{g_{00}}}v^i,\qquad
\quad h^{ik}=-g^{ik}, \\ h_0^0=h_0^i=0\,,\rule{0pt}{18pt}\qquad h_i^0=%
\displaystyle \frac{\displaystyle g_{i0}}{\displaystyle g_{00}}=-%
\displaystyle \frac{\displaystyle 1}{c\left( \displaystyle 1-\displaystyle 
\frac{\displaystyle w}{\displaystyle c^2}\right) }v_i\,,\qquad
h_k^i=-g_k^i=\delta _k^i\,. 
\end{array}
$$

\vspace*{-2pt}\noindent Definitions of the velocity of rotation of
observer's space

\vspace*{-6pt}%
$$
v_i=-c\frac{g_{0i}}{\sqrt{g_{00}}}\,,\qquad v^i=-cg^{0i}\sqrt{g_{00}}\,,\qquad
v_i=h_{ik}v^k. 
$$

\vspace*{-1pt}\noindent Derivatives with respect to space-time interval and
observable time

\vspace*{-5pt}%
$$
\frac d{ds}=\frac 1{c\sqrt{1-{\rm v}^2/c^2}}\frac d{d\tau }\,,\qquad \quad
\frac d{d\tau }=\frac{^{*}\partial }{\partial t}+{\rm v}^k\frac{^{*}\partial 
}{\partial x^k}\,. 
$$

\vspace*{-4pt}\noindent Chronometrically invariant derivatives

\vspace*{-4pt}%
$$
\frac{^{*}\partial }{\partial t}=\frac 1{\sqrt{g_{00}}}\frac \partial
{\partial t}\,,\qquad \quad \frac{^{*}\partial }{\partial x^i}=\frac
\partial {\partial x^i}+\frac 1{c^2}v_i\frac{^{*}\partial }{\partial t}\,. 
$$

\noindent Relation between determinants of physical observable
metric tensor and fundamental metric tensor

\vspace*{-5pt}%
$$
\sqrt{-g\mathstrut}=\sqrt{h}\,\sqrt{\mathstrut g_{00}}\,. 
$$

\noindent Components of fundamental metric tensor expressed through
characteristics of an accompanying frame of reference

\vspace*{-13pt}%
$$
g_{00}=\left( 1-\displaystyle \frac{\displaystyle w}{\displaystyle c^2}%
\right) ^2,      \quad g^{00}= 
\displaystyle \frac{\displaystyle 1-\displaystyle \frac{\displaystyle 1}{%
\displaystyle c^2}v_iv^i}{\displaystyle 1-\displaystyle \frac{\displaystyle w%
}{\displaystyle c^2}}\,,   \quad  g_{0i}=-\displaystyle \frac{\displaystyle v_i}{%
\displaystyle c}\left( 1-\displaystyle \frac{\displaystyle w}{\displaystyle %
c^2}\right) ,  \quad g_{ik}=-h_{ik}+\displaystyle \frac{%
\displaystyle 1}{\displaystyle c^2}v_iv_k\,. 
$$

\noindent First and second Zelmanov identities

\vspace*{-5pt}%
$$
\begin{array}{l}
\displaystyle \frac{\displaystyle ^{*}\partial A_{ik}}{\displaystyle %
\partial t}+\displaystyle \frac{\displaystyle 1}{\displaystyle 2}\left( 
\displaystyle \frac{\displaystyle ^{*}\partial F_k}{\displaystyle \partial
x^i}-\displaystyle \frac{\displaystyle ^{*}\partial F_i}{\displaystyle %
\partial x^k}\right) =0\,, \\ \rule{0pt}{22pt}\displaystyle \frac{%
\displaystyle ^{*}\partial A_{km}}{\displaystyle \partial x^i}+\displaystyle 
\frac{\displaystyle ^{*}\partial A_{mi}}{\displaystyle \partial x^k}+%
\displaystyle \frac{\displaystyle ^{*}\partial A_{ik}}{\displaystyle %
\partial x^m}+\displaystyle \frac{\displaystyle 1}{\displaystyle 2}\left(
F_iA_{km}+F_kA_{mi}+F_mA_{ik}\right) =0\,. 
\end{array}
$$

\vspace*{3pt}
\noindent Zelmanov relations between regular Christoffel
symbols and chronometrically invariant values

\vspace*{-4pt}%
$$
\begin{array}{l}
D_k^i+A_{k\cdot }^{\cdot i}= 
\displaystyle \frac{\displaystyle c}{\displaystyle \sqrt{g_{00}}}\left(
\Gamma _{0k}^i-\displaystyle \frac{\displaystyle g_{0k}\Gamma _{00}^i}{%
\displaystyle g_{00}}\right) , \\ F^k=-\displaystyle \frac{\displaystyle %
c^2\Gamma _{00}^k}{\displaystyle g_{00}}\,,\qquad g^{i\alpha }g^{k\beta
}\Gamma _{\alpha \beta }^m=h^{iq}h^{ks}\triangle _{qs}^m\,.\rule{0pt}{23pt} 
\end{array}
$$

\bigskip\medskip
\centerline{\rule{72pt}{0.4pt}}

\end{document}